\documentclass[11pt,a4paper]{article}
\pdfoutput=1
\usepackage{jheppub}
\usepackage{graphicx}
\usepackage{amssymb}
\usepackage{amsmath}
\usepackage{epsfig}
\usepackage {longtable}
\usepackage{bm}
\usepackage{slashed}
\usepackage{url}

\renewcommand{\top}    {\ensuremath{\mathrm{t}}}

\renewcommand{\wp}     {\ensuremath{\mathrm{W^+}}}
\newcommand{\wm}       {\ensuremath{\mathrm{W^-}}}

\newcommand{\beq}{\begin{equation}}
\newcommand{\eeq}{\end{equation}}
\def\bsp#1\esp{\begin{split}#1\end{split}}
\def\bal#1\eal{\begin{align}#1\end{align}}
\newcommand{\beeq}{\begin{eqnarray}}
\newcommand{\eeeq}{\end{eqnarray}}

\newcommand{\lo}       {{\rm LO}}
\newcommand{\nlo}      {{\rm NLO}}

\newcommand{\lhe}      {{\rm LHE}}
\newcommand{\pythia}   {\texttt{PYTHIA}}
\newcommand{\py}       {\texttt{PY}}
\newcommand{\herwig}   {\texttt{HERWIG}}
\newcommand{\hw}       {\texttt{HW}}
\newcommand{\powhel}   {\texttt{PowHel}}
\newcommand{\powhegbox}{\texttt{POWHEG-BOX}}

\newcommand{\helacnlo} {\texttt{HELAC-NLO}}
\newcommand{\oneloop} {\texttt{OneLOop}}

\newcommand{\helaconeloop} {\texttt{HELAC-1LOOP}}
\newcommand{\helaconeloopdd} {\texttt{HELAC-1LOOP@dd}}
\newcommand{\decayer}   {\texttt{Decayer}}
\newcommand{\helacphegas} {\texttt{HELAC-Phegas}}

\newcommand{\qd}   {\texttt{QD}}
\newcommand{\cuttools}   {\texttt{CUTTOOLS}}
\newcommand{\mint}           {\texttt{MINT}}

\newcommand{\ud}       {\mathrm{d}}

\newcommand{\as}       {\alpha_{\mathrm{s}}}
\newcommand{\tE}       {\ensuremath{\tilde{E}}}
\newcommand{\ts}       {\ensuremath{\tilde{s}}}
\newcommand{\tx}       {\ensuremath{\tilde{x}}}
\newcommand{\pt}       {\ensuremath{p_\bot}}
\newcommand{\ptmin}    {\ensuremath{p_\bot^{\min}}}
\newcommand{\kt}       {\ensuremath{k_{\bot}}}
\newcommand{\pTmiss}{\ensuremath{\slash\hspace*{-5pt}{p}_{\perp}}}
\newcommand{\W}        {\ensuremath{W}}
\newcommand{\GeV}      {\ensuremath{{\mathrm{GeV}}}}
\newcommand{\mw}       {\ensuremath{m_{\mathrm{W}}}}
\newcommand{\mt}       {\ensuremath{m_{\mathrm{t}}}}

\newcommand{\ptj}      {\ensuremath{p_{\bot,\,j}}}
\newcommand{\ptl}      {\ensuremath{p_{\bot,\,\ell}}}
\newcommand{\ptbb}     {\ensuremath{p_{\bot,\,\bq_1\bq_2}}}
\newcommand{\rad}      {\mathrm{rad}}
\newcommand{\tbWp}     {\ensuremath{\mathrm{t\to b\,W^+}}}
\newcommand{\tbWm}     {\ensuremath{\mathrm{\bar{t}\to \bar{b}\,W^-}}}
\newcommand{\Wpln}     {\ensuremath{\mathrm{W^+\to\bar{\ell}\,}\nu}}
\newcommand{\Wmln}     {\ensuremath{\mathrm{W^-\to\ell\,}\bar{\nu}}}
\newcommand{\WWbB}     {\ensuremath{\mathrm{W^+\,W^-\,b\,\bar{b}}}}
\newcommand{\epmubB}   {\ensuremath{\mathrm{e^+\,\nu_e\,\mu^-\,\bar{\nu}_\mu\,b\,\bar{b}}}}
\newcommand{\qq}       {\ensuremath{\mathrm{q}}}
\newcommand{\qaq}      {\ensuremath{\mathrm{\bar{q}}}}
\newcommand{\bq}       {\ensuremath{\mathrm{b}}}
\newcommand{\baq}      {\ensuremath{\mathrm{\bar{b}}}}
\newcommand{\tq}       {\ensuremath{\mathrm{t}}}
\newcommand{\taq}      {\ensuremath{\mathrm{\bar{t}}}}
\newcommand{\bB}       {\ensuremath{\mathrm{b\,\bar{b}}}}
\newcommand{\tT}       {\ensuremath{\mathrm{t\,\bar{t}}}}
\newcommand{\tTbB}     {\ensuremath{\mathrm{t\,\bar{t}\,b\,\bar{b}}}}
\newcommand{\pp}       {\ensuremath{p\,p(\bar{p})}}
\newcommand{\tTX}      {\ensuremath{\mathrm{t\,\bar{t}\,X}}}
\newcommand{\ptb}      {\ensuremath{p_{\bot,\,\mathrm{b}}}}
\newcommand{\ptbbar}   {\ensuremath{p_{\bot,\,\mathrm{\bar{b}}}}}
\newcommand{\ptsupp}   {\ensuremath{p_{\bot,\,\mathrm{supp}}}}
\newcommand{\sme}[1]   {\ensuremath{|{\cal M}|^2(#1)}}
\newcommand{\Phiu}     {\ensuremath{\Phi_{\mathrm{u}}}}
\newcommand{\Phid}     {\ensuremath{\Phi_{\mathrm{d}}}}

\newcommand{\muf}      {\ensuremath{\mu_\mathrm{F}}}

\newcommand{\cm}       {\ensuremath{\mathrm{CM}}}

\newcommand{\tB}       {\ensuremath{\bar{B}}}
\newcommand\Ref[1]     {Ref.\,\cite{#1}}

\newcommand\eqn[1]     {Eq.\,(\ref{#1})}
\newcommand\eqns[2]    {Eqs.\,(\ref{#1}) and~(\ref{#2})}

\newcommand\fig[1]     {Fig.\,{\ref{#1}}}
\newcommand\figs[2]    {Figs.\,{\ref{#1}} and ~\ref{#2}}
\newcommand\figss[2]   {Figs.\,{\ref{#1}}--\ref{#2}}
\newcommand\sect[1]    {Sect.\,{\ref{#1}}}

\newcommand\tab[1]     {Table~\ref{#1}}

\title{Hadroproduction of \WWbB\ at NLO accuracy
matched with shower Monte Carlo programs}
\author[a]{Maria Vittoria Garzelli,}
\author[a]{Adam Kardos}
\author[a]{and Zolt\'an Tr\'ocs\'anyi}
\affiliation[a]{Institute of Physics and MTA-DE Particle Physics Research
Group, University of Debrecen,\\
H-4010 Debrecen P.O.Box 105, Hungary}

\emailAdd{garzelli@to.infn.it}
\emailAdd{kardos.adam@science.unideb.hu}
\emailAdd{Zoltan.Trocsanyi@cern.ch}

\abstract{We present the computation of the differential cross section 
for the process $\pp \to (\WWbB \to)\;\epmubB+X$ at NLO~QCD
accuracy matched to Shower Monte Carlo (SMC) simulations using \powhel,
on the basis of the interface between \texttt{HELAC-NLO} and 
\texttt{POWHEG-BOX}. We include all resonant and non-resonant contributions.
This is achieved by fully taking into account the effect of off-shell
t-quarks and off-shell W-bosons in the complex mass scheme. We also
present a program called \texttt{DECAYER} that can be used to let the
t-quarks present in the event files for $\pp \to \tTX$ processes decay
including both the finite width of the t-quarks and spin correlations. 
We present predictions for both the Tevatron and the LHC, with emphasis
on differences emerging from three different \WWbB\ hadroproduction
computations:
(i) full implementation of the $\pp \to \WWbB$ process,
(ii) generating on-shell t-quarks pushed off-shell with a Breit-Wigner
finite width and decayed by \decayer,
and (iii) on-shell t-quark production followed by decay in the narrow
width approximation, as described by the SMC.
}


\arxivnumber{arXiv:1405.5859}

\begin{document}
\maketitle
\flushbottom

\section{Introduction}

Accurate predictions for the production of \tT-pairs alone or in
association with some hard objects, such as jets, vector and/or scalar
bosons are important for many experimental studies at hadron colliders
both aiming at better understanding of the Standard Model and searches
for new physics. However, the t-quarks and heavy bosons decay quickly
and the detectors detect their decay products. The experiments often
concentrate on leptons because the leptonic channels offer a much
cleaner final state than the hadronic ones. Thus it is important not
only to predict cross sections for the production of the heavy quarks
and bosons, but also for the spectra of the leptons that emerge in
their decays.

There are many different approximations to predict such lepton spectra.
The most precise way of treating spin correlations and off-shellness of
heavy particles is to use matrix elements that are obtained from all
Feynman graphs belonging to the leptonic final state.  The
corresponding graphs include both resonant and non-resonant ones. This
procedure in principle can be implemented in a leading-order (LO)
computation automatically using tools like \texttt{MadGraph4/MadEvent}
\cite{Alwall:2007st} or \texttt{CompHEP/CalcHEP} \cite{Pukhov:1999gg} available
for several years now.
However, in
practice, the complexity of the computation can become forbidding very
quickly. For instance, for the simple case of \tT-pair production, if
the decays are included, the final state already contains six particles
due to the t $\to W^+$b $\to \ell^+ \nu_\ell$b decay chain. Thus LO
tools like \texttt{ALPGEN} \cite{Mangano:2002ea} and \texttt{HELAC}
\cite{Cafarella:2007pc} written on the basis of recursion relations,
instead of Feynman graphs, and improved, more efficient, generation of
Feynman graphs, like in \texttt{Madgraph5} \cite{Alwall:2011uj},
have been developed to better deal with
final states with a large number of particles. These difficulties are
magnified if one wishes to perform the same computations at the
next-to-leading order (NLO) accuracy. For the process mentioned above,
one has to deal with six-point integrals, which makes the computation
numerically cumbersome.

The computation can be simplified if performed in the narrow-width
approximation (NWA), with only resonant contributions kept.  The NWA is
a good approximation if the width of the decaying particle is indeed
narrow, such as the case of decaying electroweak vector bosons, or
t-quarks that decay before hadronization.
Furthermore, selection cuts are often chosen to enhance the resonant
contributions as compared to the non-resonant ones.  Predictions for
many important production channels were computed at the next-to-leading
order (NLO) accuracy long ago. More recently, the decay of the t-quarks
was included in the NWA at NLO accuracy for three processes including
spin-correlations \cite{Melnikov:2009dn,Melnikov:2010iu,
Melnikov:2011ta,Melnikov:2011ai,Melnikov:2011qx}.
A further simplification can be made, called decay chain approximation
(DCA), if the resonant graphs are replaced by on-shell production
times decays of the heavy particles,  neglecting both off-shell effects
and spin correlations between production and decay, which may seem a
high price for simplicity.

While the NWA is a theoretically well-defined approach, it misses the
effects of parton showers and hadronization which are often
significant. Another line of research was to make improvements by
implementing the production of t-quarks at the NLO accuracy and match
those to shower Monte Carlo programs (SMCs) \cite{Kardos:2011qa,
Garzelli:2011vp,Garzelli:2011is,Garzelli:2012bn}. Within this approach
the t-quarks are produced on-shell and the decay is provided by the SMC
in the DCA at LO accuracy without any spin-correlation. As spin
correlations may affect some lepton spectra significantly, a method was
introduced in \Ref{Frixione:2007zp} to take into account the
spin-correlations in the NWA, primarily intended to be applied in
matched NLO+SMC computations.  In addition to \tT-pair production, this
method was also implemented to include the effects of spin-correlations
in single top \cite{Alioli:2009je} and \tT+jet production \cite{Alioli:2011as},
among other processes.

An important goal in this paper is to compare the predictions made
using these different approaches within the same NLO+SMC framework. For
this purpose we choose the simplest process, \tT\ hadroproduction and
the POWHEG method \cite{Nason:2004rx,Frixione:2007vw} as implemented in
the \powhel\ framework \cite{Kardos:2011qa,Garzelli:2011vp,Kardos:2011na,
Garzelli:2011is,Garzelli:2011iu,Dittmaier:2012vm} which is based on 
the \helacnlo\ \cite{Bevilacqua:2011xh} and \powhegbox\
\cite{Alioli:2010xd} programs.  The matched NLO+SMC prediction in the
DCA and NWA approximations have already been known for some time
\cite{Frixione:2003ei,Frixione:2007nu, Frixione:2007nw}, we simply
implemented those in our framework. The new ingredient of our
computation is the implementation of \WWbB-production within \powhel. 
This is among the most complex final states for which matched NLO+SMC
prediction has been considered so far.

Hadroproduction of $\WWbB$ is one of the $2\to 4$ processes included in
the Les Houches 2010 wishlist~\cite{Binoth:2010ra}.  It can be
considered a background for Higgs boson production in association with
a \bB-pair and for new physics searches, in cases where  cascades
of decays of non-SM particles may give rise to leptons and multiple
jets in combination with missing energy.

In the Standard Model (SM), most of the contribution to the inclusive
cross-section of this process comes from intermediate \tT-production,
due to the abundance of \tT-pairs expected at high-energy hadron
colliders and to the fact that t-quarks decay almost exclusively into
$W$b-pairs (with a branching ratio of about 99\%). Thus, the easiest
way to account for the bulk of this cross-section is to factorize the
computation in two parts as in the DCA.  Results of computations
according to this scheme are available since long time~\cite{Frixione:2003ei} 
and
have been included in Monte Carlo generators like MCFM~\cite{Ellis:2006ar,
Campbell:2010ff}.
Effects of spin correlations are also known at the next-to-leading
order (NLO) accuracy as well \cite{Melnikov:2009dn}.

In order to increase the accuracy of the predictions to the few-percent
level, as imposed both by the experimental high-luminosity that can be
reached at the LHC and by the need for better understanding
and measuring the t-quark properties, a full NLO QCD computation of the
$\pp \to \WWbB$ process including the effects of off-shell t-quarks, is
desirable. Such a computation was presented for the first time
in Ref.~\cite{Denner:2010jp}, performed by assigning a width to the
t-quarks in the framework of the complex mass scheme. The complex mass
scheme was originally introduced at LO accuracy and extended to the NLO
accuracy in Refs.~\cite{Denner:1999gp,Denner:2005es,Denner:2005fg}.
In this framework, besides doubly-resonant Feynman-graphs even singly-
and non-resonant graphs, and also their interference, expected to give
a contribution suppressed by powers of $\mathrm{O}({\Gamma_t}/{m_t})$,
were accounted for in the cross-section evaluation.  

As for W decays, the fully leptonic channel was considered in
Ref.~\cite{Denner:2010jp}, by treating the $W^+ \to e^+ \nu_e$
and $W^- \to \mu^- \bar{\nu}_\mu$ decays in a spin-correlated NWA.
In Refs.~\cite{Bevilacqua:2010qb, Denner:2012yc} effects of finite widths of both t-quarks and $W$ bosons were included, also including spin-correlations.
The $W$ bosons were treated in a fixed width scheme, since their decay
does not receive QCD NLO corrections. One of the main conceptual
breakthroughs of these papers was the computation of virtual correction
contributions including t-quarks in the loop with finite width, which was 
achieved by the introduction of complex masses in the one-loop scalar
integrals and in the on-shell mass renormalization counterterm. The
preservation of gauge invariance was a driving principle to check the
reliability of the implementations.  

Based on the full computation of the process $\pp\to(\WWbB\to)\;\epmubB~+X$
at NLO accuracy, as presented in Ref.~\cite{Bevilacqua:2010qb}%
\footnote{More recently, the same process has been also studied at NLO
accuracy in Refs.~\cite{Frederix:2013gra,Cascioli:2013wga,Heinrich:2013qaa}.}, in this paper we take a further step by matching the NLO QCD predictions to SMC programs
using the POWHEG method. Using an SMC allows for an evolution of the
final state to the hadron level. We provide predictions for integrated
and differential distributions at this level at both LHC and Tevatron.
We explicitly compare the results of the full high-accuracy computation
outlined above, with the ones arising in simplified cases, where
(a) t-quark production is factorized with respect to their decay,
performed either in the DCA, neglecting spin-correlations (as
implemented in the SMC), or
(b) according to a more advanced approach, where events are produced
with on-shell t-quarks, which are pushed off-shell according to a
Breit-Wigner distribution in a second step, and then decayed at a mixed
LO/NLO accuracy \cite{Frixione:2007zp}, thus including
spin-correlations (as implemented in the code \decayer)%
\footnote{An independent alternative program, called {\texttt{MadSpin}}, 
was provided to the \texttt{MadGraph5} framework in \Ref{Artoisenet:2012st}.}.

After the implementation of several $2 \to 3$ processes at NLO+PS
accuracy, the first $2 \to~4$ process, the production of a
$W^+$ pair in association with two jets, was implemented in the
\powhegbox\ \cite{Melia:2011gk,Jager:2011ms}. Although seemingly
similar, this latter process is much simpler than our case as far as
the singularity structure is concerned. The squared matrix element of
the leading-order (LO) cross section is finite for $W^+\,W^+\,j\,j$
production. For the \WWbB-production, as we consider massless b-quarks,
the squared matrix elements are singular even at LO, which necessitates
the suppression of such singularities. In this respect our computation
is more similar to \tTbB-production with massless  b-quarks, 
which was studied at the hadron
level by including NLO QCD corrections matched to a SMC program by
means of the POWHEG matching procedure \cite{Kardos:2013vxa}.  The
squared matrix element for this process is also singular already at LO.

The main outcomes of our computations are Les Houches events (LHEs),
i.e.~parton-level events in files according to the Les Houches accord
\cite{Boos:2001cv}.  For \tT-production such events contain a \tT-pair
and possible further one parton corresponding to first emission. In the
case of \tT-production processed by \decayer, as well as for
\WWbB-production the events contain four leptons, a \bB-pair (we chose
the \epmubB\ channel) and possible first emission.  We perform analysis
of these events with some standard selections cuts, but we emphasize
that the events are available upon request for almost arbitrary
experimental analysis. Our first results on this process on the basis of
a preliminary set of events were presented 
in \Ref{Garzelli:2012zfa, Kardos:2012nza}.

\section{\WWbB-production}
\label{sec:WWbb}

An important goal in this paper is to check the precision one can
achieve with the decay of heavy particles in the decay-chain
and narrow-width approximations.  In order to make unquestionable
statements, we aim at comparing the distributions obtained with
approximations to those obtained with an exact computation at the NLO
accuracy, when both resonant and non-resonant t-quarks are taken into
account in the matrix elements. This amounts to considering all
Feynman-graphs to the $\pp \to \WWbB$ process that contribute at the
NLO accuracy.

The starting point for our computations is the factorization theorem,
which gives the cross section for the collision of hadrons $A$ and $B$
as a convolution,
\beq
\sigma_{AB} = \sum_{a,b}
\int_0^1\!\ud x_a \,f_{a/A}(x_a,\mu_F^2)
\int_0^1\!\ud x_b\, f_{b/B}(x_b,\mu_F^2)
\,\sigma_{ab}(x_a p_A, x_b p_B;\mu_F^2) \,,
\label{eq:factorization}
\eeq
of the parton density functions (PDFs)
$f_{p/P}(x_{p},\mu_F^2)$ ($p = a$, $b$ and $P = A$, $B$)
with the partonic cross section for the collision of partons $a$ and $b$,
$\sigma_{ab}(p_a,p_b;\mu_F^2)$.  Although we set the b-quarks massless
when computing the hard-scattering matrix elements, we neglect b-quarks
in the initial state, so the summation over $a$ and $b$ runs over u-,
d-, c-, s-quarks and the gluon.

\begin{figure}[t]
\centering
\includegraphics[width=0.6\textwidth]{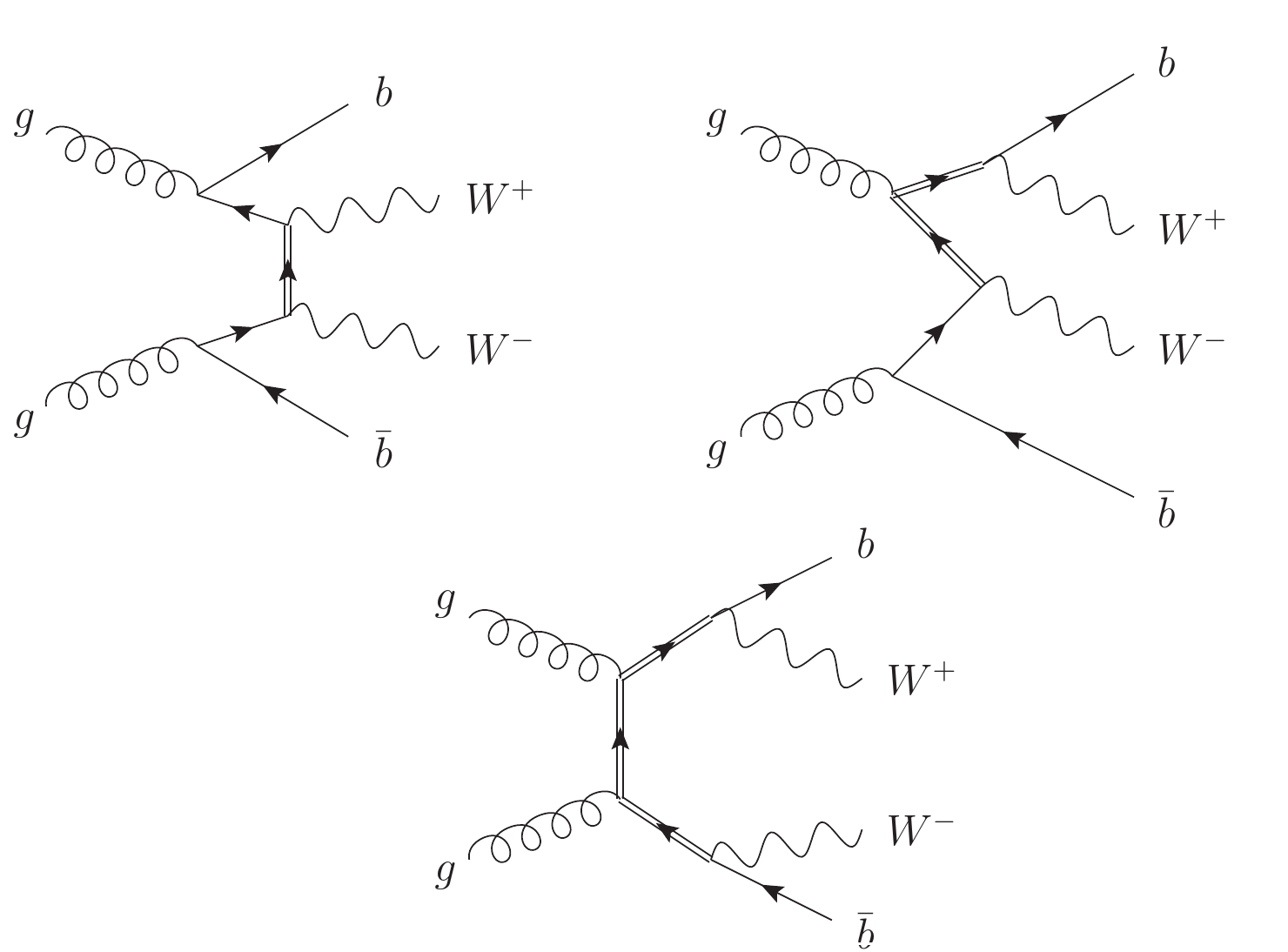}
\caption{\label{fig:Borngraphs} Sample Feynman graphs for the 
three types of \WWbB\ production at Born level: non-resonant,
singly-resonant and doubly-resonant ones.  }
\end{figure}

The partonic cross section has the perturbative expansion
\beq
\sigma_{ab}(p_a,p_b;\mu_F^2) =
\sigma^{\lo}(p_a,p_b)+\sigma^{\nlo}(p_a,p_b;\mu_F^2)
\equiv \sigma_{\nlo}(p_a,p_b;\mu_F^2)
\label{eq:sigmaexpansion}
\eeq
up to NLO accuracy. The LO contribution,
\beq
\sigma^{\lo}(p_a,p_b) = {1\over2s} \int\!\ud\Phi_B
\overline{|\mathcal{M}_B^{(0)}|^2}
\,,
\label{eq:sigmaLO}
\eeq
is the integral of the spin- and color-averaged squared Born matrix
element $\overline{|\mathcal{M}_B|^2}$ over the phase space
$\Phi_B$ of the six final-state particles
(b, \baq, $\ell$, $\bar{\nu}_{\ell}$ $\bar{\ell'}$, $\nu_{\ell'}$),
multiplied with flux normalization, where $s=x_ax_bS$, with
$S=(p_A+p_B)^2$ being the square of the hadronic center-of-mass energy.  
We show three sample Feynman-graphs (the decays of the vector bosons being
suppressed) of the three types in \fig{fig:Borngraphs}: (a) non-resonant,
(b) singly-resonant and (c) doubly-resonant $W$b production. There are 38
graphs for the $gg$-channel and 14 graphs for the $q\bar{q}$-channel (with
undecayed vector bosons in the final state).

\begin{figure}[t]
\centering
\includegraphics[width=0.6\linewidth]{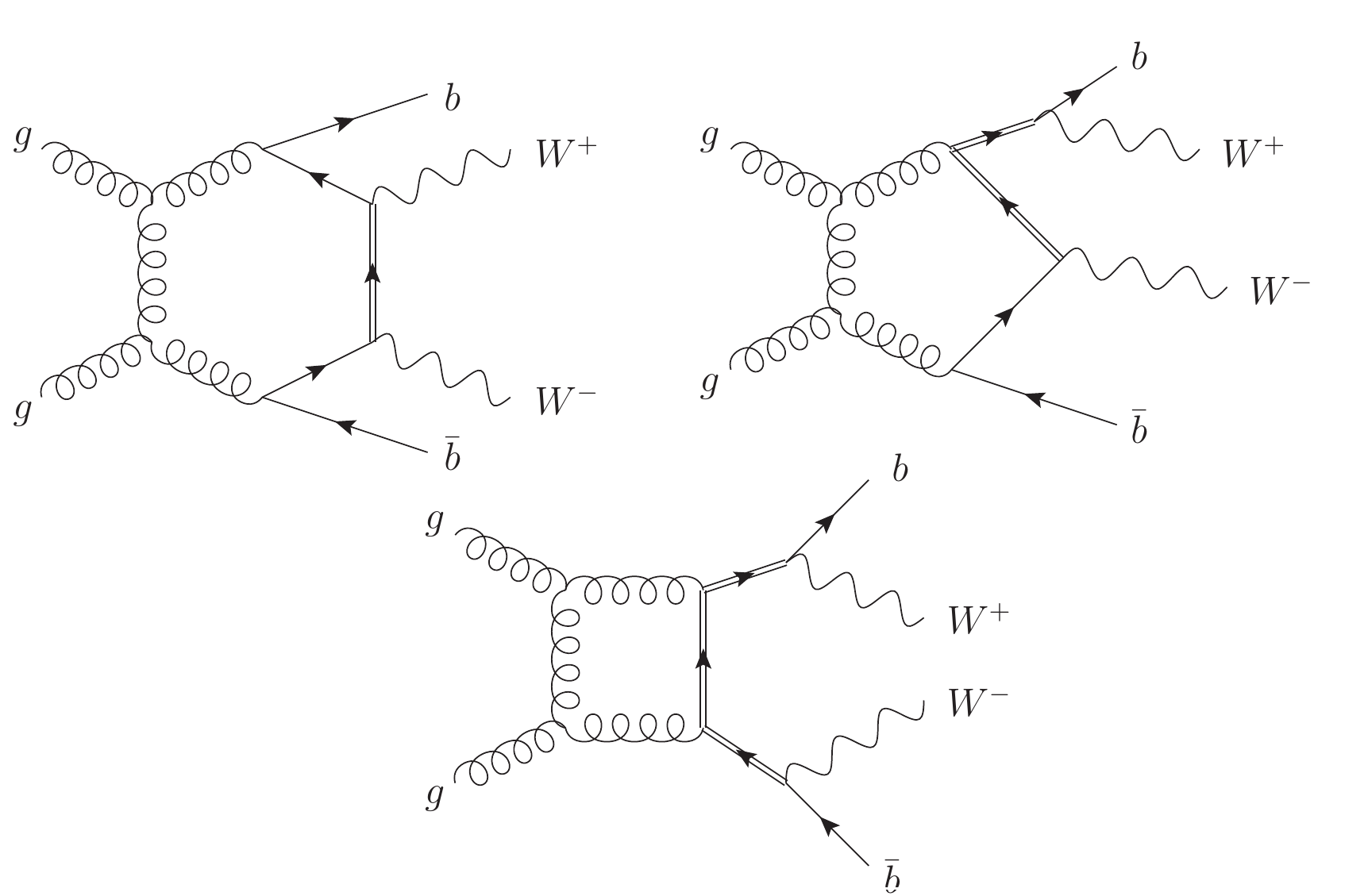}
\caption{\label{fig:loopgraphs} Sample Feynman graphs for the 
three types of \WWbB\ production at one-loop level.  }
\end{figure}

The NLO correction is a sum of two separately divergent contributions,
the real and virtual corrections, with finite sum for infrared safe
observables.  The squared matrix element $R=\overline{|\mathcal{M}_R|^2}$
for the real and $\overline{2 {\rm Re}(\mathcal{M}_B^*\mathcal{M}_V)}$
for the virtual contributions are obtained from $2\to 7$ tree graphs
and $2\to 6$ one-loop graphs, respectively. Sample Feynman-graphs for
the three basic types are shown in \figs{fig:loopgraphs}{fig:realgraphs},
with vector-boson decay suppressed again. There are 795 one-loop graphs in
the $gg$ channel and 294 one-loop graphs in the $q\bar{q}$-channel. For
the real corrections the $qg$-channel, leading to an extra quark in the
final state, opens too.
\begin{figure}[t]
\centering
\includegraphics[width=0.6\linewidth]{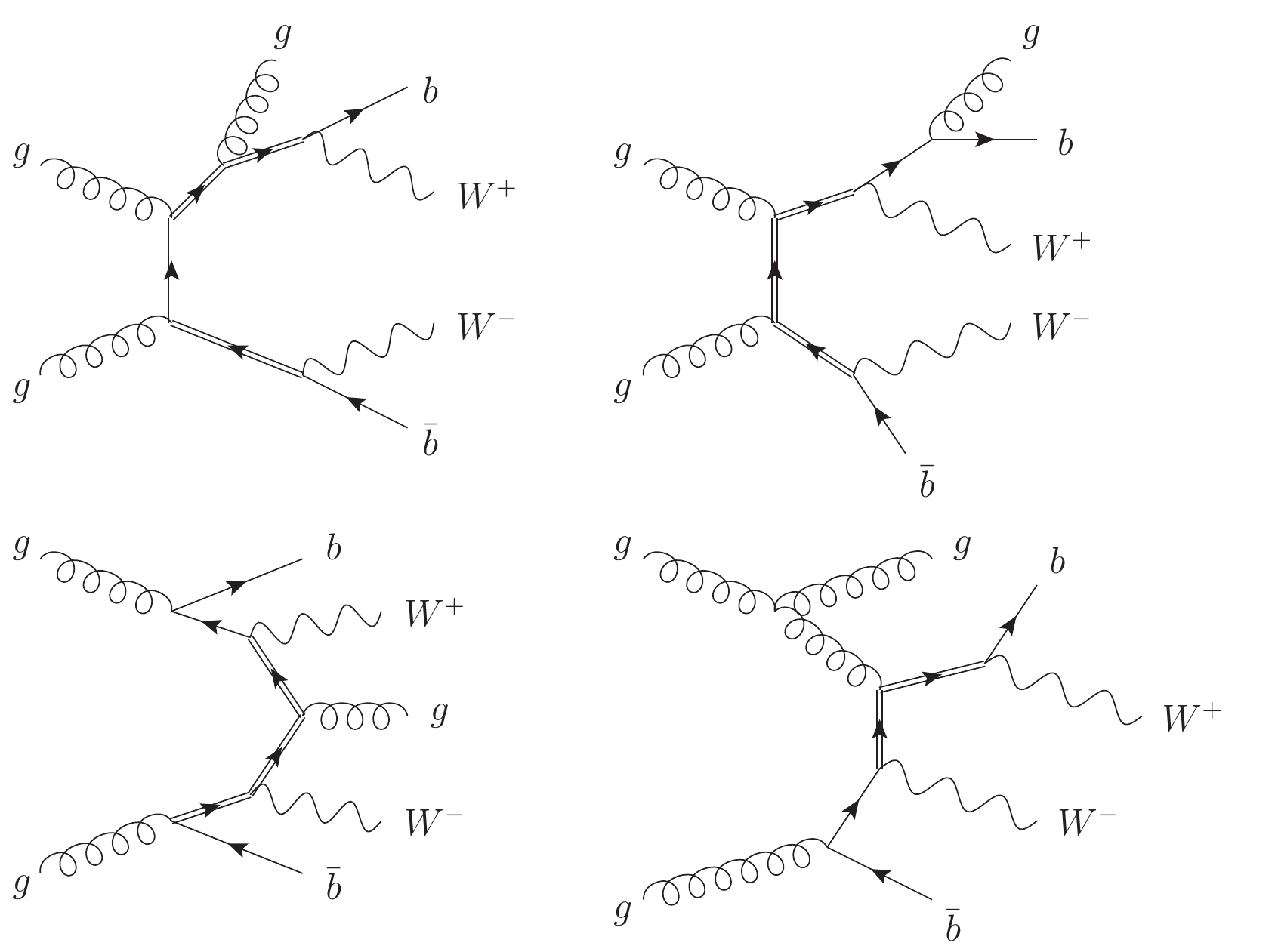}
\caption{\label{fig:realgraphs} Sample Feynman graphs for the 
three types of \WWbB+$g$ production at tree level.  }
\end{figure}

The finite sum of the two types of radiative corrections can be
rewritten as the sum of three finite terms in the following general form:
\bal
\sigma^{\nlo}(p_a,p_b;\mu_F^2) &=
  \sigma^{\nlo\{V\}}(p_a,p_b;\Phi_B)
+ \sigma^{\nlo\{R\}}(p_a,p_b;\Phi_R)\nonumber\\
&+\int_0^1\!\ud x \Bigl[
  \sigma^{\nlo\{C\}}(x;xp_a, p_b;\Phi_B;\mu_F^2) +
  \sigma^{\nlo\{C\}}(x;p_a,x p_b;\Phi_B;\mu_F^2)\Bigr]
\,,
\label{eq:sigmaNLO}
\eal
where $\sigma^{\nlo\{R\}}$ is the regularized real correction (real
radiation minus subtraction terms) and $\sigma^{\nlo\{V\}}$ is the
regularized virtual corrections (one-loop corrections plus integrated
subtraction terms),
\beq
\sigma^{\nlo\{V\}} = \int\!\ud\Phi_B V(\Phi_B)
\,,\qquad
\sigma^{\nlo\{R\}} = \int\!\ud\Phi_R \hat{R}(\Phi_R)
\,,
\eeq
written in the usual POWHEG terminology \cite{Frixione:2007vw}. 
In the same terminology $\sigma^{\lo}$ in \eqn{eq:sigmaLO} is the
integral of $\overline{|\mathcal{M}_B|^2} \equiv B(\Phi_B)$. The
third term in \eqn{eq:sigmaNLO} contains the finite remainders from the
cancellation of the $\epsilon$-poles of the initial-state collinear
counterterms, in the POWHEG terminology usually denoted by
$G_{\oplus}(\Phi_{B,\oplus})$ and $G_{\ominus}(\Phi_{B,\ominus})$,
\beq
\sigma^{\nlo\{C\}}(x p_a, p_b) =
\int\!\ud\Phi_B\frac{1}{x}G_{\oplus}(\Phi_B)
\,,\qquad
\sigma^{\nlo\{C\}}(p_a, x p_b) =
\int\!\ud\Phi_B\frac{1}{x}G_{\ominus}(\Phi_B)
\,,
\eeq
where $x$ is the parton-in-parton momentum fraction in the collinear
factorization term.

There are standard techniques to compute the finite contributions in
\eqn{eq:sigmaNLO} discussed in the literature in full detail, therefore,
we do not deal with this problem here. The necessary formulas for the
NLO and PS matching in the two popular subtraction schemes
\cite{Catani:1996vz,Frixione:1995ms} are given in  
\Ref{Frixione:2007vw}. In the \powhegbox\ framework used in the present
computation, the FKS subtraction scheme \cite{Frixione:1995ms} is
implemented in a process independent way. To obtain the POWHEG cross
section for the generation of an event with first (hardest) emission
included, one first parametrizes the real-emission phase space in terms
of the Born phase space and three more variables that describe the
radiation process, leading to 
\beq
\ud\Phi_R = \ud\Phi_B \ud\Phi_{\rad}
\,,
\label{eq:PSfact}
\eeq 
where $\ud\Phi_{\rad}$ includes the Jacobian of this change of integration
variables. Next one defines the NLO-corrected fully differential cross
section belonging to the underlying Born configuration
\beq
\tB(\Phi_B) =
B(\Phi_B) + V(\Phi_B)
+ \int\!\ud\Phi_{\rad}\hat{R}(\Phi_R) +
\int\!\frac{\ud x}{x}\Big[G_{\oplus}(\Phi_B)+G_{\ominus}(\Phi_B)\Big]
\,,
\label{eq:tB}
\eeq
and the POWHEG Sudakov form factor
\beq
\Delta(\Phi_B, \pt) = \exp
\left\{-\int \frac{\ud \Phi_{\rad} R(\Phi_R) \Theta(\kt(\Phi_R)-\pt)}
{B(\Phi_B)}\right\}
\,.
\eeq
The function $\kt(\Phi_R)$ is equal to the transverse momentum
of the emitted parton relative to the emitting one.  Then the POWHEG
fully differential cross section (the cross section that can be
obtained from the LHEs) is defined as
\beq
\ud \sigma_{\lhe} = \tB(\Phi_B)\ud \Phi_B
\left[\Delta(\Phi_B, \ptmin) + \ud \Phi_{\rad}
\Delta\Big(\Phi_B, \kt(\Phi_R)\Big) \frac{R(\Phi_R)}{B(\Phi_B)}
\Theta(\kt(\Phi_R)-\ptmin)\right]
\,.
\label{eq:sigmaLHE}
\eeq
The advantage of this formula is that it can be used to generate equal
weight events with Born configuration (first term) or including first
radiation (second term). Computing the derivative of the Sudakov form
factor with respect to $\pt$, one can prove the unitarity relation
\cite{Frixione:2007vw}
\beq
\int\!\ud\Phi_{\rad}
\Delta\Big(\Phi_B, \kt\Big) \frac{R(\Phi_R)}{B(\Phi_B)}
\Theta(\kt(\Phi_R)-\ptmin)
= 1-\Delta(\Phi_B, \ptmin)
\,.
\label{eq:unitarity}
\eeq
Substituting into \eqn{eq:sigmaLHE}, one finds that the total POWHEG
cross section is equal to that at NLO accuracy,
\beq
\int\!\ud \sigma_{\lhe} = \int\!\ud\Phi_B\,\tB(\Phi_B) = \sigma_{\nlo}
\,,
\label{eq:sigmaLHE-NLO}
\eeq
provided the LO cross section is finite. If not, the total POWHEG
cross section can only be defined with some technical cuts
\cite{Alioli:2010qp,Kardos:2013vxa} (see
\sect{sec:powhel} for details in our implementation). In such cases,
instead of checking the total cross section one can but check the
consistency of the differential distributions obtained from the POWHEG
formula with those at NLO accuracy.  

The differential distribution of the observable $O$ can be computed as
\beq
\frac{\ud \sigma_\lhe}{\ud O} =
  \int\!\ud \Phi_B \tB(\Phi_B)\delta(O(\Phi_B)-O)
\,.
\eeq
Using \eqn{eq:PSfact} and the explicit form of $\ud \sigma_{\lhe}$ in
\eqn{eq:sigmaLHE}, we can rewrite it as
\bal
\bsp
\frac{\ud \sigma_\lhe}{\ud O} &=
  \int\!\ud \Phi_B \tB(\Phi_B)\delta(O(\Phi_B)-O)
\\[2mm]
&\qquad\times
\left[\Delta(\Phi_B, \ptmin) + \ud \Phi_{\rad}
\Delta\Big(\Phi_B, \kt(\Phi_R)\Big) \frac{R(\Phi_R)}{B(\Phi_B)}
\Theta(\kt(\Phi_R)-\ptmin)\right]
\\[2mm]
&+ \int\!\ud \Phi_R
\Delta\Big(\Phi_B, \kt(\Phi_R)\Big) \frac{\tB(\Phi_B)}{B(\Phi_B)}
\Theta(\kt(\Phi_R)-\ptmin)
\\[2mm]
&\qquad \times R(\Phi_R)
\Big[\delta(O(\Phi_R)-O) - \delta(O(\Phi_B)-O)\Big]
\,.
\label{eq:LHEacc1}
\esp
\eal
We can again use the unitarity relation in \eqn{eq:unitarity} to
simplify the first integral, and use the expansions of the Sudakov form
factor and the $\tB$ functions in the strong coupling \cite{Frixione:2007vw},
\beq
\Delta\Big(\Phi_B, \kt(\Phi_R)\Big) \frac{\tB(\Phi_B)}{B(\Phi_B)} =
1+{\rm O}(\as)
\,,
\label{eq:expansion}
\eeq
to obtain the relation
\bal
\bsp
\frac{\ud \sigma_\lhe}{\ud O} &=
  \int\!\ud \Phi_B \tB(\Phi_B)\delta(O(\Phi_B)-O)
\\[2mm]
&+ (1+{\rm O}(\as))\int\!\ud \Phi_R R(\Phi_R)
\Big[\delta(O(\Phi_R)-O) - \delta(O(\Phi_B)-O)\Big]
\Theta(\kt(\Phi_R)-\ptmin)
\,.
\label{eq:LHEacc2}
\esp
\eal
Using \eqns{eq:tB}{eq:LHEacc2}, we find that
\beq
\frac{\ud \sigma_\lhe}{\ud O} = \frac{\ud \sigma_\nlo}{\ud O}
+  {\rm O}(\as) \int\!\ud \Phi_R R(\Phi_R)
\Big[\delta(O(\Phi_R)-O) - \delta(O(\Phi_B)-O)\Big]
\,,
\label{eq:LHEacc3}
\eeq
where $\Phi_R$ is parametrized as in \eqn{eq:PSfact} and we dropped the
$\Theta(\kt(\Phi_R)-\ptmin)$ function, its effect being suppressed by
\ptmin\ \cite{Frixione:2007vw}.
Thus any observable has the NLO accuracy up to higher order
corrections multiplying the real emission contribution, where the
magnitude of the correction is set by the expansion in
\eqn{eq:expansion}. Therefore, when the NLO K-factor is large, the
distributions obtained from $\ud \sigma_{\lhe}$ may differ from the NLO
cross section significantly (see \Ref{Alioli:2008tz} for more detailed
discussion). It is possible to decrease this difference by writing the
real correction as a sum, $R = R_S + R_R$, where $R_S$ is constrained to
near singular regions, hence called {\em singular} contribution, while
$R_R$ is called the remnant~\cite{Alioli:2008tz}. Using such a
decomposition, the size of the difference between the POWHEG and NLO
cross sections is controlled by $R_S$:
\beq
\frac{\ud \sigma_\lhe}{\ud O} = \frac{\ud \sigma_\nlo}{\ud O}
+  {\rm O}(\as) \int\!\ud \Phi_R R_S(\Phi_R)
\Big[\delta(O(\Phi_R)-O) - \delta(O(\Phi_B)-O)\Big]
\,.
\label{eq:LHEacc4}
\eeq
 
\subsection{\powhel\ implementation}
\label{sec:powhel}

The \powhegbox\ provides a general framework to compute the POWHEG
cross section in \eqn{eq:sigmaLHE}.  In this framework, the following
ingredients are needed:
\begin{itemize}
\itemsep -2pt
\item
The flavor structures of the Born and real radiation emission
subprocesses, listed in \tab{tab:wwbbsubprocesses}.
\item
The Born-level phase space, that we generate to emphasize the
doubly-resonant kinematics (see below).
\item
We obtain the squared matrix elements for the Born and the real-emission
processes and color-correlated Born amplitudes with all incoming momenta
using \helacnlo~\cite{Bevilacqua:2011xh} (in particular, \helaconeloop\
\cite{vanHameren:2009dr}, based on the OPP method~\cite{Ossola:2006us}
complemented by Feynman-rules for the computation of the QCD $R_2$
rational terms~\cite{Draggiotis:2009yb}, 
and \texttt{Helac-Dipoles}~\cite{Czakon:2009ss}).  
The matrix elements in the
physical channels were obtained by crossing.  In order to treat the
numerical instabilities, we implemented double-double precision
({\tt dd}-precision means 30 digit accuracy) numerics by
developing a \helaconeloopdd\ version of the \helaconeloop\ program.
\item
We project spin-correlated Born amplitudes from the helicity basis
to the Lorentz one by using the polarization vectors.
\end{itemize}
\begin{table}[t]
\centering
\begin{tabular}{|c|c|}
\hline
\hline
$q \, \bar{q}\, e^- \, \bar{\nu}_e \, \mu^+ \, \nu_\mu \, \bq \, \baq  \to 0$
&
$g \, g\, e^- \, \bar{\nu}_e \, \mu^+ \, \nu_\mu \, \bq \, \baq  \to 0$
\\
\hline
$q \, \bar{q}\, e^- \, \bar{\nu}_e \, \mu^+ \, \nu_\mu \, \bq \, \baq\, g \to 0$
&
$g \, g\, e^- \, \bar{\nu}_e \, \mu^+ \, \nu_\mu \, \bq \, \baq\, g \to 0$
\\
\hline
\hline
\end{tabular}
\caption{\label{tab:wwbbsubprocesses} 
Born-level (first row) and real emission (second row) subprocesses.
}
\end{table}

With this input in the \powhegbox\ we can generate hadronic events. In
principle we can choose any SMC program for generating parton showers,
decays of heavy particles and hadronization. There is however, a caveat
when choosing the PS. We generate events with hardest emission measured
by its transverse momentum.  If the ordering variable
in the PS is the transverse momentum, such as in \pythia\ \pt-ordered, 
then the final-state system after first radiation is evolved using the 
SMC and any emission with transverse momentum higher than that of the 
first emission is impossible (automatically vetoed shower).  If the 
ordering variable in the PS is different from the transverse 
momentum of the parton splitting (e.g. angular-ordered showers in 
\herwig~\cite{Corcella:2002jc}, 
or virtuality-ordered showers in \pythia~\cite{Sjostrand:2006za}), 
then the
hardest emission is not necessarily the first one. In such cases
these SMC codes discard in subsequent splittings shower 
emissions with larger transverse momentum than that in the real
emission correction (vetoed showers). In addition, a truncated shower simulating
wide-angle coherent soft emission before the first emission is also needed in
principle, but its effect was found small \cite{LatundeDada:2006gx}. As
there is no implementation of truncated shower in \herwig\ using
external LHE event files, the effect of the truncated showers is absent
from our predictions. To quantify these differences we make predictions 
in \sect{sec:pheno} with both \pt-ordered and virtuality-ordered 
\pythia, as well as with \herwig.

\begin{figure}[t]
\centering
\includegraphics[width=6cm]{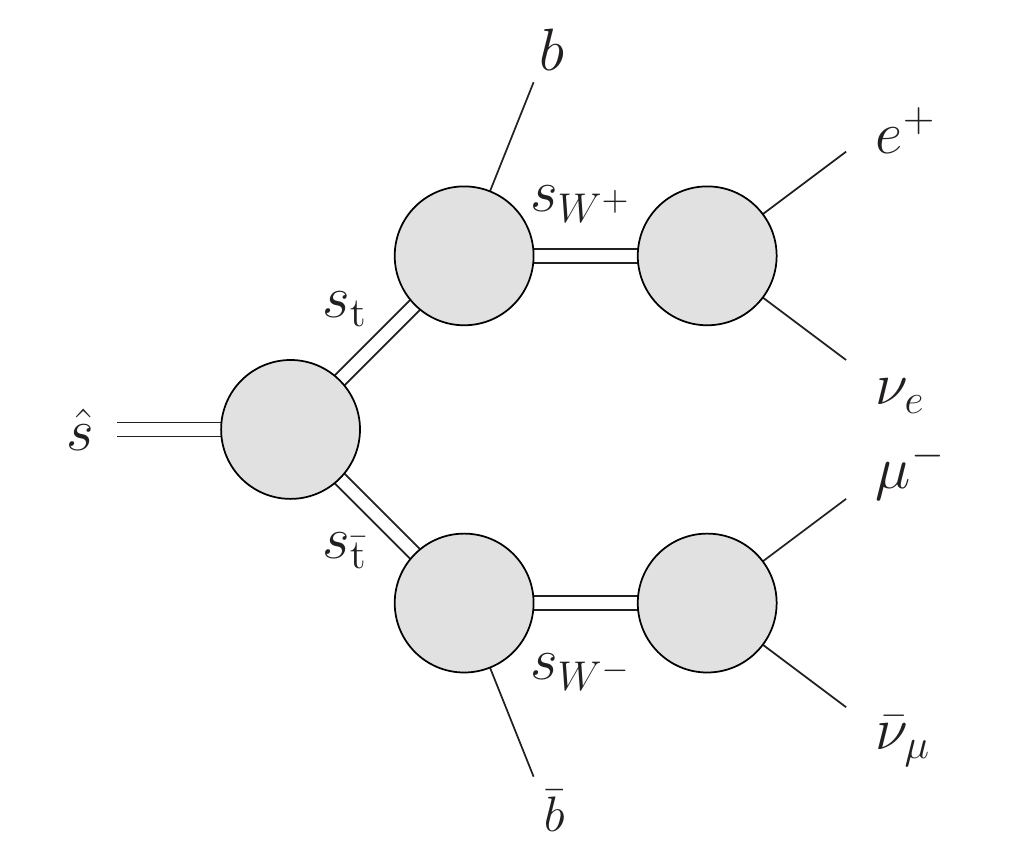}
\caption{\label{fig:doublyresonant} Schematic representation of the phase
space for the doubly-resonant channel in off-shell top and $W$
production. On each double line we depicted the corresponding
kinematic invariant.}
\end{figure}

We take into account the finite width of the t-quarks and $W$-bosons in
the complex mass scheme \cite{Denner:1999gp,Denner:2005fg} where the
Feynman propagator factor is substituted as
\begin{align}
\frac{1}{p^2 - m^2 + i\varepsilon} \to \frac{1}{p^2 - \mu^2 +
i\varepsilon}
\,,
\quad
\mu^2 = m^2 -im\Gamma
\,.
\end{align}
The presence of these propagators results in Breit-Wigner factors in the
squared matrix elements. Thus it is expected that the dominant
contributions come from those configurations where the two-particle
kinematic invariant assigned to the off-shell particle, e.g.\
$s_\top = (p_\bq+p_{\W^+})^2$, is around its pole mass, e.g.\
$s_\top \simeq m^2_\top$. In order to integrate over these resonances
efficiently, we built the phase space emphasizing the doubly-resonant
channel as shown in \fig{fig:doublyresonant}.

The \powhegbox\ uses \mint\ as its integrator, which performs well in
the absence of resonances.  For our case with four invariants having
Breit-Wigner peaks in the physical region the integrator fails to set
up an integration grid, therefore, we performed the optimization for
these variables, described as follows. Let us consider a massive
particle with momentum $p$, mass $m$, and width $\Gamma$, then the
kinematic invariant can be written as $s = p^2$, hence the Breit-Wigner
factor is proportional to
\begin{align}
\mathcal{F}_{\mathrm{BW}}(s) =
\frac{1}{(s - m^2)^2 + m^2\Gamma^2}
\,.
\end{align}
We generate the phase space such that the integration variable is the
invariant $s$ itself, and introduce a new integration variable 
\beq
\rho(s) = \int^s\!\ud s'\,\mathcal{F}_{\mathrm{BW}}(s')
     = \frac{1}{m\Gamma}\tan^{-1}\left(\frac{s - m^2}{m\Gamma}\right)
\,.
\eeq
The Jacobian $1/\mathcal{F}_{\mathrm{BW}}(s)$ cancels the factor
$\mathcal{F}_{\mathrm{BW}}(s)$ in the squared matrix element. The
limits on $\rho$ are related to the limits on $s$ by
\beq
\rho_{\min/\max} =
\frac{1}{m\Gamma}\tan^{-1}\left(\frac{s_{\min/\max} - m^2}{m\Gamma}\right)
\,.
\eeq
For the present calculation we used $s_{\min} = 25\,\GeV^2$ and $s_{\max}=s$.
Finally, we map $\rho$ to $\xi\in[0,1]$ using
\begin{align}
\rho &=  \rho_{\mathrm{min}}+(\rho_{\mathrm{max}}-\rho_{\mathrm{min}})\xi
\,,
\end{align}
hence the original integration measure over $s$ is expressed through $\xi$
as
\begin{align}
\ud s = \frac{\rho_{\max} -
\rho_{\min}}{\mathcal{F}_{\mathrm{BW}}(s)}\ud\xi
\,.
\end{align}
After performing the change of the integration variable from $s$ to
$\xi$, the integrand is flat in the new variable, and \mint\ can produce
the integration grid.

The generation of the matrix elements is straightforward using
\helacnlo. During integration we had to solve two more problems.
The first one is that for vanishing transverse momentum of
the b-quarks or vanishing invariant mass of the \bB-pair the Born cross
section becomes singular. While this can never happen in a LO computation
due to the application of selection cuts, it is a problem in the POWHEG method because
the selection cuts can only be applied after event generation. The
traditional way of treating this problem is the introduction of 
generation cuts \cite{Alioli:2010qp,Kardos:2013vxa}. We use 
$\ptb \geq 2$\,GeV for both the b-  
and $\bar{\mathrm{b}}$-quark 
and a cut $m_{\bB}\geq 1$\,GeV. With these cuts the LO cross
section becomes finite, but the generation of the events is still
rather inefficient because most of the events are generated in the
region of small \ptb, and are thus lost when the physical selection cuts
(usually much higher, in the region of 20 -- 30\,GeV) are applied. In
order to make the generation of events more efficient, we introduce
suppression factors \cite{Alioli:2010xa}. As we want to suppress the
region of small \ptb, the natural choice for the suppression is
\begin{align}
\mathcal{F} = \left(\frac{\ptb^2}{\ptb^2 + \ptsupp^2}
\;\frac{\ptbbar^2}{\ptbbar^2 + \ptsupp^2}\right)^i
\,,
\end{align} 
where we use $i=3$ in our calculation.

The second problem is a purely numerical one and is related to the
numerical computation of one-loop amplitudes through \cuttools\
\cite{Ossola:2007ax}, as implemented in \helaconeloop.  In order to
control numerical instabilities, we developed a dd-precison version of
the program, the \helaconeloopdd\ code~\cite{Kardos:2013vxa}, which is
a straightforward extension of \helaconeloop\ to double-double
precision using \qd\ \cite{hida00}. 

\section{Production of decaying \tT-pairs}
\label{sec:ttdecay}

The simplest way of implementing the decay of heavy particles in a NLO
computation matched with SMC is to generate on-shell particles and then
let them decay by the SMC, corresponding to decay in the DCA. This
procedure has been criticized strongly \cite{Melnikov:2011qx} because it
neglects spin correlations. The other extreme is to include the
off-shell effects of the heavy particles throughout the computation,
which however can easily lead to very cumbersome computations.

\subsection{Angular correlations in particle production}

In \Ref{Frixione:2007zp} a method was presented to implement
decays in a way such that angular correlations between decay
products of a heavy particle and the rest of the event are taken into
account. It is an improvement to the DCA with including 
angular correlations at LO accuracy.  This method is based on a
hit-and-miss technique through the following steps:
\begin{enumerate}
\item
Generate LHEs with heavy particles in the final state (\tT-pair in
our case). We denote the collection of the kinematic variables that
characterize such an event with \Phiu. Compute the squared matrix element
\sme{\Phiu} for the event.
\item
For each LHE, generate the four-momenta of the decay products uniformly
within the decay phase space of the corresponding parent particle. The
corresponding kinematics is described by \Phid, with
$\ud \Phid = \ud \Phiu \prod_i \ud \Phi_{h_i\to l_{i_1}l_{i_2}\dots}$.
Compute the squared matrix element \sme{\Phid} for the event with decay
products in the final state.
\item
Generate a uniform random number
$r \in [0,U_\ud(\{\Phi_{h_i\to l_{i_1}l_{i_2}\dots}\})]$, where
$U_\ud(\{\Phi_{h_i\to l_{i_1}l_{i_2}\dots}\})$ is an upper bound for
the ratio $R = \sme{\Phid}/\sme{\Phiu}$. If $R \leq r$, then discard
\Phid\ and return to step 2.
\item
Otherwise, replace the momenta of the heavy particles by those of their
decay products as obtained in step 2 ($\Phiu \to \Phid$).
\end{enumerate}
We implemented this method in a program called \decayer, together with
a further improvement which takes into account the off-shell effects of
the decaying particles, as described in the next subsection.

\subsection{Reinstating off-shell effects}
\label{subsec:offshell}

The program \powhel\  generates events with the undecayed phase space
$\Phiu$ (specified explicitly in \eqn{eq:Phiu}). We would like to
reweight these events with a Breit-Wigner distribution.  In the
following we describe the method implemented in the program \decayer\
to perform the decay of the on-shell heavy quarks, generated by
\powhel, with off-shell effects also taken into account.  We denote the
momenta of the heavy particles in $\Phiu$ by $\{q_i\}_{i=1}^n$. In
addition to the $n$ heavy particles, we also allow for $m$ light
particles with momenta $\{p_j\}_{j=1}^m$ in the undecayed event.  Our
aim is to generate a decayed event, with a corresponding phase space
point in $\Phid$ with momenta $p_i^\mu$ and $\{k_{i_1}, k_{i_2}\}_{i=1}^n$,
where $k_{i_1}^\mu + k_{i_2}^\mu = \tq_i^\mu$, with off-shell
momenta $\tq_i^\mu$ obtained from the momenta of the undecayed
particles using the procedure described below.

The phase space $\Phid$ for total incoming four-momentum $Q^\mu$ can be
written in a factorized form,
\beq
\ud \Phi_{2n+m}\Big(Q;\{k_{i_1}, k_{i_2}\}_{i=1}^n, \{p_j\}_{j=1}^m\Big) =
\ud \Phi_{n+m}\Big(Q;\{\tq_i\}_{i=1}^n, \{p_j\}_{j=1}^m\Big)
\prod_{i=1}^n \frac{\ud \tq_i^2}{2 \pi} \ud \Phi_2(\tq_i; k_{i_1}, k_{i_2})
\,.
\label{eq:PS}
\eeq
In \eqn{eq:PS} the two-particle phase space measures can be written as
\beq
\ud\Phi_2(\tq;\,k_1\,,k_2) = \frac{1}{32\pi^2}
\frac{\lambda(\tq^2,k_1^2,k_2^2)}{\tq^2}\ud\Omega
\label{eqn:twopartPS}
\,,
\eeq
where $\ud\Omega = \ud\phi\,\ud\cos\theta$ is the surface element of the
unit sphere. Here $\lambda$ is the triangle function
\beq
\lambda(a,b,c) = \sqrt{a^2 + b^2 + c^2 - 2ab - 2ac - 2bc}
\,,
\eeq
which can be used to express the magnitude of the three-momentum for
one of the decay products, $|\boldsymbol{k}_1^{(\tq)}|$, in the rest
frame of $\tq$ (hence the upper index $(\tq)$) as
\beq
|\boldsymbol{k}_1^{(\tq)}| =
\frac{1}{2\sqrt{\tq^2}}\lambda(\tq^2,k_1^2,k_2^2)
\,.
\label{eqn:twopartPSmom}
\eeq
In this frame $\boldsymbol{k}_1^{(\tq)}+\boldsymbol{k}_2^{(\tq)}=0$. 

The squared matrix elements contain a Breit-Wigner factor
$\mathcal{F}_{\mathrm{BW}}(s_i)$ for the propagator of each heavy
particle $i$. In the narrow width approximation $\Gamma_i/m_i \to 0$, and
we can make the replacement
\beq
\lim_{\Gamma_i/m_i\to 0} \mathcal{F}_{\mathrm{BW}}(s_i) =
\frac{\pi}{m_i \Gamma_i} \delta(s_i-m_i^2)
\,,
\eeq
($s_i = \tq_i^2$), which puts particle $i$ onto its mass-shell.  With
this replacement we can perform the integrations over all $\ud \tq_i^2$,
and obtain
\bal
\bsp
\ud \Phi_{2n+m}\Big(Q;\{k_{i_1}, k_{i_2}\}_{i=1}^n, \{p_j\}_{j=1}^m\Big)
&\to \ud \Phi_{n+m}\Big(Q;\{q_i\}_{i=1}^n, \{p_j\}_{j=1}^m\Big)
\\[2mm] &\times 
\prod_{i=1}^n \frac{1}{2\pi} \ud \Phi_2 (m_i; k_{i_1}, k_{i_2})
\,,
\label{eq:PSDCA}
\esp
\eal
where all $q_i$ are on shell ($q_i^2 = m_i^2$). The phase space on the
right hand side of \eqn{eq:PSDCA} is that in DCA. The first factor 
corresponds to the undecayed phase space $\Phiu$,
\bal
\bsp
\ud \Phi_{n+m}\Big(Q;\{q_i\}_{i=1}^n, \{p_j\}_{j=1}^m\Big)
& = \prod_{i=1}^n \frac{\ud^3 q_i}{(2\pi)^3 2\omega(q_i)}
  \prod_{j=1}^m \frac{\ud^3 p_j}{(2\pi)^3 2\omega(p_j)}
\\[2mm] &\times
(2\pi)^4 \delta^{(4)}\Big(\sum_i q_i + \sum_j p_j - Q\Big)
\,,
\label{eq:Phiu}
\esp
\eal
with $\omega(p) = \sqrt{p^2+\boldsymbol{p}^2}$, while the second one is
\beq
\prod_{i=1}^n \frac{1}{2\pi} \ud \Phi_2 (m_i; k_{i_1}, k_{i_2})
= \prod_{i=1}^n \frac{1}{(4\pi)^3}
\frac{\lambda(m_i^2,k_{i_1}^2,k_{i_2}^2)}{m_i^2}\ud\Omega_i
\,.
\eeq

We now turn to the reweighting of the undecayed events with momenta
$q_i^\mu$ with a Breit-Wigner function.  In going from $\Phiu$ to
$\Phid$, we reverse the arrow in \eqn{eq:PSDCA} by inserting
\beq
1=\prod_{i=1}^n \int\!\ud \tq_i^2 \frac{m_i \Gamma_i}{\pi} 
\mathcal{F}_{\mathrm{BW}}(\tq_i^2)
\,,
\eeq
and multiplying by the factor
\beq
\mathcal{F}_1 = \prod_{i=1}^n
\frac{\omega(q_i)}{\omega(\tq_i)}\,
\frac{\lambda(\tq_i^2,k_{i_1}^2,k_{i_2}^2)}
{\lambda(m_i^2,k_{i_1}^2,k_{i_2}^2)}\,
\frac{m_i^2}{\tq_i^2}
\,.
\label{eq:f1factor}
\eeq
Of course, inserting the first factor is formal.  We actually generate
the virtualities $\tq_i^2$ according to the Breit-Wigner distribution,
following the procedure described in \sect{sec:WWbb}. Using the
virtualities $\tq_i^2$, we construct the four momenta by keeping the
three-momenta of the original event ($\boldsymbol{q_i}$) and changing
the energy components $E_i$ of the heavy particles according to
\beq
\tE_i = \sqrt{\tq_i^2 + \boldsymbol{q_i}^2}
\,.
\eeq
This way three-momentum conservation is maintained, but energy
conservation is lost, 
\beq
E_\cm = \sum_{i=1}^n E_i +  \sum_{j=1}^m E_j
\ne \tE_\cm = \sum_{i=1}^n \tE_i +  \sum_{j=1}^m E_j
\,.
\eeq
We reinstate energy conservation by rescaling the momentum fractions of
the incoming partons $x_a$ and $x_b$ uniformly,
\beq
\tx_a = \sqrt{\frac{\ts}{s}}\,x_a
\,,\qquad
\tx_b = \sqrt{\frac{\ts}{s}}\,x_b
\,,
\eeq
where $\sqrt{s} = E_\cm$ and $\sqrt{\ts} = \tE_\cm$.  Thus the total
partonic center of mass energy is 
\beq
\sqrt{\tx_a \tx_b S} = \sqrt{\ts}
\,,
\eeq
and energy conservation is recovered.
The original \powhel\ weights include the PDFs in
\eqn{eq:factorization}. The rescaling of the momentum
fractions induces a rescaling of the event weights by a factor 
\beq
\mathcal{F}_2 =
\frac{f_{a/A}(\tx_1,\muf)\,f_{b/B}(\tx_2,\muf)}
     {f_{a/A}(x_1,\muf)\,f_{b/B}(x_2,\muf)}
\label{eq:f2factor}
\,.
\eeq
A practical application of this method to the case of \tT-decays is
provided in Appendix A. 

\section{Checks \label{sec:checks}}

The \powhel+SMC framework allows for predictions at different levels  
of evolution during the collision. First, to check correct
implementation of the computations, we compared our \powhel\
predictions at the NLO accuracy to those computed in
Ref.~\cite{Bevilacqua:2010qb} for the LHC at center-of-mass energy
$\sqrt{s} = 7$\,TeV, and found agreement.  In making these
comparisons, we adopt the PDG \cite{Nakamura:2010zzi} values for the
parameters, already used by Ref.~\cite{Bevilacqua:2010qb}:
\beq
\begin{tabular}{lll}
 $m_W=80.398\,\GeV$ &$m_z=91.1876\,\GeV$ &$m_\tq=172.6\,\GeV$
\\[2mm]
 $\Gamma_W=2.141\,\GeV$ &$\Gamma_Z=2.4952\,\GeV$ &$\Gamma_\tq=1.35\,\GeV$\,.
\end{tabular}
\eeq
However, when making new predictions we changed the values of $m_\tq$ and
$\Gamma_\tq$ as specified in \sect{sec:pheno}.  The renormalization and
factorization scales were chosen equal to \mt, and we used the
\texttt{CTEQ6m} parton distribution functions from \texttt{LHAPDF}.
The strong coupling $\alpha_s$ was computed with 2-loop running with
$\Lambda_5 = 226$\,MeV.  These are the choices adopted in
\Ref{Bevilacqua:2010qb}.

Next, to check the
reliability of the matching procedure, we compared our \powhel\
predictions from the LHEs (i.e.~distributions obtained from events
including no more than the first radiation emission) with the NLO
ones.  These checks were performed on all observables studied by
Ref.~\cite{Bevilacqua:2010qb}, by exactly applying the same system
of cuts presented in that paper. In general, the overall agreement 
turned out to be acceptable, even if slightly worse as compared with the
typical values obtained in our previous papers in the study of less
complicated 2 $\rightarrow$ 3 hard scattering processes matched to
SMC~\cite{Kardos:2011qa,Garzelli:2011vp,Garzelli:2011is}. The
differences can be ascribed to the larger NLO K-factor, which allows
for less agreement between the NLO and LHE predictions (see
\eqn{eq:LHEacc3}).  

\begin{figure}
\begin{center}
\includegraphics[width=0.49\textwidth]{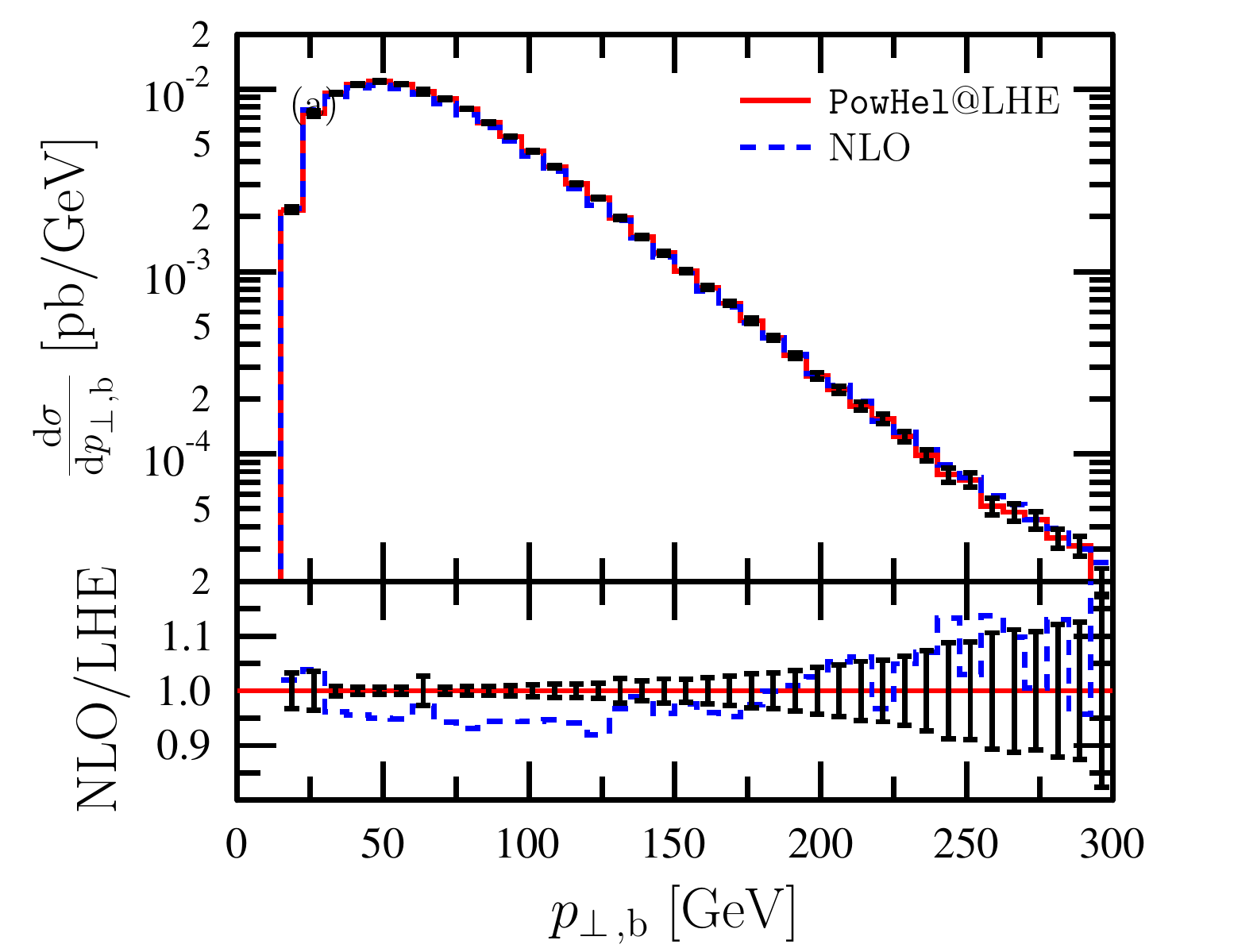}
\includegraphics[width=0.49\textwidth]{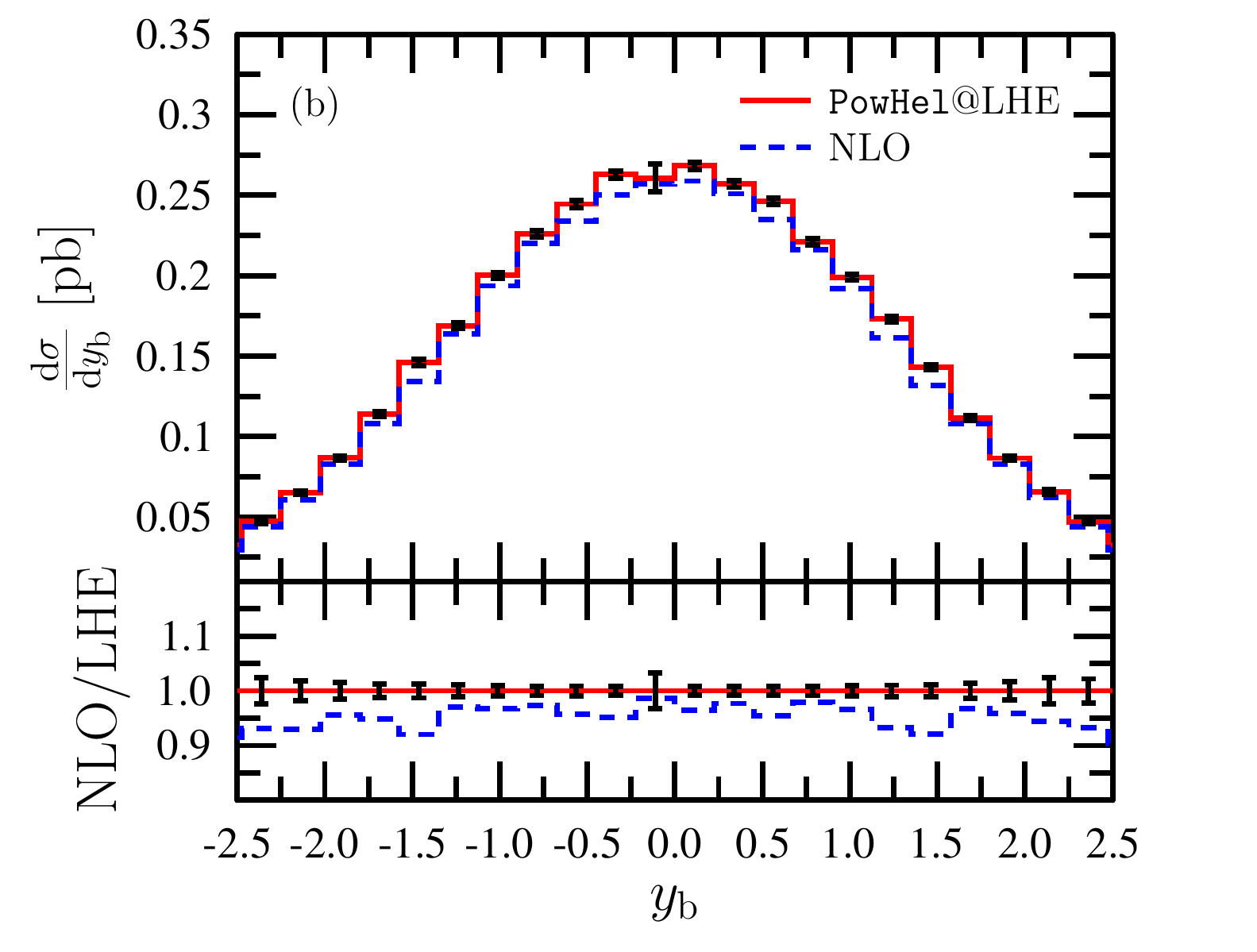}
\includegraphics[width=0.49\textwidth]{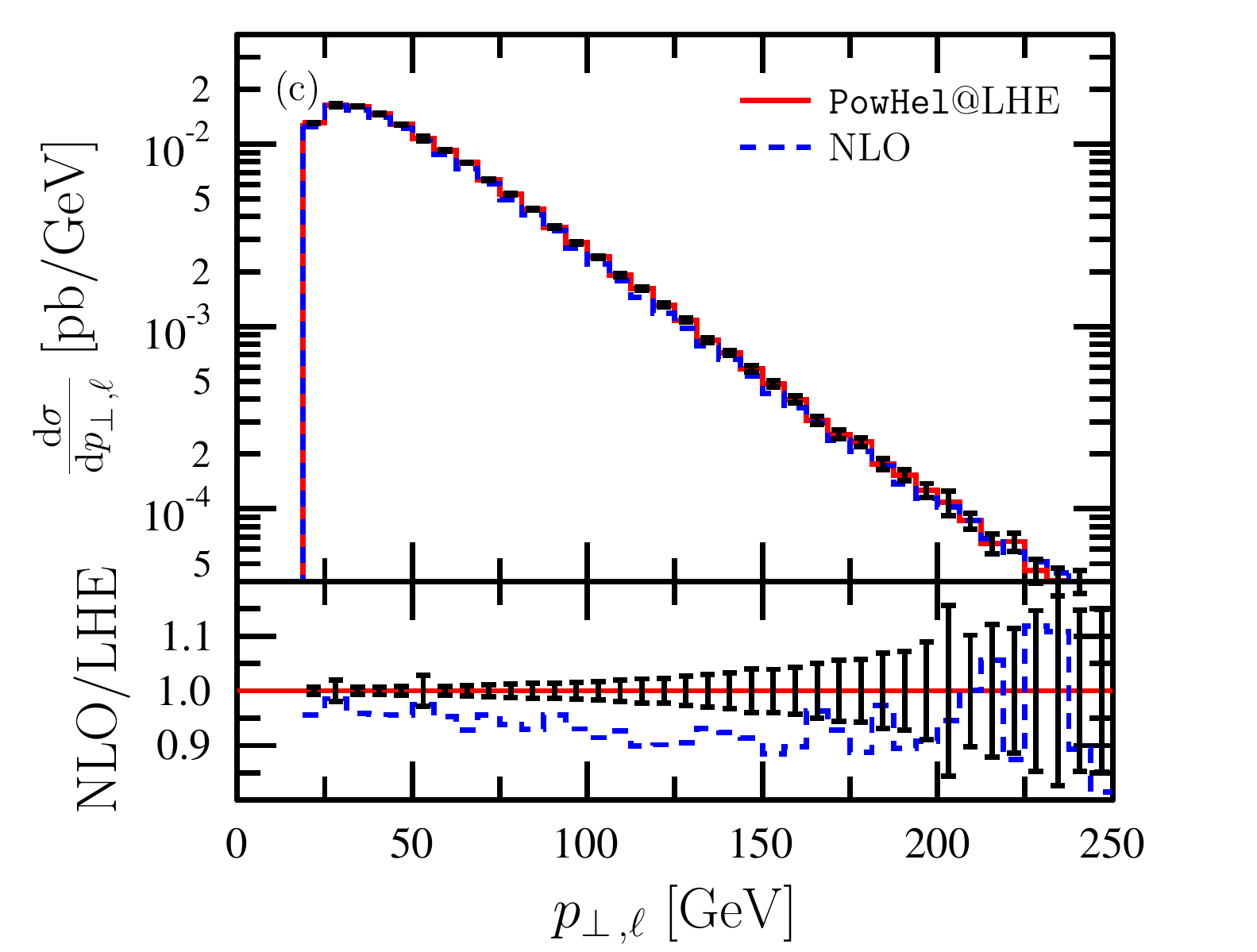}
\includegraphics[width=0.49\textwidth]{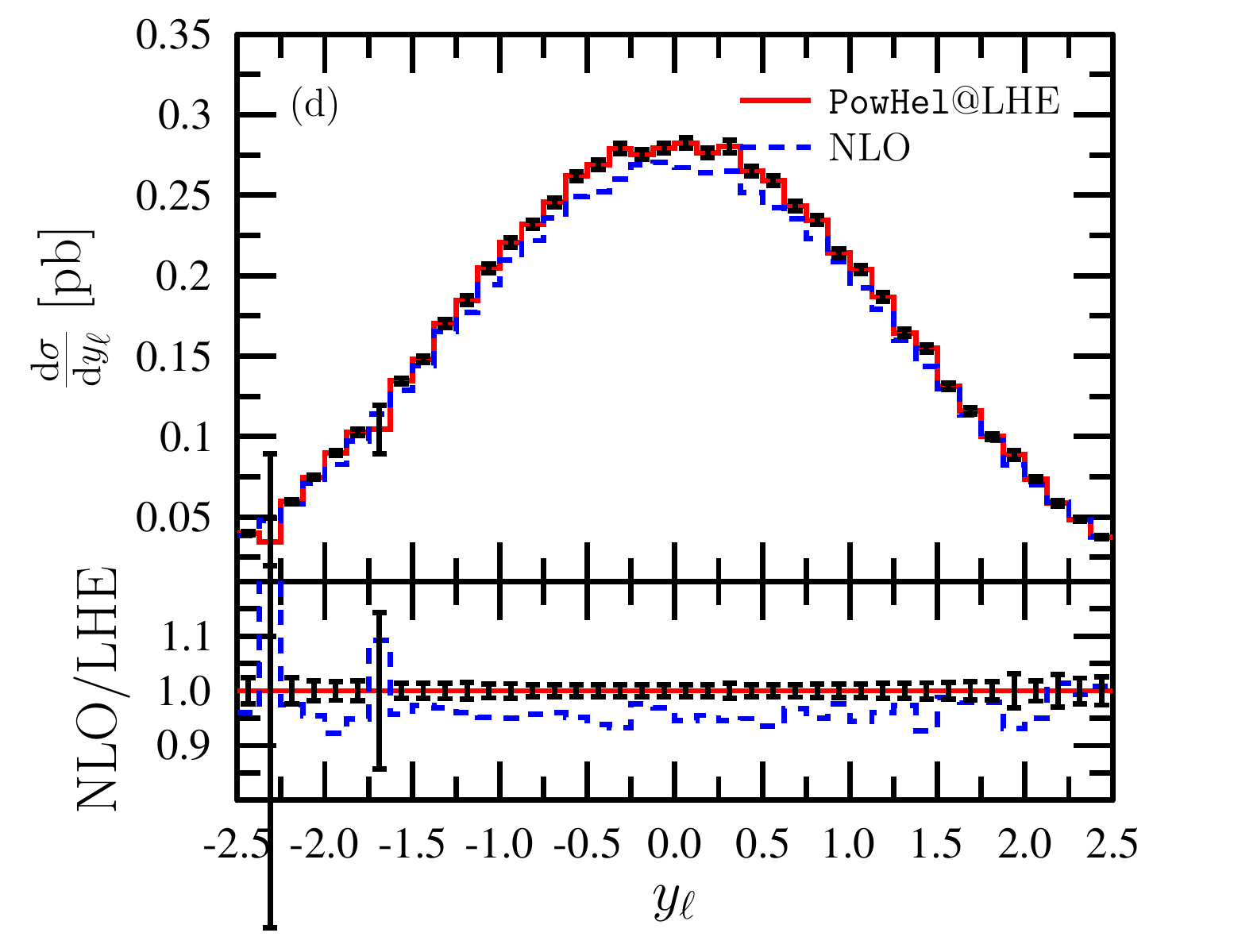}
\includegraphics[width=0.49\textwidth]{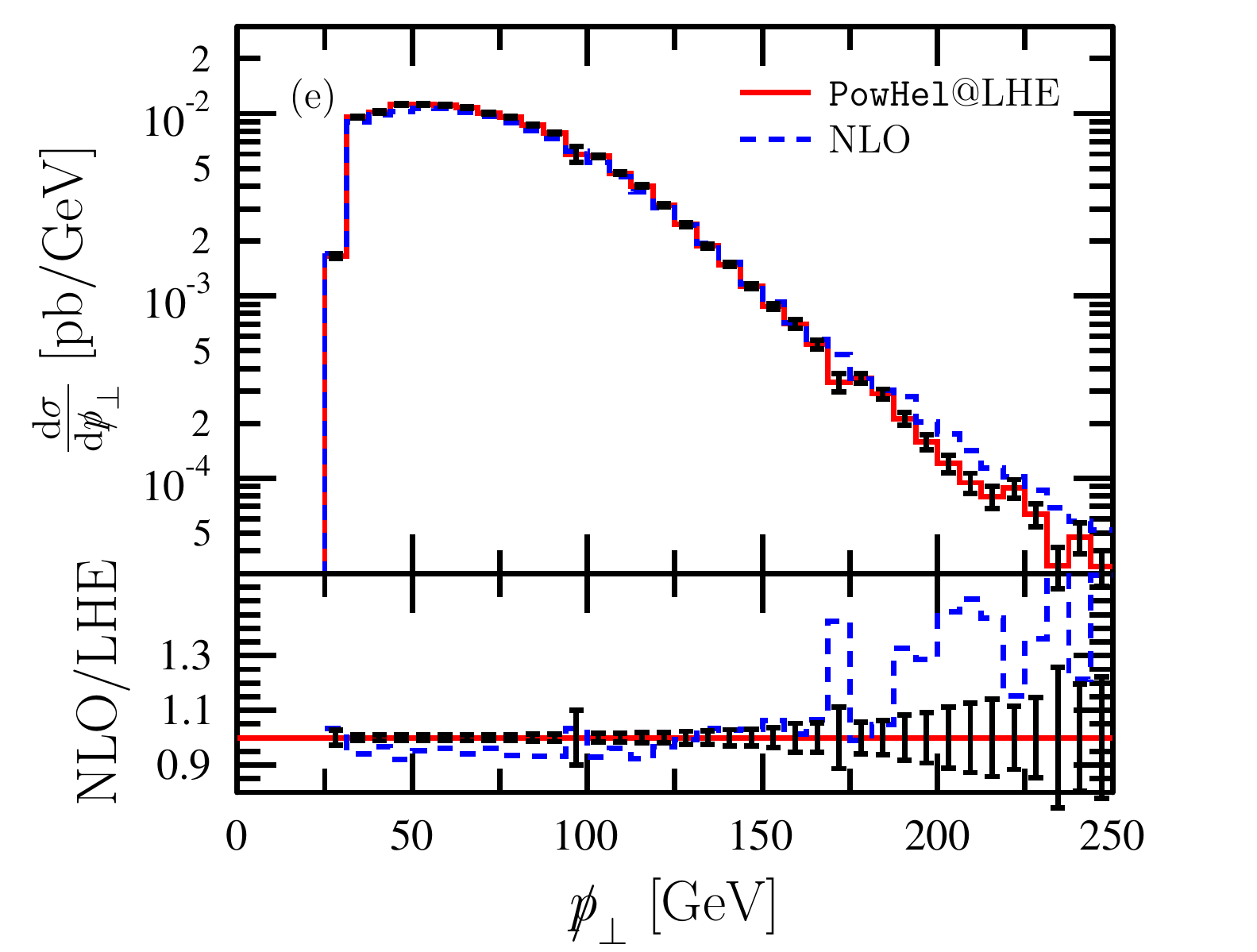}
\includegraphics[width=0.49\textwidth]{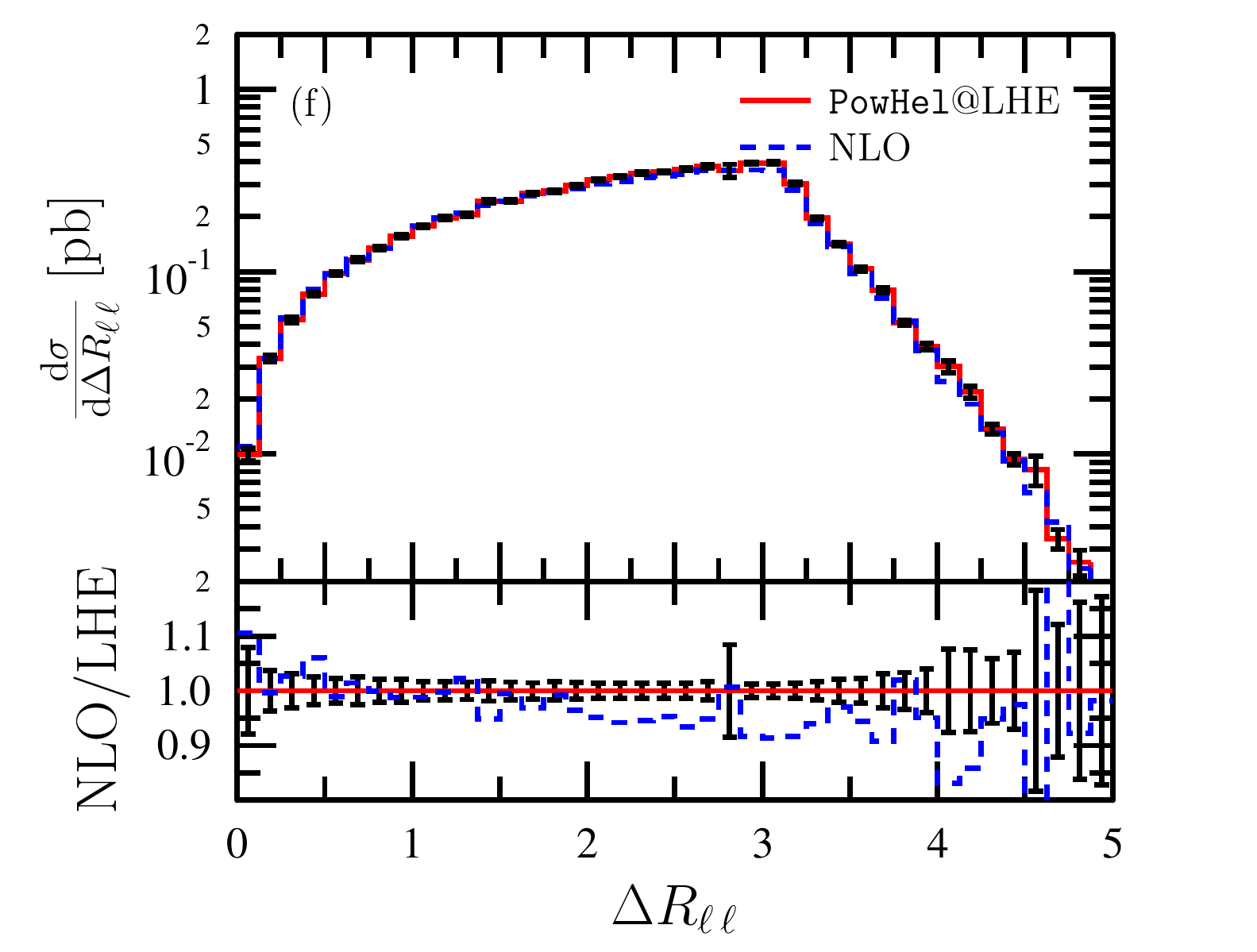}
\caption{\label{fig:nlo-lhef}Distributions of
a) transverse momentum,
b) rapidity of the b-quark,
c) transverse momentum,
d) rapidity of the charged lepton,
e) missing transverse momentum and of 
f) $\Delta R$-separation of the lepton--anti-lepton
pair at NLO and at LHE accuracy.  The lower inset shows the ratio
LHE/NLO with statistical uncertainties of the \powhel\ predictions.}
\end{center}
\end{figure}

The comparisons are shown in \fig{fig:nlo-lhef}, where the transverse
momentum (a) and rapidity of the b-quark (b) and the charged
lepton (c, d) are plotted together with the $\sigma_\nlo/\sigma_\lhe$
ratios shown in the lower inset of each figure. By looking at the
figures it is clear that the rapidity from the LHEs overestimates the
rapidity at the NLO accuracy by 3--5\,\%, whereas the shape of the
distribution is unaffected.  The shapes of the \pt-distributions
from the LHEs are slightly distorted (i.e. within a few percent) with
respect to the NLO ones (taking into account the statistical
uncertainties of both computations).  These results are within the
band of NLO scale dependence and compatible with the expectation about
the perturbative accuracy of LHE predictions given in \eqn{eq:LHEacc3}
that allows for increasing difference with increasing NLO K-factor
\cite{Alioli:2008tz,Oleari:2011ey}. For the present case the inclusive
NLO K-factor is close to 1.5 \cite{Bevilacqua:2010qb}, which gives
the small but noticeable difference between the NLO and LHE predictions.

In \fig{fig:nlo-lhef} we show two more plots comparing the LHE and NLO
predictions. The first is the missing transverse momentum in
\fig{fig:nlo-lhef}.e that is relevant for new-physics searches
based on missing transverse energy plus jets and leptons.  We see
similar level of agreement as for the case of the other transverse
momenta up to about 150\,GeV. In the hard tail the LHEs give
significantly smaller cross section than the prediction at NLO. 
This difference may be due to the higher order corrections exhibited
on the right hand side of \eqn{eq:LHEacc3}, or simply to the lower
statistics in the tail. The NLO K-factor is very
large in this region (in the range of 2--4) \cite{Maestre:2012vp},
therefore, this difference is well within the NLO scale-dependence. 
Nevertheless, in order to have a quantitative understanding of
the accuracy of our predictions for the missing transverse momentum
above 150\,GeV we would need much more events, which is beyond the
scope of our computational resources at present.
This lack of quantitative understanding of
the accuracy is also true for the tail of the \pt-distribution of the
hardest \bB-dijet system, not shown here. Finally, we show the $\Delta
R$-separation of the charged lepton-antilepton pair in \fig{fig:nlo-lhef}.f. This
distribution is sensitive to the spin correlations between production and
decay of the charged leptons. In this case, we find that the agreement
between the distributions from the LHEs and at the NLO accuracy is
very good over the whole kinematic range.

\section{Phenomenology \label{sec:pheno}}

As mentioned before, we consider and compare three different cases:
\begin{enumerate}
\item
one is the computation of the $\pp \to (\WWbB \to)\; \epmubB+X$
process at NLO accuracy, including t-quarks with finite width
throughout, as well as $W$ bosons. The $W$ bosons decay into
lepton pairs through a propagator involving the complex $W$ mass and
the lepton current.
\item
another is the computation of the $\pp \to \tT$ process at NLO
accuracy.  The generated t-quarks are on shell and their decays,
$\bar{{\rm t}} \to W^-\,\bar{{\rm b}}$ and t $\to W^+$\,b are computed
by the SMC in the DCA neglecting spin correlations. The decays of the
$W$ bosons are also performed by the SMC.
\item
a third corresponds to the computation of the $\pp \to \tT$ process at NLO
accuracy. In this case, the heavy quarks, produced on shell, are pushed
off shell and decayed by means of the \decayer\ program, taking into
account spin correlation effects, as described in \sect{sec:ttdecay}. 
The decays of the $W$ bosons are also performed by \decayer.
\end{enumerate}

We make predictions at the hadron level.  As mentioned earlier, for
making predictions we adopted the value of the t-quark
mass $m_\tq = 173.2$\,GeV \cite{Lancaster:2011wr}. Correspondingly, we
use $\Gamma_\tq^\nlo = 1.32$\,GeV for case 1 and $\Gamma_\tq^\lo =
1.41$\,GeV for cases 2 and 3 \cite{Worek:private}.  As for the SMC,
to simulate the PS and hadronization, we used \pythia~6.4.28
\cite{Sjostrand:2006za}.  We used both the untuned version
(denoted by \py1), providing a virtuality-ordered PS, and a further
version, tuned to the Perugia 2011 set of values~\cite{Skands:2010ak},
one of the recent tunes, providing a \pt-ordered PS, updated on the
basis of recent LHC data (denoted by \py2). We also used 
\herwig~6.520~\cite{Corcella:2002jc} (denoted by \hw). In order to simplify
the analysis, we kept $\pi^0$s, that can be easily reconstructed from
their $\gamma\gamma$ decay products, stable. All other particles and
hadrons were allowed to be stable or to decay according to the default
implementation of the SMC.  In \herwig\ (anti-)muons were also set 
stable, as in \pythia\ by default.

While the b-quark masses were set to zero in generating the hard-scattering
amplitudes, these were kept to their default values implemented in
\pythia\ in the SMC evolution to $B$-hadrons. Quark masses in \herwig\
were set to the same values as in \pythia\ default.  In the
configuration of the SMC, for all other mass and width parameters,
including light quark masses, we used the default values already
implemented. We use the \texttt{CTEQ6.6m} parton distribution functions
from \texttt{LHAPDF} with $\Lambda_5 = 226$\,MeV and strong coupling
$\alpha_s$ computed with 2-loop running.

For cases 2 and 3 we can potentially consider all $W$ decay channels
(leptonic and hadronic ones). In order to be able to compare our
predictions to those given in \Ref{Bevilacqua:2010qb} (without
including any SMC effects), we nevertheless forced the decay into the
\epmubB\ channel, adopting in both cases a $W$ branching ratio in $e^+$
and $\mu^-$ corresponding to that implemented in \pythia\
($B(W^+ \to e^+ \nu_e) = B(W^- \to \mu^- \bar{\nu}_\mu) =0.108$, also
enforced in \herwig). 

All hadronic tracks with $|\eta| < 5 $ were used to build hadronic
jets. We used the anti-\kt\ algorithm \cite{Cacciari:2008gp} as
implemented in FastJet \cite{Cacciari:2011ma}, with $R=0.4$ and
$R=1.2$, to study the effect of different jet reconstruction
strategies.  While in collider experiments b-jets are reconstructed
with finite tagging efficiencies from the tracks pointing towards
vertices displaced with respect to the primary interaction vertex, in
our theoretical simulations b-jets were tagged by means of the
information included in the {\texttt{MCTRUTH}} parameter available in
the SMC, tracking back their evolution up to their origin as b-quarks,
and b-jet tagging efficiencies were neglected.

In the event generation we apply technical cuts of $\ptb,\:\ptbbar>2$\,
GeV and $m_{\bB} > 1$\,GeV (see \sect{sec:powhel} for details). Taking
into account that physical cuts should always be well above the
technical ones, we consider the following set of physical cuts:
\begin{enumerate}
\itemsep=-2pt
\item
Each jet is required to have transverse momentum $\ptj > 20$\,GeV
and pseudorapidity $|\eta_j| < 5$, otherwise it is not counted among the
jets.
\item
Each of the jets satisfying the 1st condition, to be classified as a
b- or $\bar{{\rm b}}$-jet, is required to be b-tagged and have
$|\eta_{\rm b}| < 3$, due to the geometry of the tracking system.
\item
We require at least one b-jet and one $\bar{{\rm b}}$-jet.
\item
Each charged lepton is required to have $\ptl > 20$ GeV and
$|\eta_\ell| < 2.5$, otherwise it is not counted among the leptons.
\item
We require at least one charged lepton and one charged anti-lepton, that
are isolated from all jets by requiring $\Delta R(\ell,j) > 0.4\,\,(1.2)$
in the azimuthal angle--pseudorapidity plane. If there are more leptons
that pass cut 4, those are kept without isolation from the jets.
\item
We require a minimum missing transverse momentum $\pTmiss > 30$\,GeV.
\end{enumerate}
These cuts present some modifications with respect to those in
\Ref{Bevilacqua:2010qb} providing predictions at the NLO accuracy.

In addition to the three cases listed at the beginning of this section,
we can also compare predictions after parton shower (PS) and after full
SMC (PS + hadronization) simulations, obtained with either \pythia, or
\herwig.  This makes many options for comparison. We decided to present
four groups of plots:
\begin{enumerate}
\itemsep=-2pt
\item
comparison of predictions for \WWbB-production (case 1) from the LHEs,
after parton shower (PS) and after full SMC with \py1 (here we employ
an additional jet veto to the selection cuts (1--6), see below);
\item
comparison of predictions of the three cases from the LHEs;
\item
comparison of predictions of the three cases after full SMC with \py1;
\item
comparison of predictions for case 1 after full SMC for different SMC
programs: \py1, \py2 and \hw.
\end{enumerate}
In order to make connections among these groups of plots, we always
include predictions at the hadron level generated with \py1.

We produced at least 30 distributions for each possible case and each
combination of comparisons. In order to be able to compare the various
effects, we selected six standard plots that we show for each group:
\begin{enumerate}
\itemsep=-2pt
\item
distribution of the transverse momentum of the hardest b-jet (jet
originating from a b-quark), $p_{\bot,\bq_1}$,
\item
distribution of the transverse momentum of the hardest isolated positron,
$p_{\bot,e^+}$,
\item
distribution of the pseudorapidity of the hardest b-jet, $\eta_{\bq_1}$,
\item
distribution of the pseudorapidity of the hardest isolated positron,
$\eta_{e^+}$,
\item
distribution of the invariant mass of the hardest b-jet and the hardest
isolated positron,
$m_{\bq_1 e^+}$
\item
distribution of the azimuthal separation of the hardest isolated
positron and muon $\Delta \phi_{e^+ \mu^-}$.
\end{enumerate}
These fall into three categories:
(i) distributions of the b-quarks only,
(ii) those of the lepton sector, and
(iii) those involving a b-quark and a charged lepton.

\subsection{Predictions at the LHC \label{sec:LHC}}

In order to study the effect of the SMC, first we compare predictions
for \WWbB-production at different levels of theoretical description:
obtained from the LHEs, after PS and after full SMC. For the PS and SMC
we use \py1. In addition to the selection cuts (1--6) we employ a jet
veto on the non b-jets, too. The reason for this jet veto is that in
the LHEs there can be at most one extra jet besides the b- and \baq-jets,
while after PS and SMC there are usually many more (less energetic
ones). Thus the selection cuts affect the latter much more,
decreasing the cross sections significantly, which is more a
consequence of the selection cuts than the effect of the PS and
hadronization. In \tab{tab:sigmaLHC} we show the inclusive cross sections
after selection cuts. The 10\,\% decrease of the cross section after PS
(with jet veto) is mainly due to the different effect of the selection
cuts when applied at different stages of the evolution of the events,
and very similar decrease is observed on the distributions below. The
additional 2\,\% decrease after full SMC however, appears in the
distributions very differently.
\begin{table}
\begin{center}
\begin{tabular}{|c|c|c|}
\hline
\hline
 & cuts (1--6) & cuts (1--6) + jet veto \\
\hline
$\sigma_\lhe$ (fb) & $847\pm 4$ & $442\pm 3$ \\
$\sigma_{\rm PS}$ (fb) & $686\pm 4$ & $398\pm 3$ \\
$\sigma_{\rm SMC}$ (fb) & $630\pm 4$ & $390\pm 3$ \\
\hline
\hline
\end{tabular}
\caption{\label{tab:sigmaLHC}Cross-sections from the LHEs, after PS and
at the hadron level at the LHC after cuts (1--6) (first column) and
after an additional jet veto (second column).
The quoted uncertainties are statistical only.
The PS and SMC predictions are obtained with the \py1 SMC.}
\end{center}
\end{table}

\begin{figure}
\begin{center}
\includegraphics[width=0.49\textwidth]{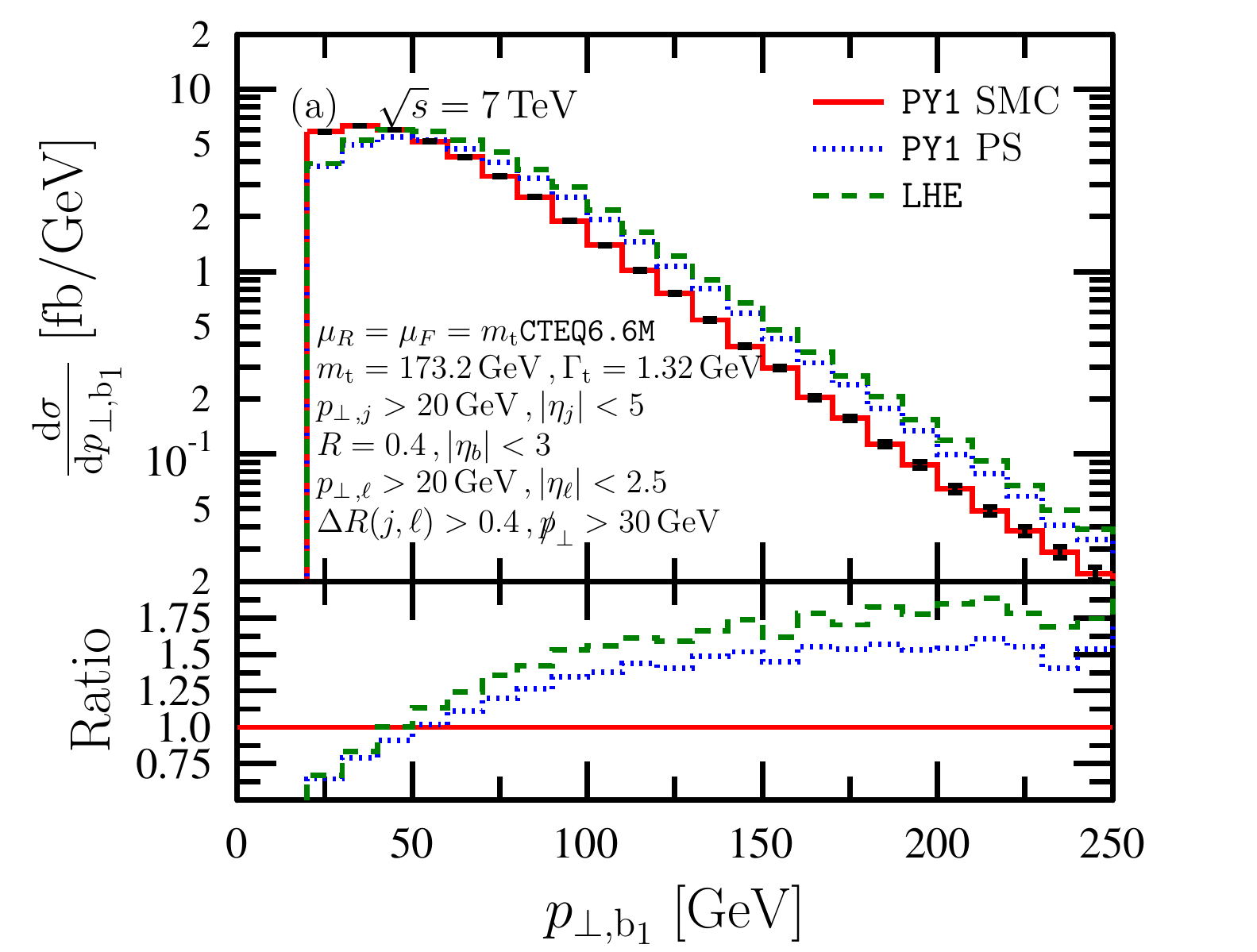}
\includegraphics[width=0.49\textwidth]{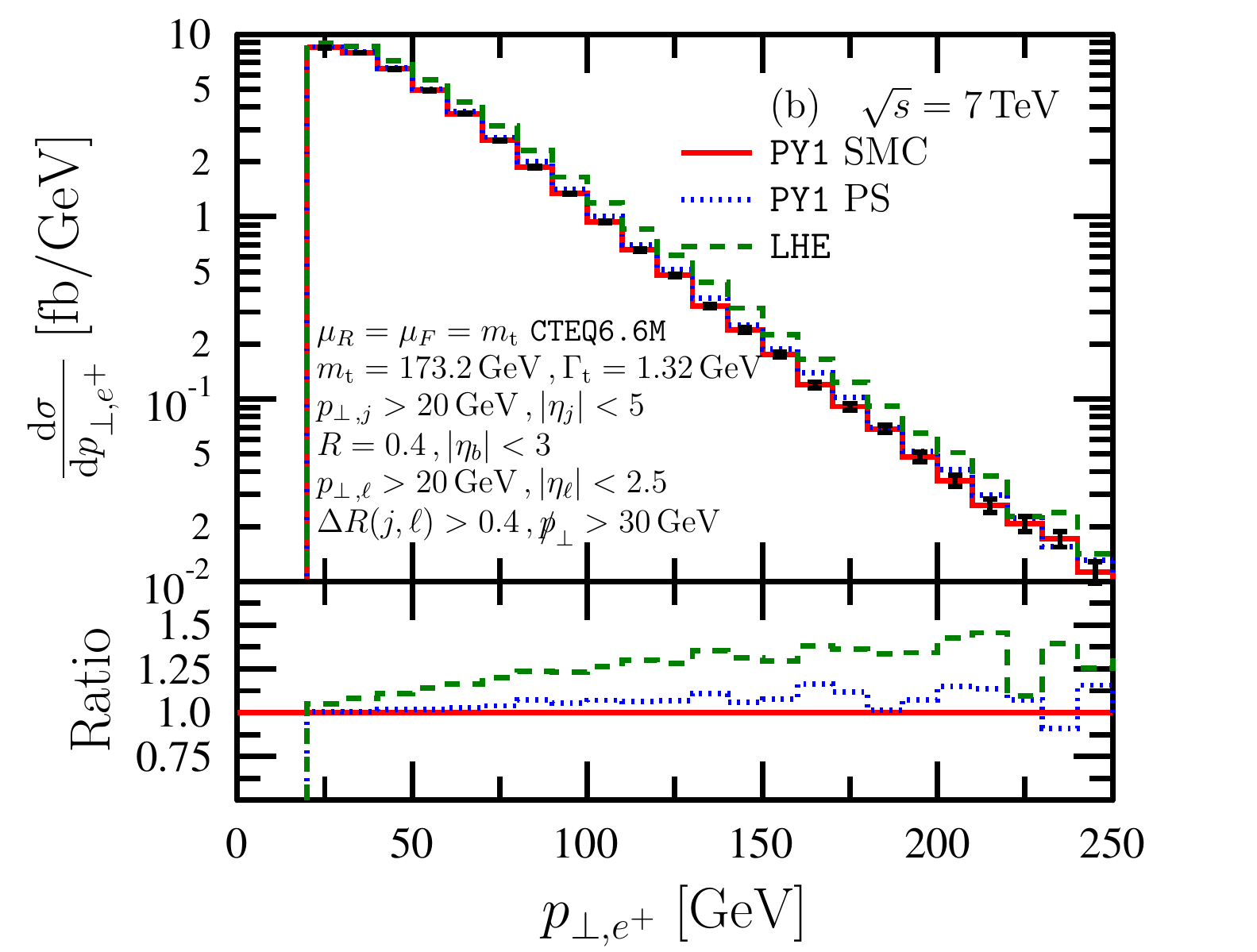}
\caption{\label{fig:pt-jveto} Distributions of transverse momentum of a) 
the hardest b-jet and b) the hardest isolated positron after LHE, after PS
and after full SMC with \py1.
The lower inset shows the ratio of the predictions to the full SMC case.}
\end{center}
\end{figure}

\begin{figure}
\begin{center}
\includegraphics[width=0.49\textwidth]{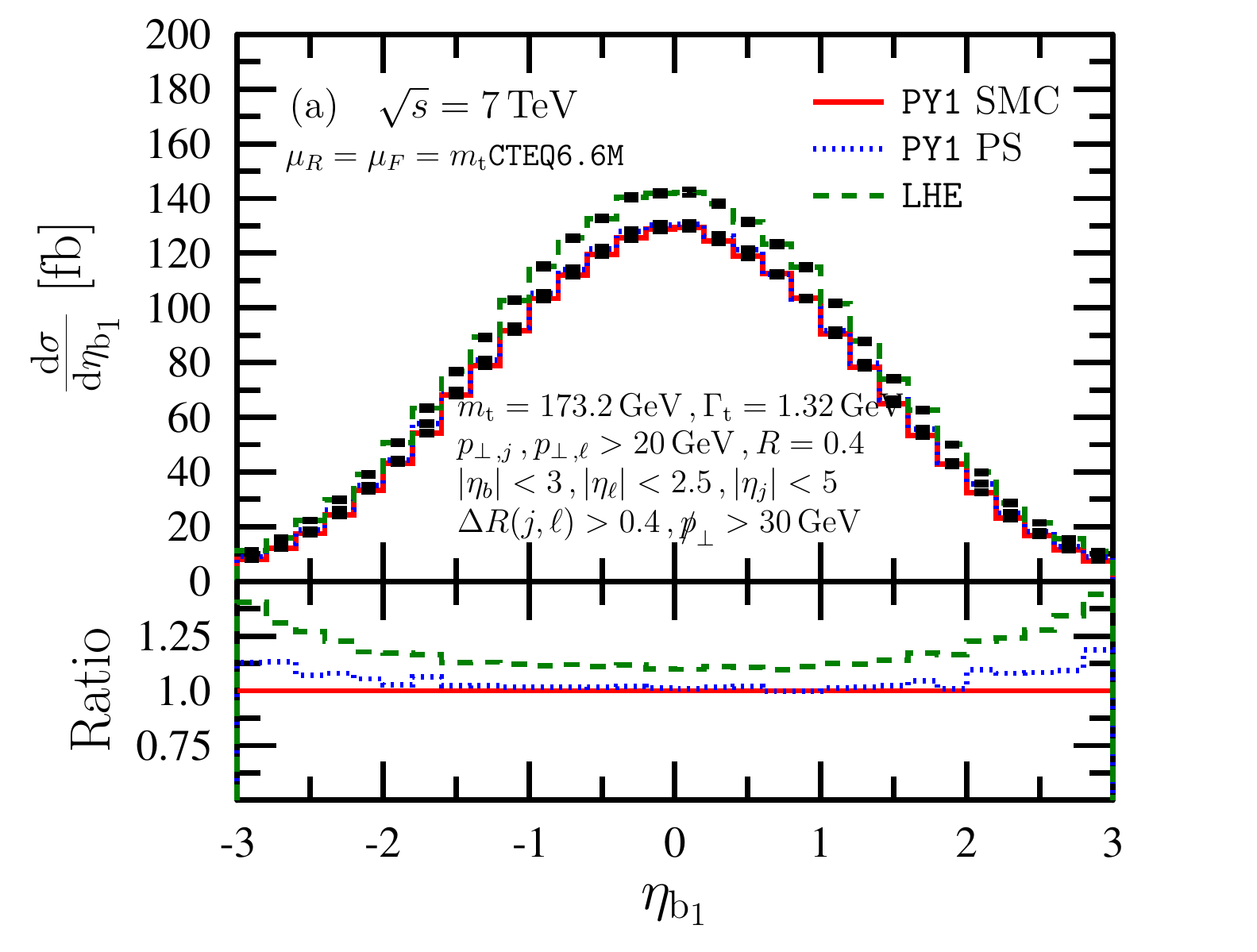}
\includegraphics[width=0.49\textwidth]{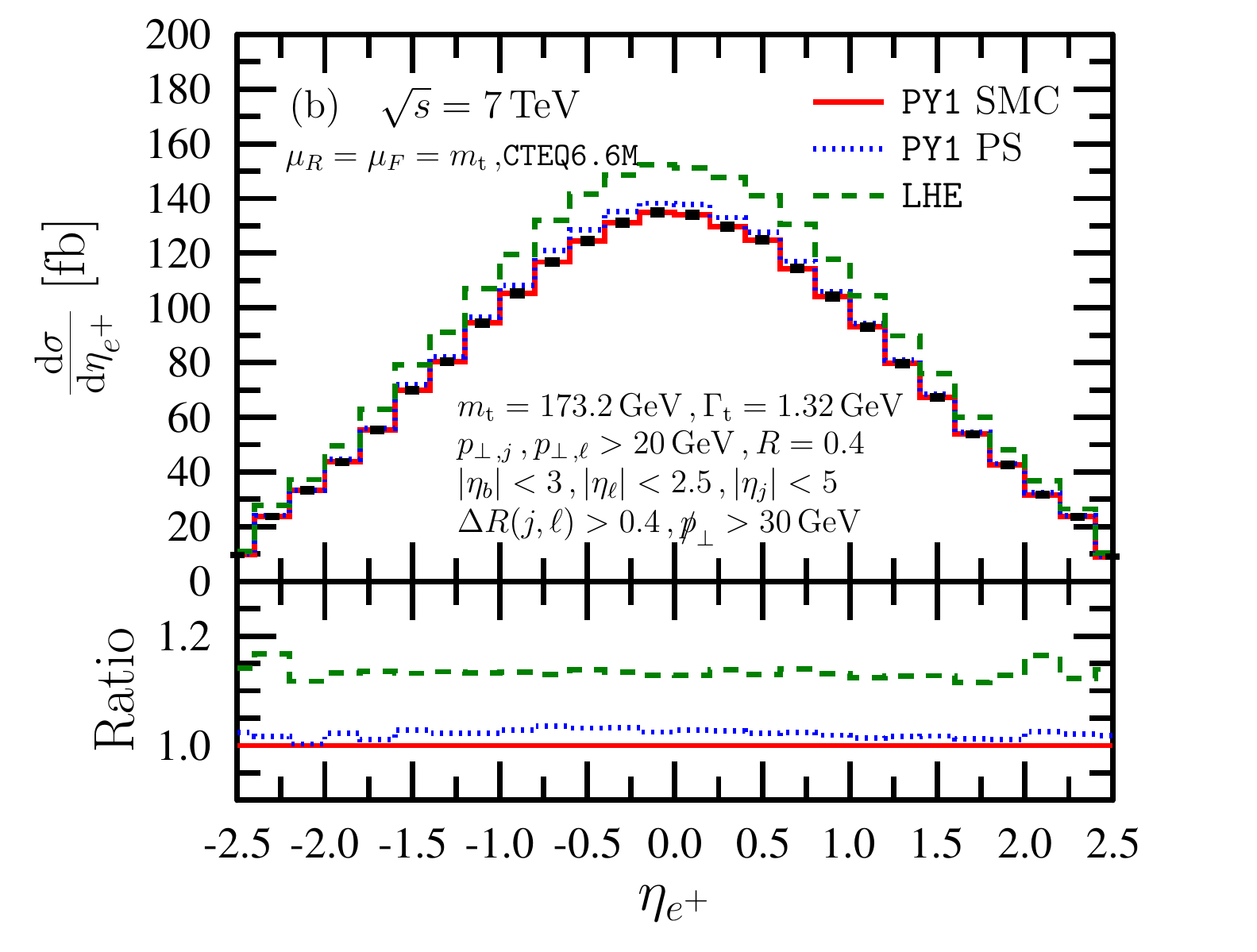}
\caption{\label{fig:eta-jveto} Pseudorapidity distributions of a) 
the hardest b-jet and b) the hardest isolated positron after LHE, after PS
and after full SMC with \py1.
The lower inset shows the ratio of the predictions to the full SMC case.}
\end{center}
\end{figure}

In \fig{fig:pt-jveto} we show the distributions of $p_{\bot,\bq_1}$ and
$p_{\bot,e^+}$ , while in \fig{fig:eta-jveto} we show the pseudorapidity
distributions of the same objects. We find that the effect of the
parton shower is a fairly uniform decrease, it is in the range of
0--20\,\%, independently of the observable, and in general it is about
10\,\%, similarly to the decrease of the inclusive cross section.
The effect of hadronization however, depends on the observable
strongly. For the $p_{\bot,\bq_1}$-distribution it can reach 50\,\% for
\pt\ above 150\,GeV.  Inspection of the curve shows that the large
effect is mainly due to the softening of the jets, a shift of the
distribution by about 25\,GeV. One important reason for this softening
is the decay of the unstable hadrons, which often transforms partonic
energy to electromagnetic and missing energy. In the case of the
positron transverse momentum such decays are absent, and the effect of
hadronization is much smaller, at most 5\,\% apart from the statistical
fluctuations. For pseudorapidity distributions this transformation of
energy does not influence the direction of the jet significantly,
therefore, the effect of hadronization is negligible except for
$|\eta_{\bq_1}| > 2$.

\begin{figure}
\begin{center}
\includegraphics[width=0.49\textwidth]{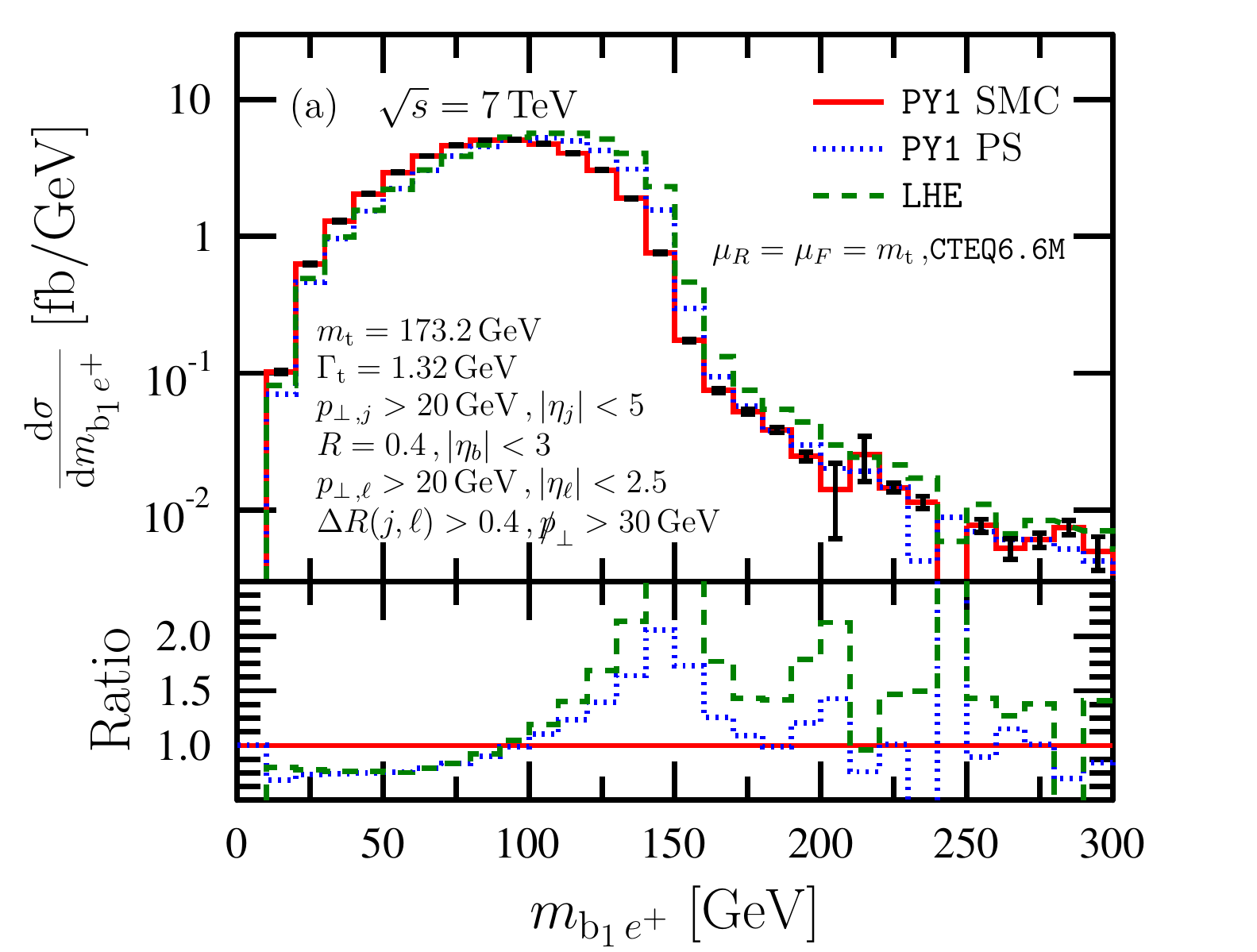}
\includegraphics[width=0.49\textwidth]{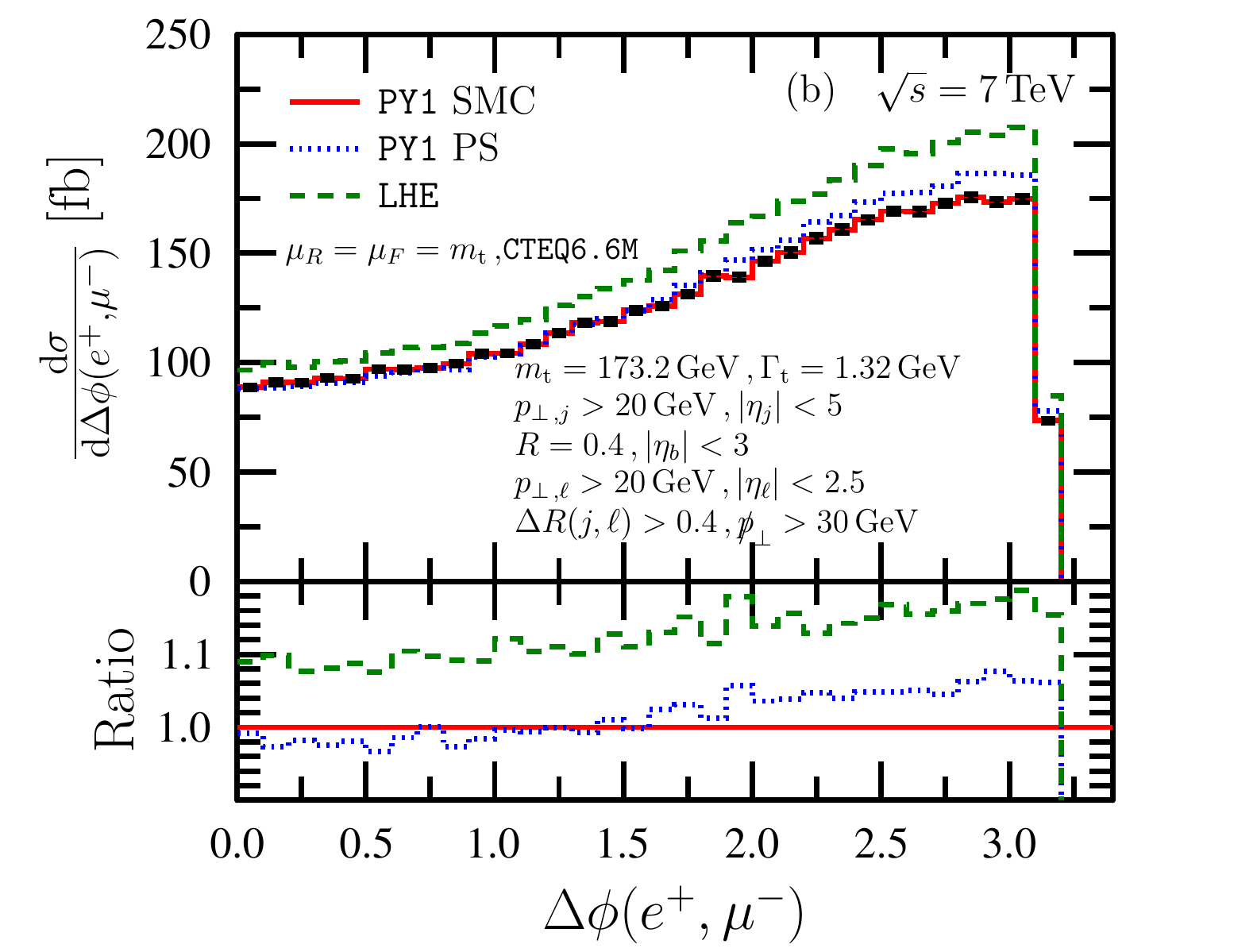}
\caption{\label{fig:mb1ep-jveto} Distributions of
a) invariant mass of hardest b-jet and the hardest isolated
positron and 
b) azimuthal separation between the hardest isolated positron and muon
after LHE, after PS and after full SMC with \py1.
The lower inset shows the ratio of the predictions to the full SMC case.}
\end{center}
\end{figure}

An interesting observable is the invariant mass of the combination of
the hardest b-jet and the hardest isolated positron shown in
\fig{fig:mb1ep-jveto}.a. For decaying on-shell t-quarks into
$W^+ + \bq \to  e^+\, \nu_e$\,b, neglecting the masses of all final decay
products, we have
\beq
m_\tq^2 = p_\tq^2 = m_{W^+}^2 + 2 p_{e^+} p_\bq  + 2 p_{\nu_e} p_\bq 
\,,
\eeq
so $m_{e^+ \bq} \leq \sqrt{m_\tq^2-m_W^2 - m_{\nu_e \bq}^2}$. Thus, at
lowest order in \tT-production, there is a strict kinematic limit for
the invariant mass of the b-quark and the positron at
$\sqrt{m_\tq^2-m_W^2}\simeq 153$\,GeV, that is quite sensitive to
$m_\tq$, which hints that this distribution is useful to measure $m_\tq$
\cite{Biswas:2010sa}. For off-shell t-quarks (e.g.~in a computation at
NLO accuracy) this kinematic limit is smeared, nevertheless there is a
sharp fall of the cross section in the fixed order predictions.%
\footnote{As we show below the singly- and non-resonant contributions
have a significant effect above this limit.}
In \fig{fig:mb1ep-jveto}.a we show that the main effect of the PS
and also that of the hadronization is to smear this sharp edge observed
in the fixed-order computation, as expected. Apart from this region
around 150\,GeV, the corrections of the SMC are modest. 
We show the azimuthal separation between the hardest isolated positron
and muon in \fig{fig:mb1ep-jveto}.b. For this observable, the main
effect of the PS is almost unaffected by the hadronization effects, and
the effect of PS is also modest, varying slightly around 10\,\%.

\begin{figure}
\begin{center}
\includegraphics[width=0.49\textwidth]{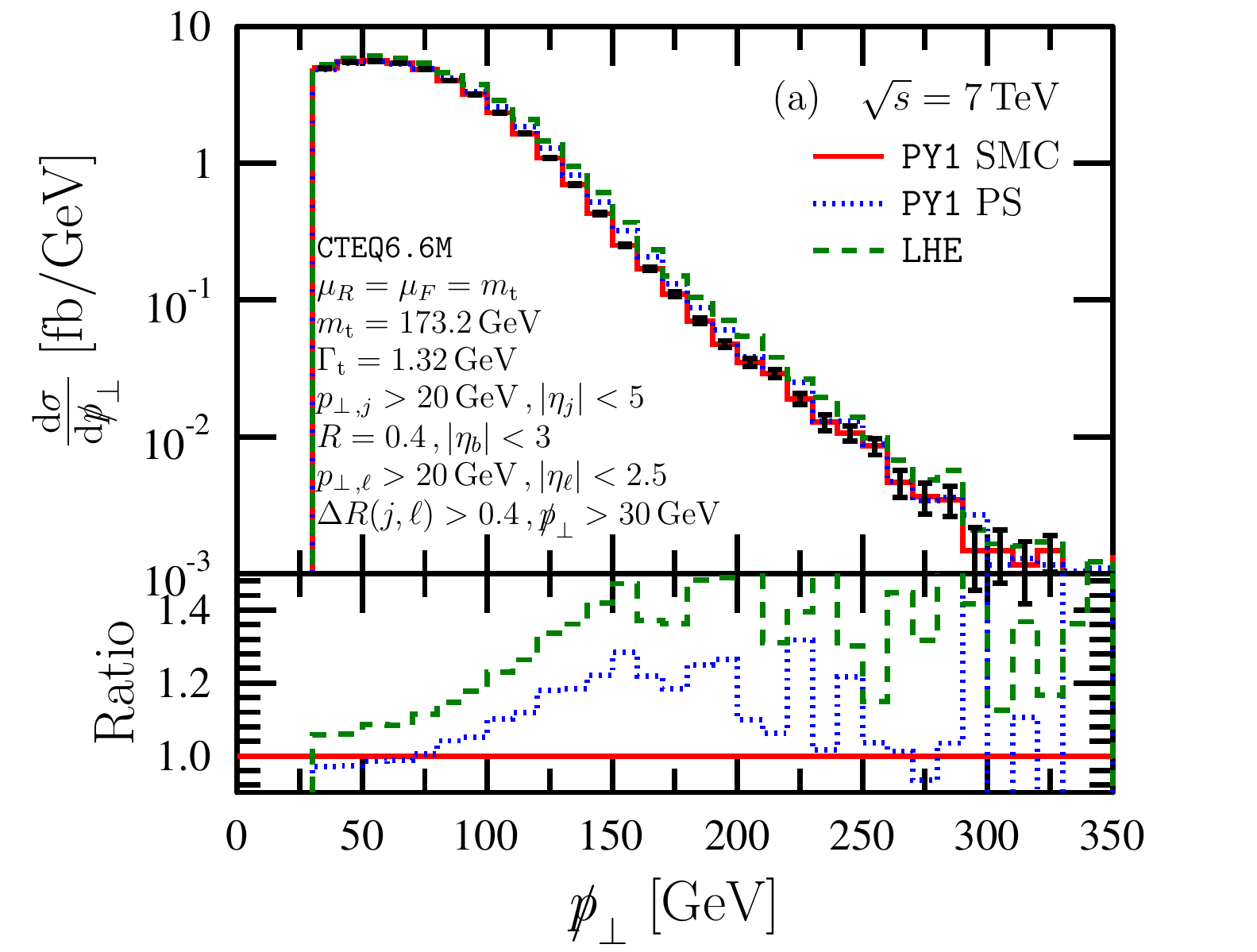}
\includegraphics[width=0.49\textwidth]{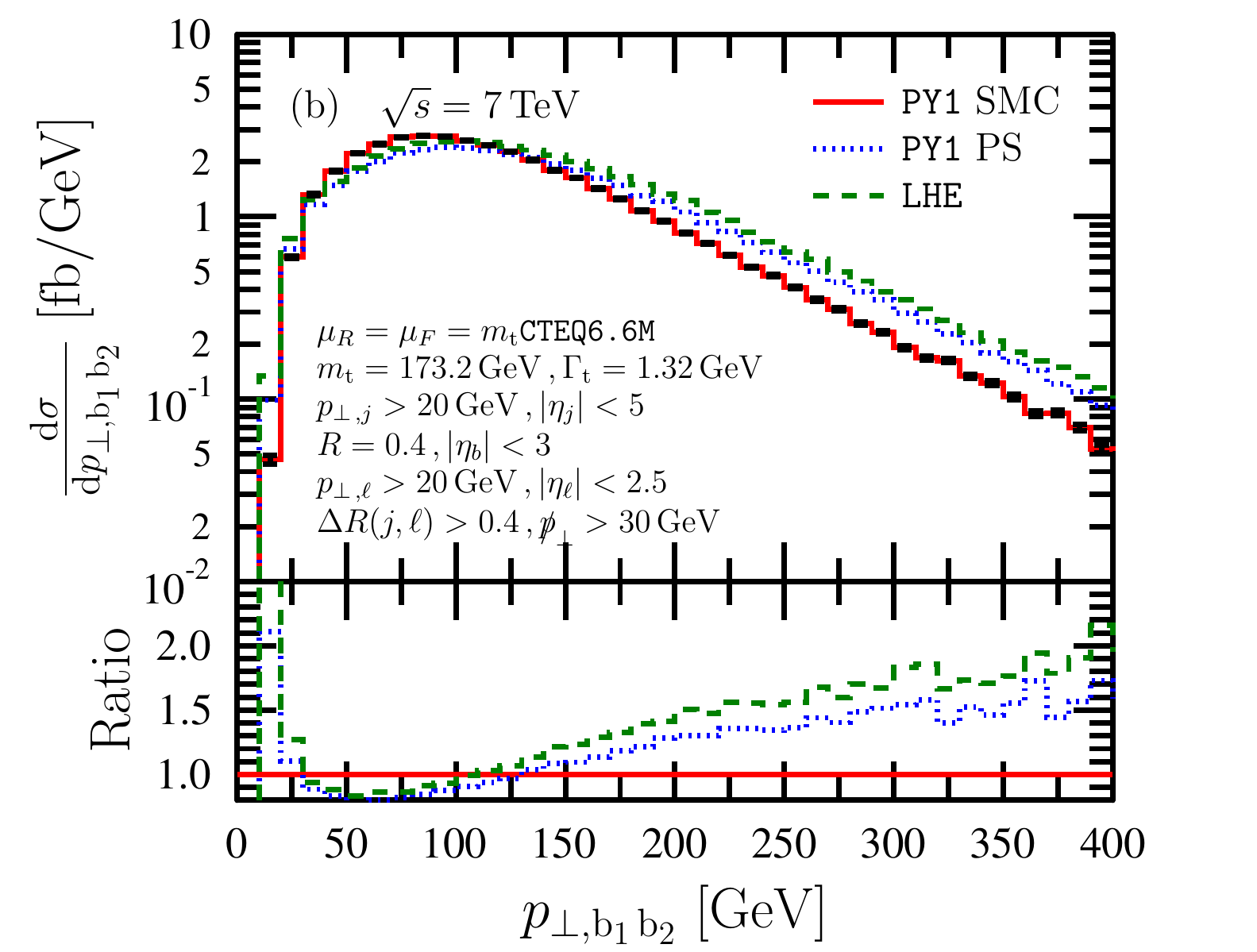}
\caption{\label{fig:ptmiss-jveto} Distributions of
a) missing transverse momentum and of
b) \pt\ of the hardest \bB\ dijet system
after LHE, after PS and after full SMC with \py1.
The lower inset shows the ratio of the predictions to the full SMC case.}
\end{center}
\end{figure}

We show the effect of PS and SMC for two more variables: the
distribution of the missing transverse momentum, \pTmiss, and that of
$\ptbb = \sqrt{(p_{x,\bq_1}+p_{x,\bq_2})^2+(p_{y,\bq_1}+p_{y,\bq_2})^2}$
of the hardest b- and \baq-jets in \fig{fig:ptmiss-jveto}.  For \pTmiss\ we
find notable changes above 100\,GeV where the effect of PS is up to
20\,\% (resulting entirely from the decrease of the
cross section due to the different effect of the selection cuts on the
LHEs and after PS), and that of the hadronization peaks at $\pTmiss
\simeq 150$\,GeV where it reaches 25\,\%. For \ptbb\ the decrease due
to the PS is again between 0--20\,\% for the same reason. However, the
effect of the hadronization is large, between 30 and 50 \,\% above
150\,GeV, the range which is important in boosted-Higgs searches with a
large \tT\ background \cite{Butterworth:2008iy}. Similarly to the case
of \pt-distribution of the hardest b-jet this large effect is due to
the transformation of the hadronic energy into electromagnetic and
missing energy during hadronization.

We can draw the following general conclusions for the given set of
selection cuts:
\begin{itemize}
\itemsep=-2pt
\item
The effect of the parton shower is usually a 0--20\,\% decrease because
the jets become softer and fewer events pass the selection cuts.
\item
The effect of hadronization is small (less than 10\,\%) in the hard
leptonic sector, except for \pTmiss.  These distributions are
determined by the hard-scattering process and the decay of the heavy
particles, and not influenced by hadronization effects. In fact, even
if other leptons can be emitted at lower energy scales, in particular
by hadron decays, the isolation criterion is quite effective in
disentangling just the positron coming from the $W^+$ decay (and the
muon coming from the $W^-$ decay).
\item
The effect of hadronization can be up to 50\,\% in the \pt-spectra in
the hadronic sector (and negative) due to the transformation of
hadronic energy into electromagnetic and missing energy during
hadronization.
\item
The \pt-spectra are softened by PS and further by hadronization,
while the angular spectra are modestly influenced by PS and hardly by
hadronization, as expected.
\end{itemize}

Next we compare distributions for the three cases obtained from the
LHEs. As the NWA approximation with NLO decays is known to describe
most kinematic distributions fairly precisely (the inclusive cross
section in NWA is about 0.5\,\% smaller than the prediction of the full
calculation) \cite{Denner:2012yc}, this comparison gives information
mainly about the importance of the NLO decays as compared to the LO
decays in the two approximations: DCA and the one obtained with the
program \decayer, that includes the treatment of both spin correlations
and off-shell effects.  In order to compare the importance of the SMC
effects to the effect of including the decays, we also show the full
SMC predictions for case~1 computed with \py1. We use the selection
cuts (1--6).  

\begin{figure}
\begin{center}
\includegraphics[width=0.49\textwidth]{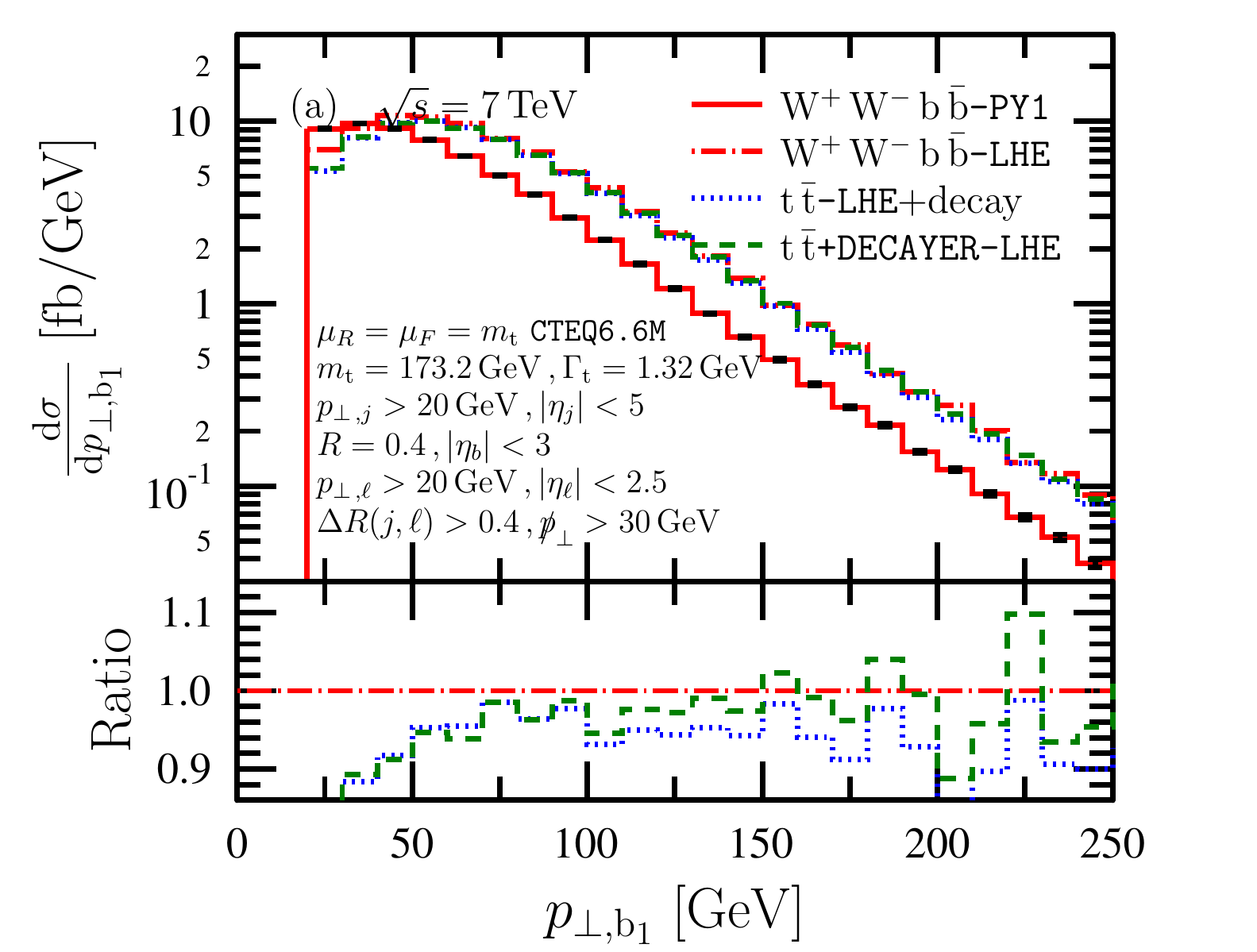}
\includegraphics[width=0.49\textwidth]{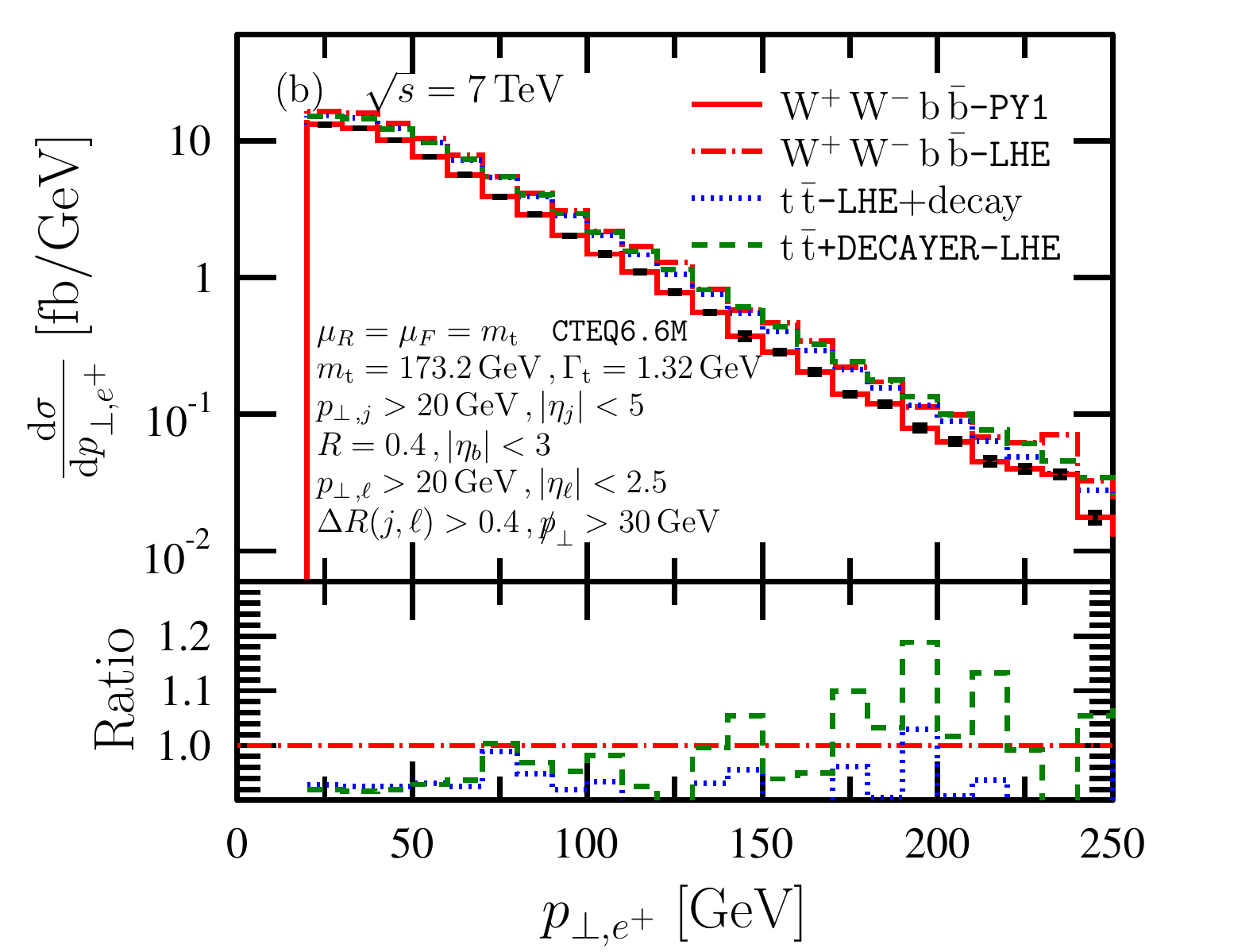}
\caption{\label{fig:pt-lhe} Distributions of transverse momentum of
a) the hardest b-jet and
b) the hardest isolated positron from the LHEs for the three cases.
The lower inset shows the ratio of the predictions with decay
to the \WWbB-prediction.}
\end{center}
\end{figure}

As seen in \fig{fig:pt-lhe} including the decays gives a good
description over the whole \pt-range for the distributions of both
$p_{\bot,\bq_1}$ and $p_{\bot,e^+}$, with \decayer\ performing slightly
better at large \pt.  The effect of the decays is in general smaller
than the effect of the full SMC. In fact, when comparing
\figs{fig:pt-jveto}{fig:pt-lhe} we see that the effect of the
hadronization increases significantly if we allow for non-b jets that
are also taken into account in the selection cuts (an effect due to
different particle content).

\begin{figure}
\begin{center}
\includegraphics[width=0.49\textwidth]{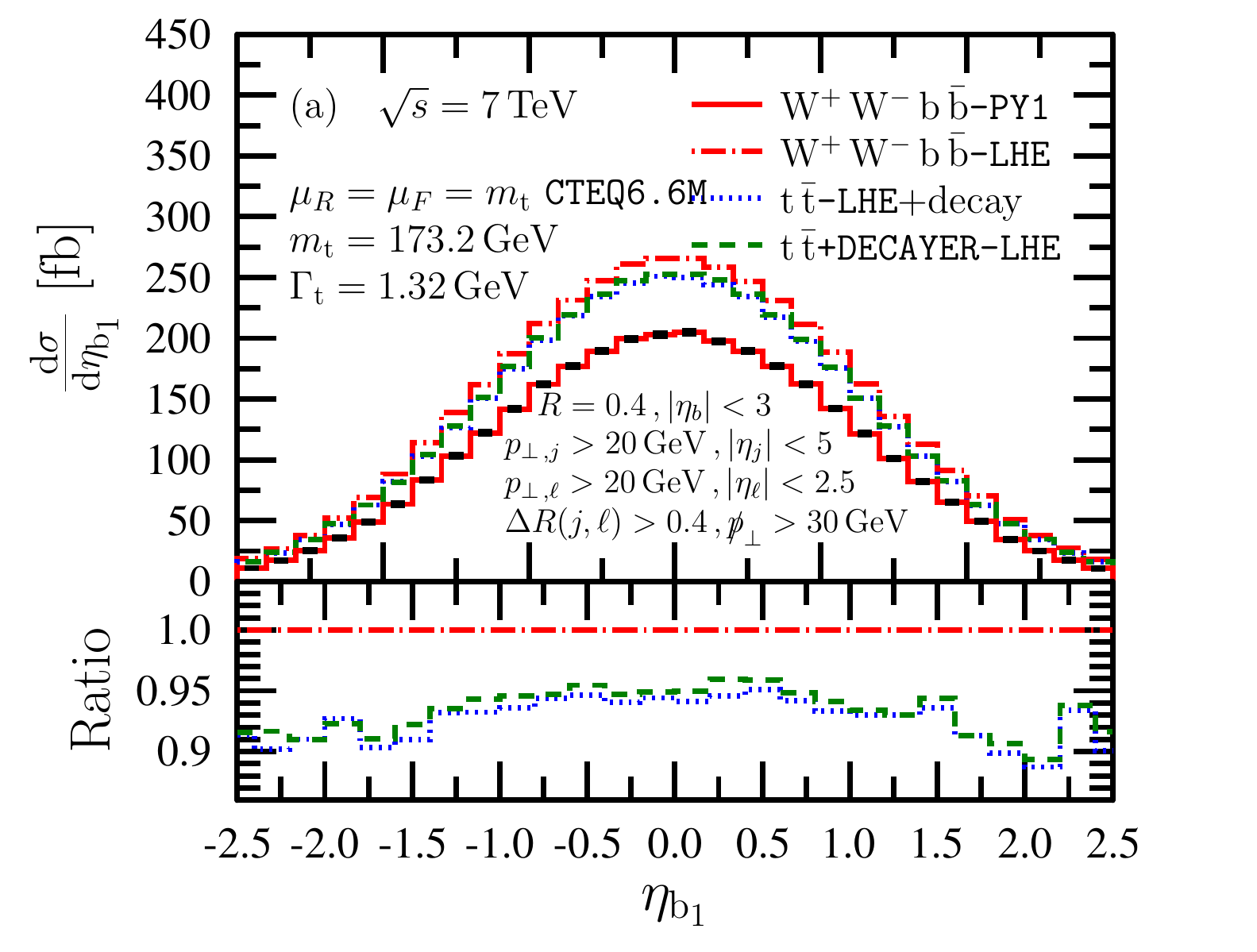}
\includegraphics[width=0.49\textwidth]{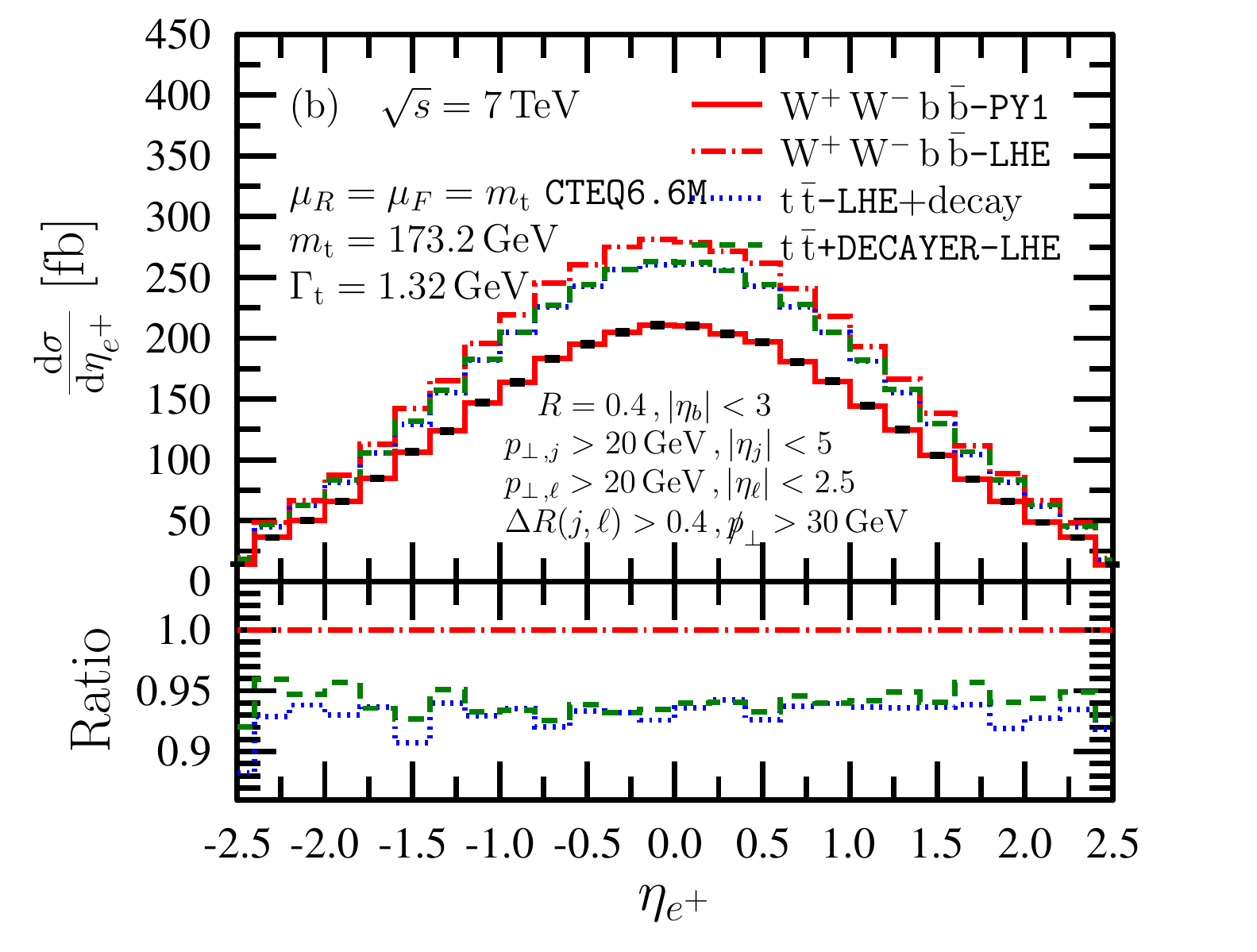}
\caption{\label{fig:eta-lhe} Pseudorapidity distributions of
a) the hardest b-jet and
b) the hardest isolated positron from the LHEs for the three cases.
The lower inset shows the ratio of the predictions with decay
to the \WWbB-prediction.}
\end{center}
\end{figure}

Similar, although less drastic effects can be observed in the pseudorapidity
distributions (see \fig{fig:eta-lhe}), where we see a small effect of
the approximate treatment of decays, an almost uniform downwards shift
amounting to 5--10\,\%. We attribute these differences to the missing NLO
corrections in the decays.

\begin{figure}
\begin{center}
\includegraphics[width=0.49\textwidth]{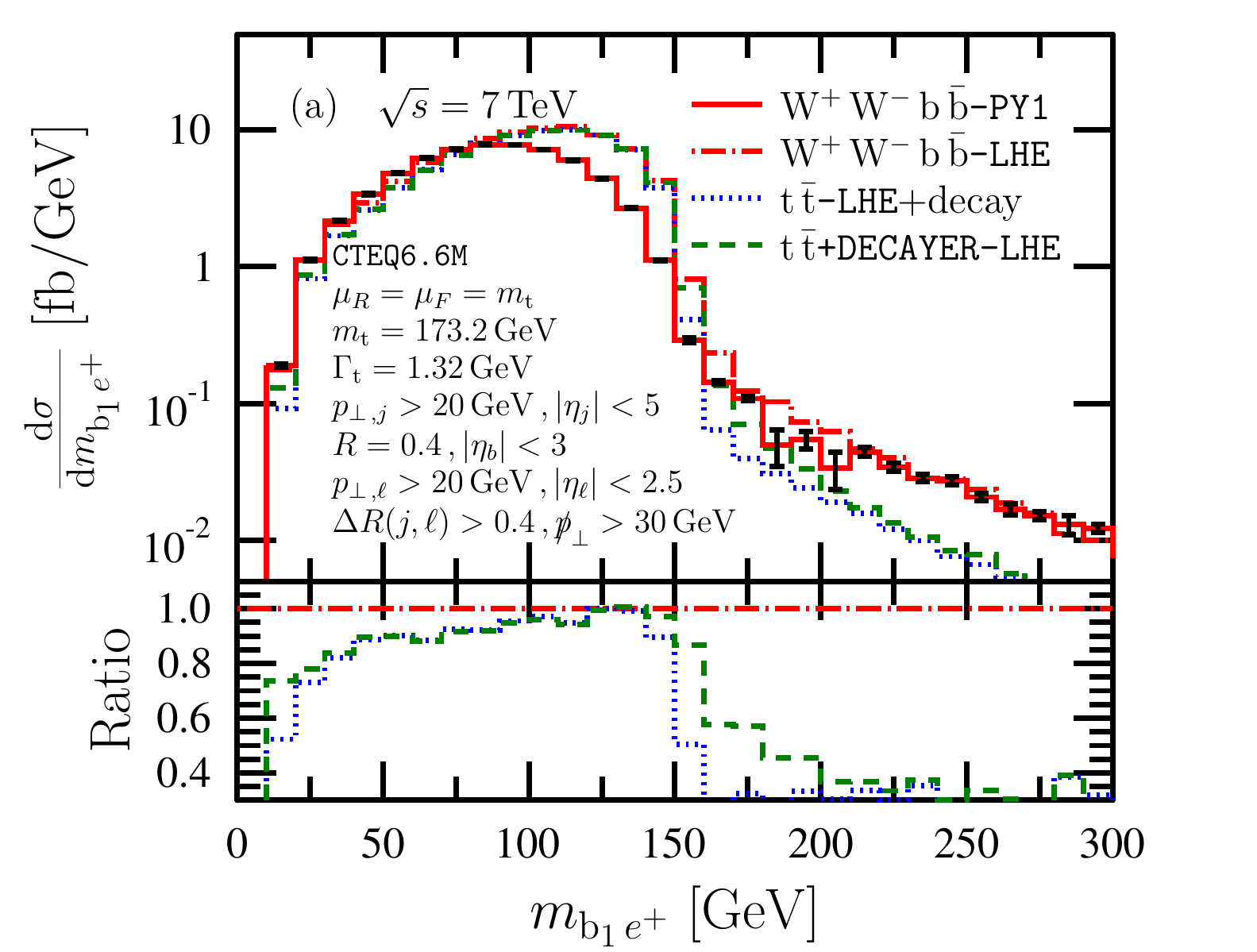}
\includegraphics[width=0.49\textwidth]{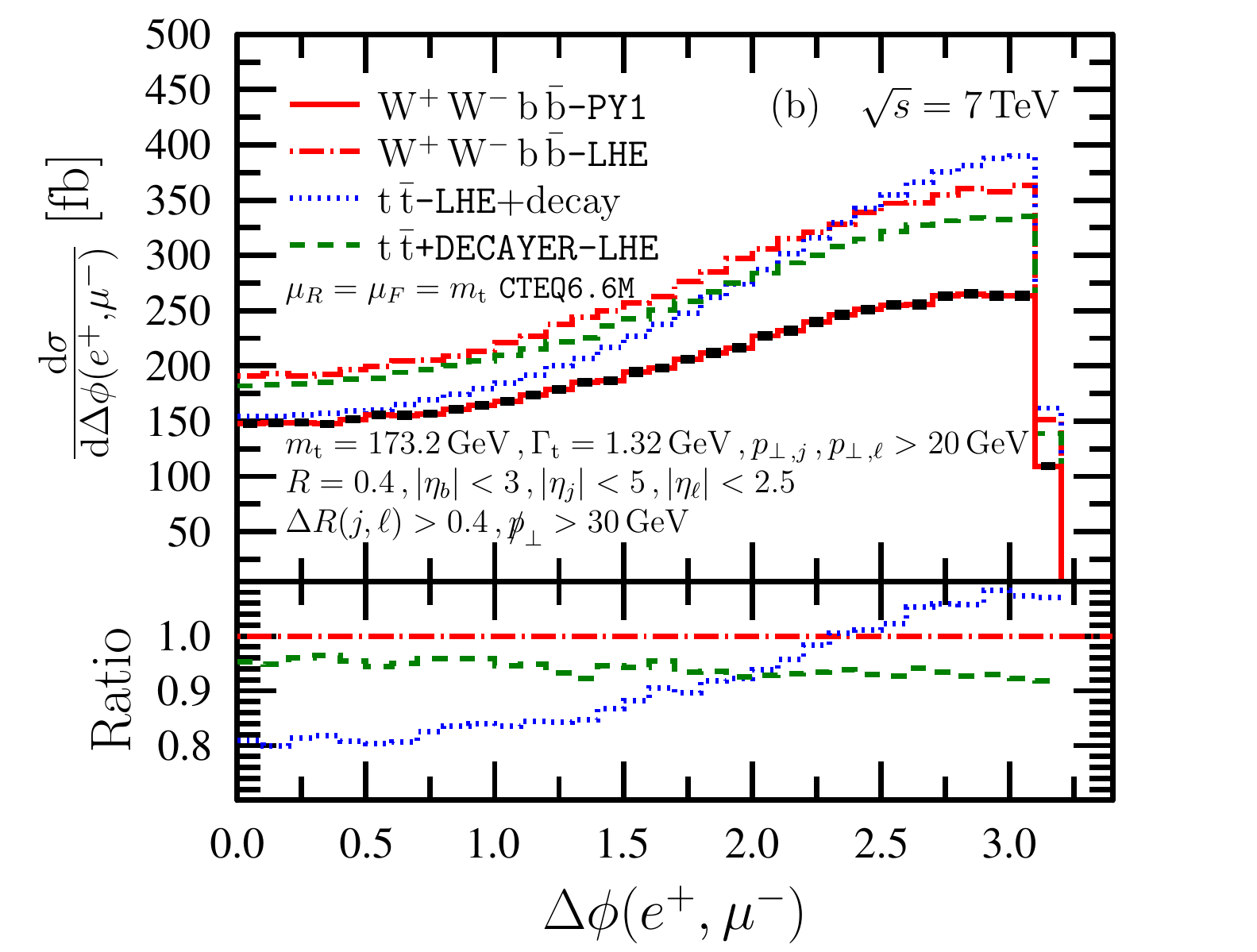}
\caption{\label{fig:mb1ep-lhe} Distributions of
a) invariant mass of hardest b-jet and the hardest isolated positron and
b) azimuthal separation between the hardest isolated positron and muon
from the LHEs for the three cases.
The lower inset shows the ratio of the predictions with decay
to the \WWbB-prediction.}
\end{center}
\end{figure}

The decays of the heavy particles are well described by both DCA and
\decayer\ up to 150\,GeV in the $m_{\bq_1 e^+}$-distribution, see
\fig{fig:mb1ep-lhe}.a. For larger values DCA and \decayer\ fail,
leading to an underestimate up to 50\,\%, which is a result of the
missing singly- and non-resonant contributions mostly (about 40\,\%
\cite{Denner:2012mx}), and of the missing NLO corrections in the
decays to much less extent (about 10\,\%). Also the two approximate
predictions differ in the range of 150--200\,GeV due to the off-shell
effects in \decayer, missing from the DCA.  The effect of the full SMC
is small (within 10\,\%) below 100\,GeV and above 175\,GeV, but can
become very large in between, especially at 150\,GeV, where the LHE
cross section has a sharp drop. As one expects, this sharp drop is
smeared by the PS and hadronization (see \fig{fig:mb1ep-jveto}.a).

The $\Delta \phi_{e^+\mu^-}$-distribution, shown in
\fig{fig:mb1ep-lhe}.b, is an example where the differences between the
three cases are clearly visible.  The prediction by \decayer\ is within
10\,\% of the full one, with slightly increasing difference towards the
separation by 180$^\circ$.  Nevertheless, the shapes of the LHE
predictions are similar, implying that \decayer\ gives a fairly good
approximation to describe spin correlations. The difference in
normalization is due to the missing NLO corrections in \decayer.
The DCA approximation has a different shape due to the absence of spin
correlations. Similar pattern as seen between the DCA and \decayer\
approximations was already observed in the parton-level calculation of
\Ref{Mahlon:2010gw} performed with and without spin correlations. For
this distribution the full SMC decreases the cross section by about
30\,\% almost uniformly.

We conclude that the predictions with decays give the shapes correctly,
the NLO corrections in the decays in general cause only a
uniform increase up to 10\,\% except for distributions involving the
charged lepton emerging from the decay of the $W$-bosons (usually the
hardest isolated charged lepton).  The DCA does not describe the shape
of the $\Delta \phi_{e^+\mu^-}$-distribution due to the lack of spin
correlations and both approximations fail in the hard tail of the
$m_{\bq_1 e^+}$-distribution due to the lack of singly- and
non-resonant contributions.

We now turn to make predictions at the hadron level that are more
interesting from the experimental point of view. For this kind of
predictions the selection cuts (1--6) were applied after shower,
hadronization, hadron decay and the application of jet algorithms. 
For the three cases, the integrated cross-sections after cuts are collected
in \tab{tab:table1}, for different jet sizes (anti-\kt\ with $R=0.4$
versus anti-\kt\ with $R=1.2$). The cross-section in case 1, the most
accurate one, is larger than the cross-section in case 2 and 3 by
$\sim$ 10\% due to the NLO accuracy in the decays and, to less extent, to the
non-resonant contributions. The cross-section  decreases in all cases
significantly by using a larger jet radius because we also use a more 
severe jet-lepton isolation with $R=1.2$. By comparing cases 2 and 3 it
turns out that the effect of spin-correlations in t-quark and $W$-boson
decays increases the cross section by a couple of percent because our
selection cuts affect the spin-correlated decays slightly differently.
\begin{table}
\begin{center}
\begin{tabular}{|c|c|c|c|}
\hline
\hline
R/case & \WWbB\ & DCA & \decayer\ \\
\hline
$\sigma(R=0.4)$ (fb) & $630\pm 4$ & $573\pm 1$ & $582\pm 1$\\
$\sigma(R=1.2)$ (fb) & $300\pm 3$ & $253\pm 1$ & $261\pm 1$\\
\hline
\hline
\end{tabular}
\caption{\label{tab:table1}Cross-sections at the hadron level at the LHC
after cuts (1--6) for the three cases as a function of the $R$
parameter of the anti-\kt\ algorithm used to identify jets. The quoted
uncertainties are statistical only.
The predictions are obtained with the \py1 SMC.}
\end{center}
\end{table}
\begin{figure}
\begin{center}
\includegraphics[width=0.49\textwidth]{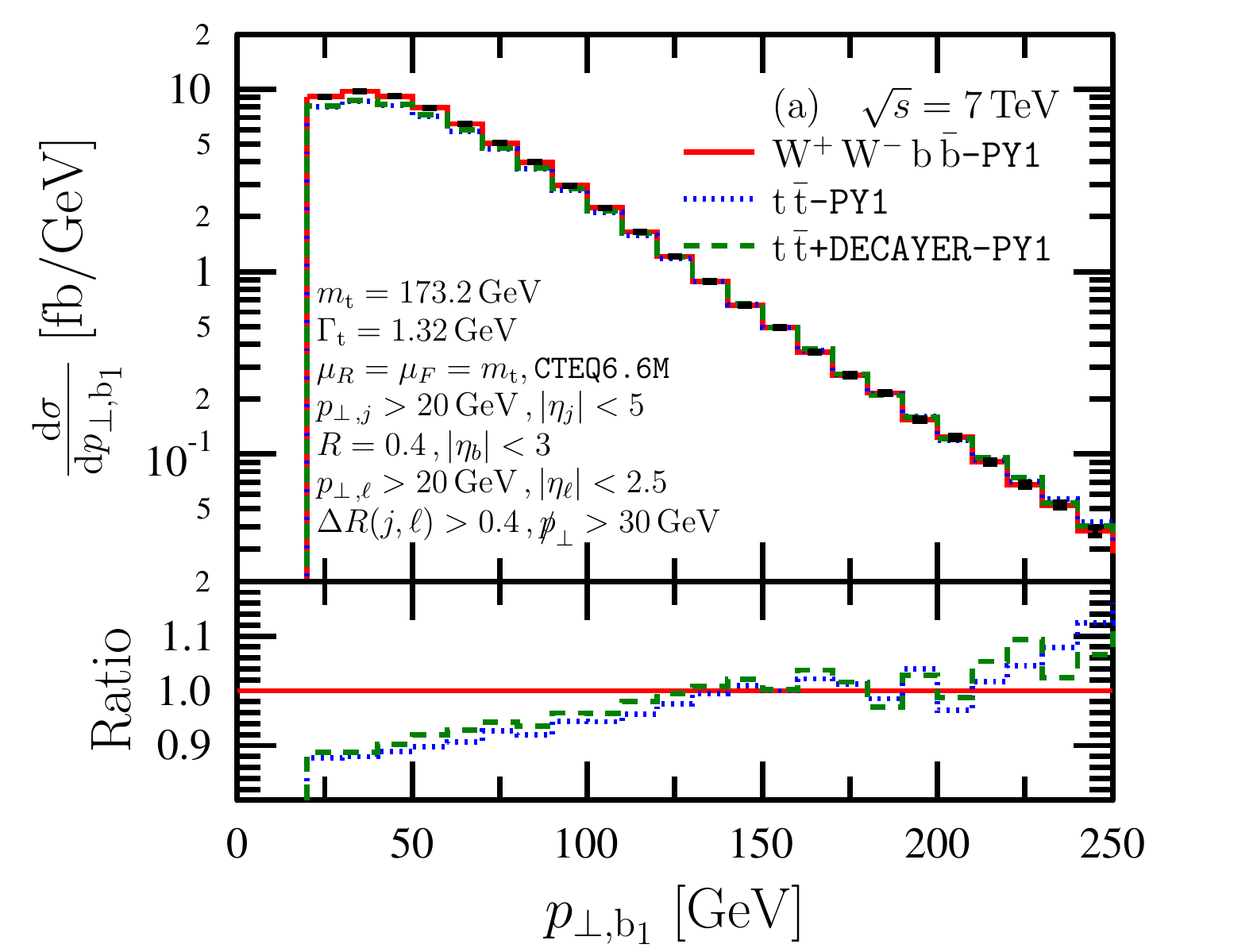}
\includegraphics[width=0.49\textwidth]{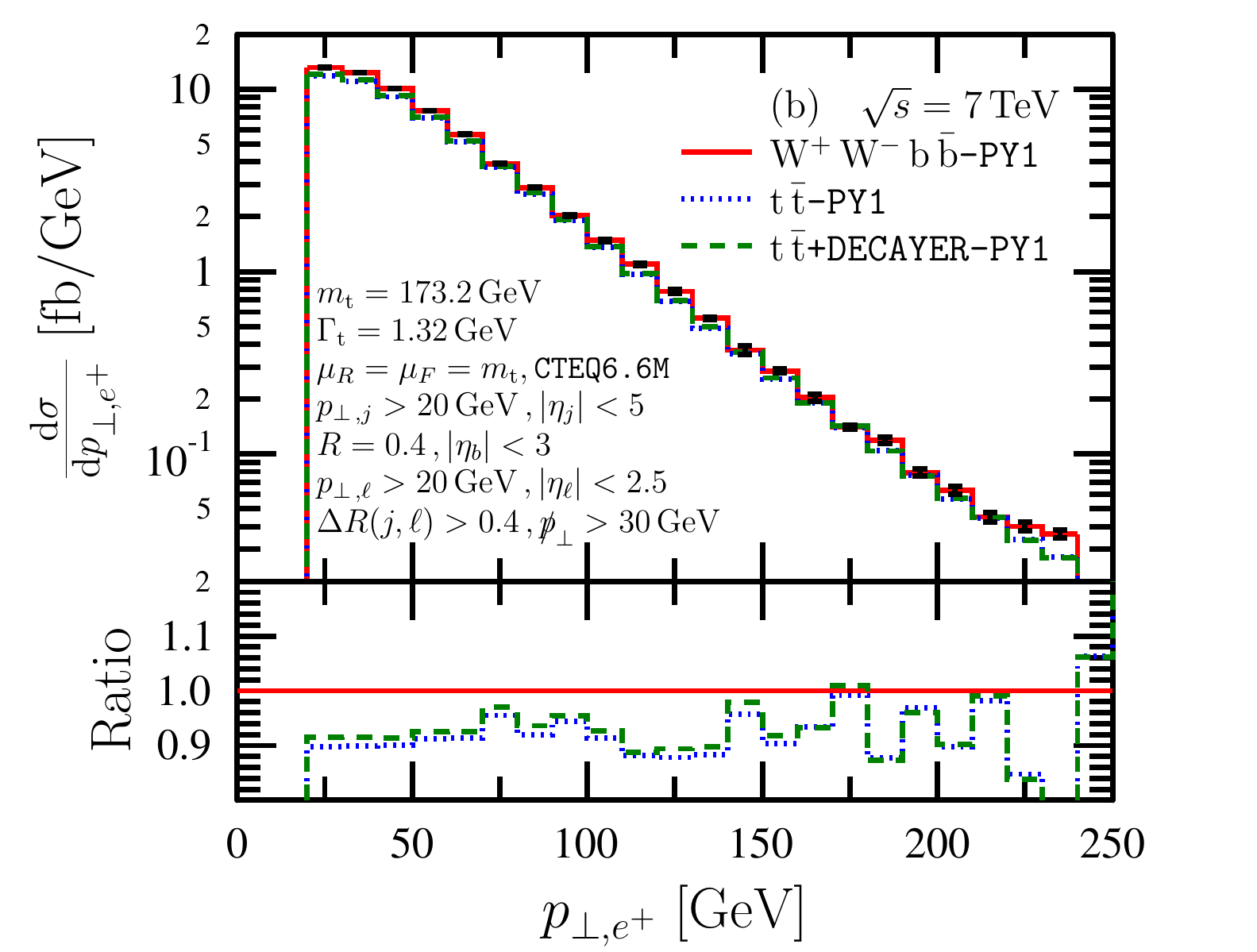}
\caption{\label{fig:pt-py} Distributions of transverse momentum  of
a) the hardest b-jet and
b) the hardest isolated positron
after full SMC for the three cases.  The lower inset shows the ratio of
the predictions with decays of the t-quarks in the DCA and
\decayer\ as compared to the complete \WWbB\ computation.}
\end{center}
\end{figure}

We also compared almost 50 distributions belonging to the three cases,
our standard selection is shown in \figss{fig:pt-py}{fig:mb1ep-py}.
In the inset in the lower part of each figure we plotted the ratio of the
cases 2 and 3 (with decays of heavy particles) to the
default one (\WWbB-production). We see in \figs{fig:pt-py}{fig:eta-py}
the general trend that \decayer\ and DCA give very similar predictions
both in shape and normalization for \pt- and $\eta$-distributions,
while the full \WWbB-computation followed by the SMC differs, but
within 10\,\% of these. These differences are due to the lack of NLO
corrections in the decays, and are present already at the
LHE level, as discussed above (see for instance \fig{fig:pt-lhe}.b).
The SMC does not change the main features already seen in the LHEs.
\begin{figure}
\begin{center}
\includegraphics[width=0.49\textwidth]{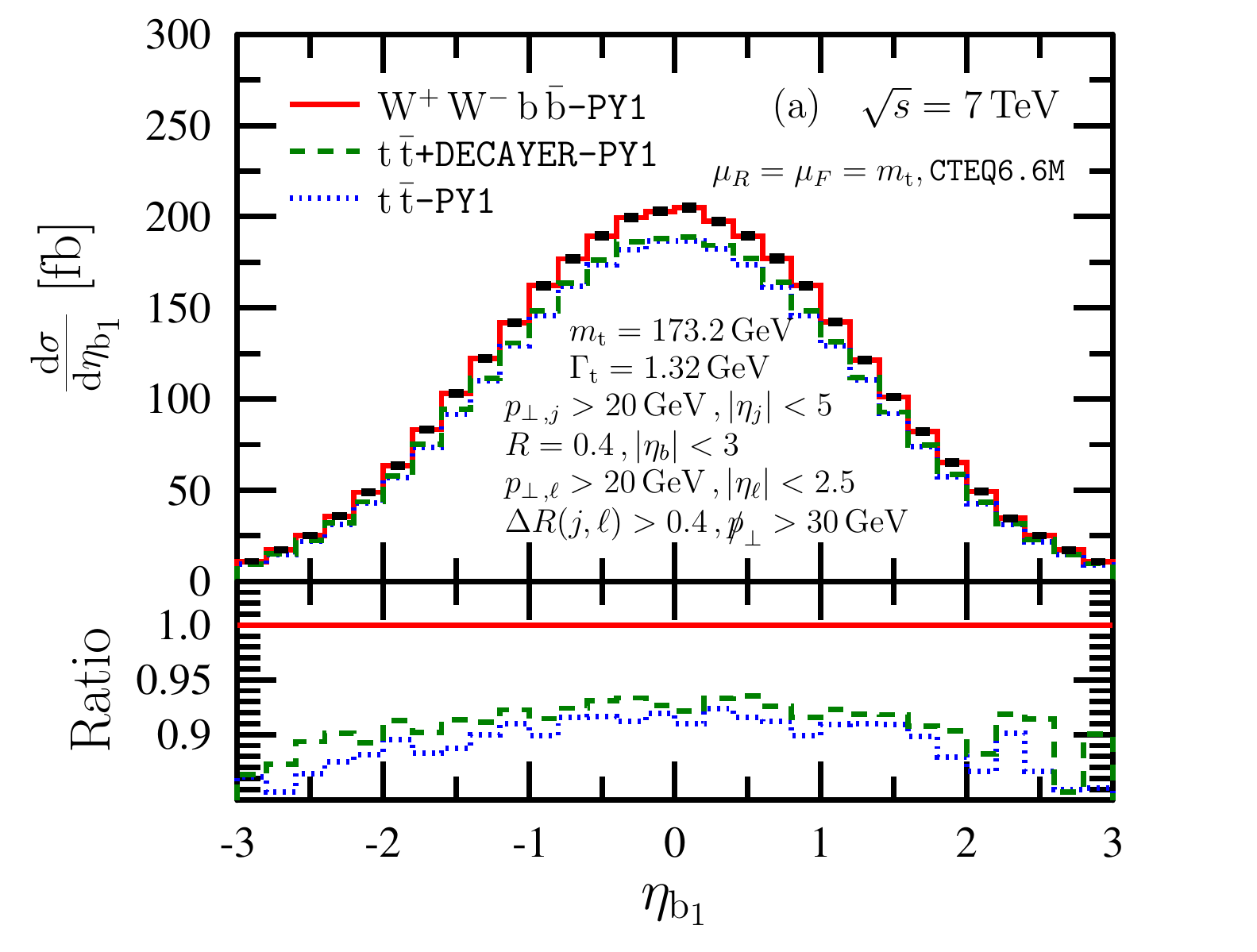}
\includegraphics[width=0.49\textwidth]{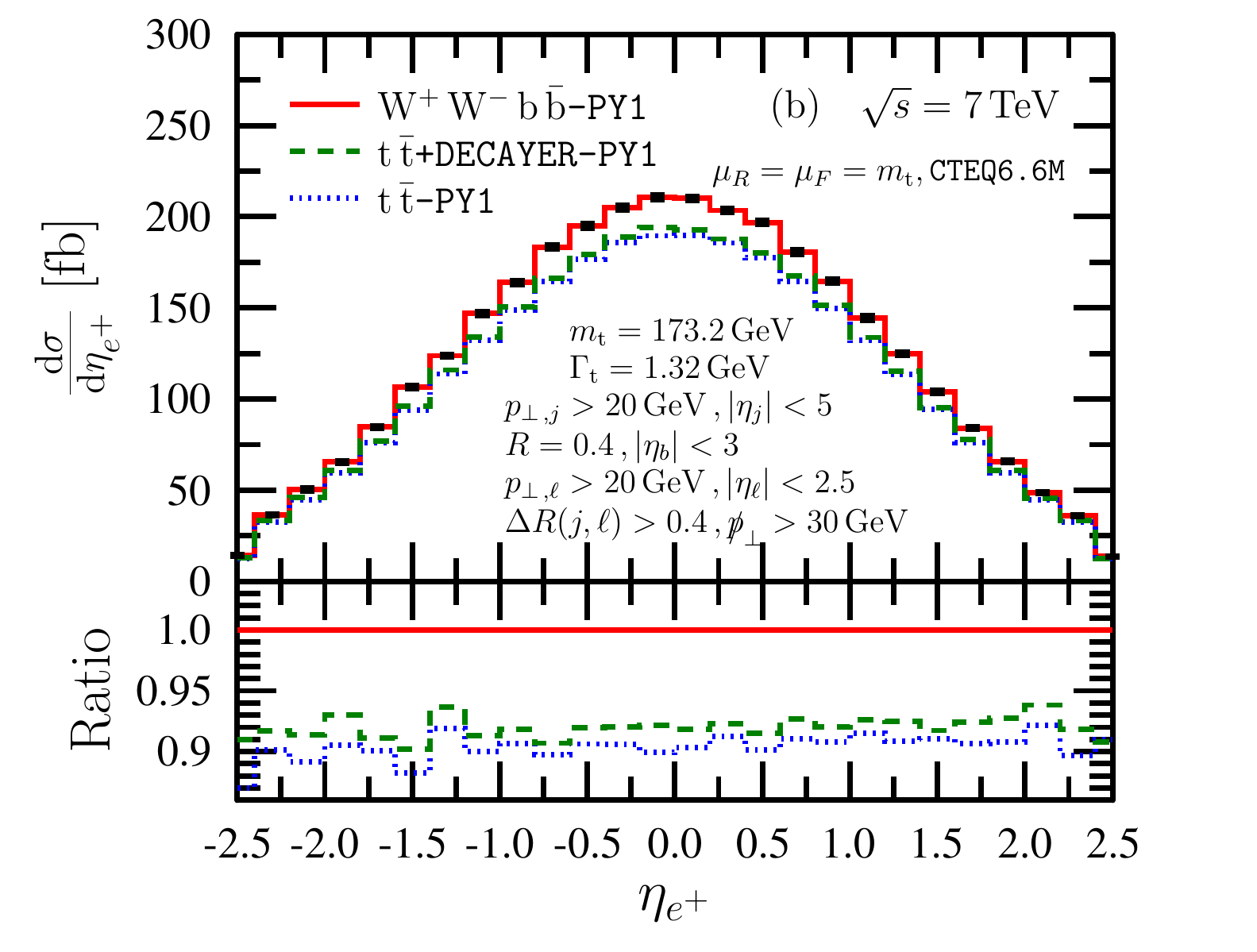}
\caption{\label{fig:eta-py} Pseudorapidity distributions of
a) the hardest b-jet and
b) the hardest isolated positron
after full SMC for the three cases.  The lower inset shows the ratio of
the predictions with decays of the t-quarks in the DCA and
\decayer\ as compared to the complete \WWbB\ computation.}
\end{center}
\end{figure}

\begin{figure}
\begin{center}
\includegraphics[width=0.49\textwidth]{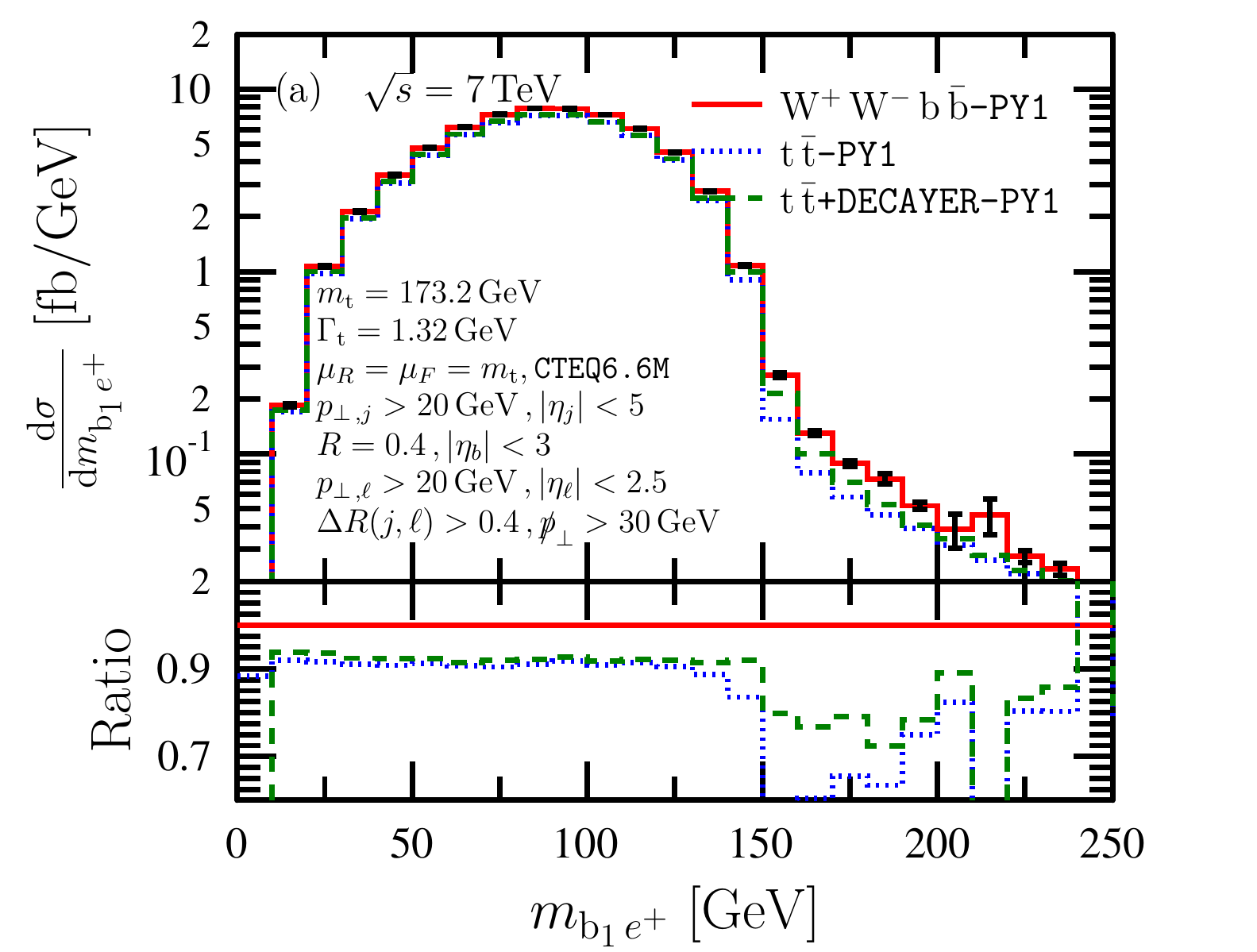}
\includegraphics[width=0.49\textwidth]{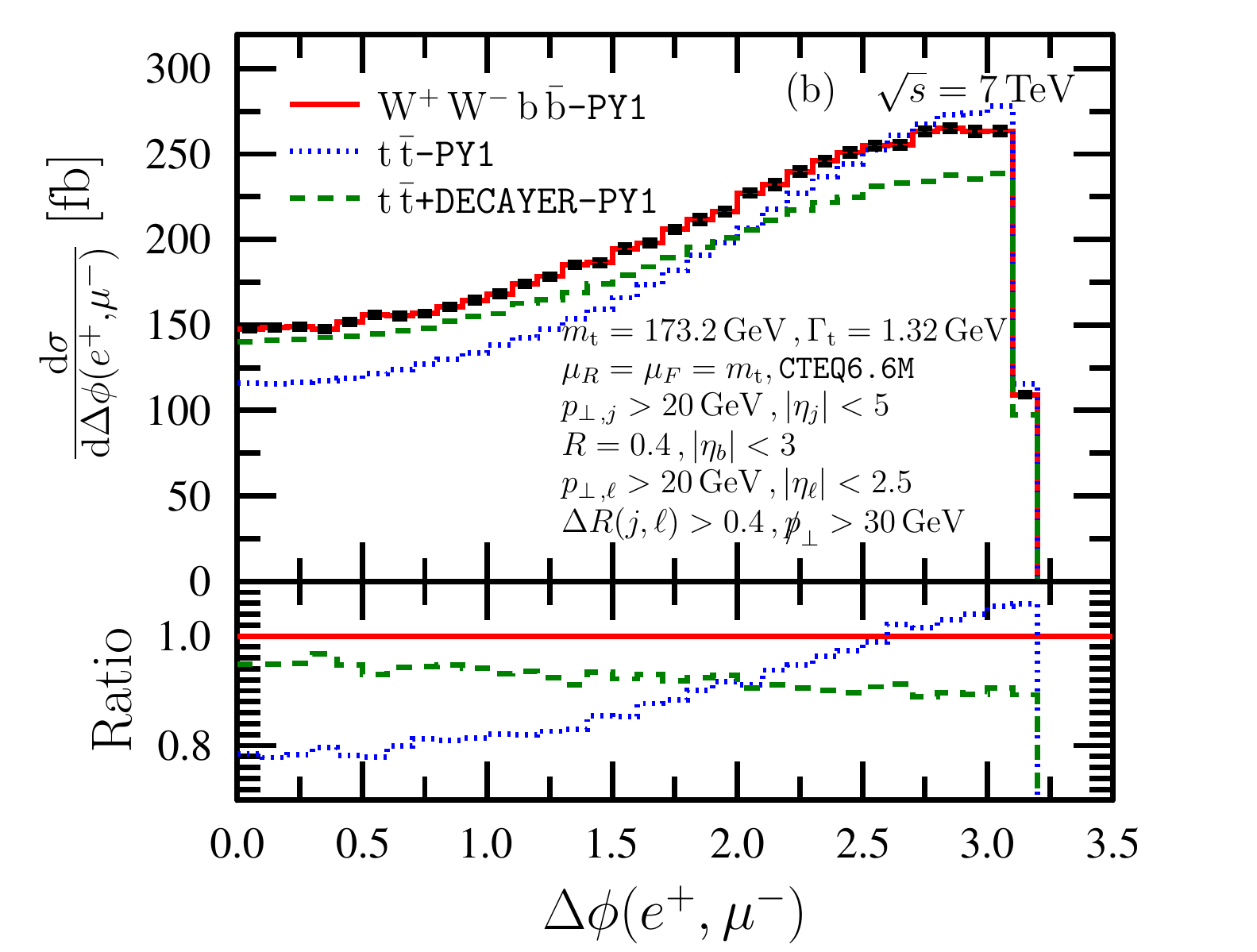}
\caption{\label{fig:mb1ep-py} Distributions of 
a) invariant mass of hardest b-jet and the hardest isolated positron and of
b) azimuthal separation between the hardest isolated positron and muon
after full SMC for the three cases.  The lower inset shows the ratio of
the predictions with decays of the t-quarks in the DCA and
\decayer\ as compared to the complete \WWbB\ computation.}
\end{center}
\end{figure}

There are some exceptions to this general trend, some shown in
\figs{fig:mb1ep-py}{fig:ptj1-py}. The $m_{\bq_1 e^+}$-distribution shows a
similar pattern as seen in the LHEs (cf. \fig{fig:mb1ep-lhe}) but the
large effect of the singly- and non-resonant graphs above 150\,GeV
becomes much reduced after SMC, and also the sharp drop in the cross
section at 150\,GeV is smeared (already seen in \fig{fig:mb1ep-jveto}.a).

In \fig{fig:mb1ep-py}.b we present the $\Delta \phi_{e^+\mu^-}$-distribution.
This distribution is an example where the differences between the three
cases were clearly visible in the LHEs. These differences are only
slightly altered by the PS, or the full SMC. In particular, the effect
of including the spin-correlations leads to an increase of the
distribution for small azimuthal separation $\Delta\phi_{e^+\mu^-}$,
where the distribution of case 3 approaches that of case 1, both
including spin correlations.  These are both 15--20\,\% larger than the
distribution of case 2, where spin correlations are neglected. At large
separations however, after SMC the latter becomes even larger than the
predictions from case 1.

\begin{figure}
\begin{center}
\includegraphics[width=0.49\textwidth]{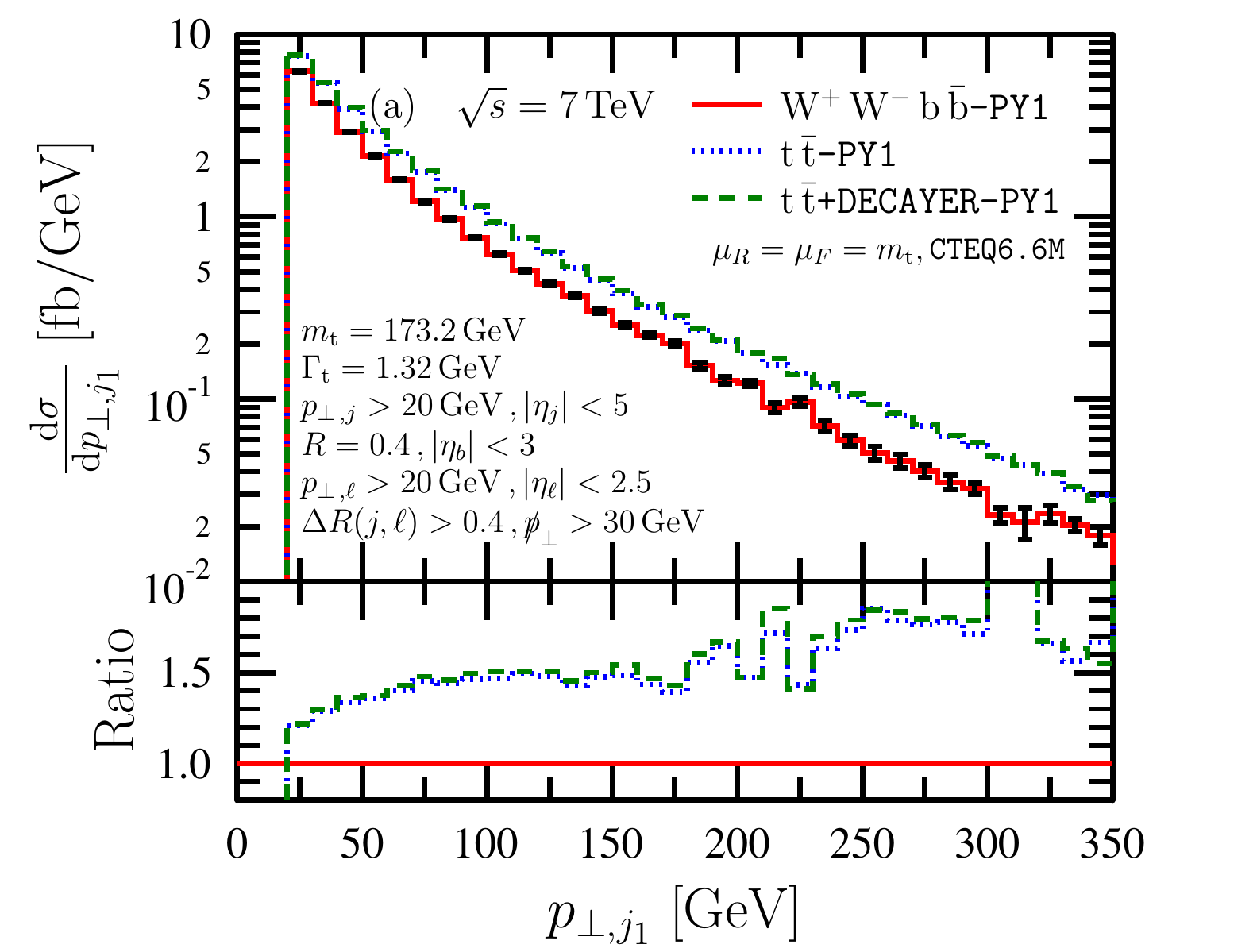}
\includegraphics[width=0.49\textwidth]{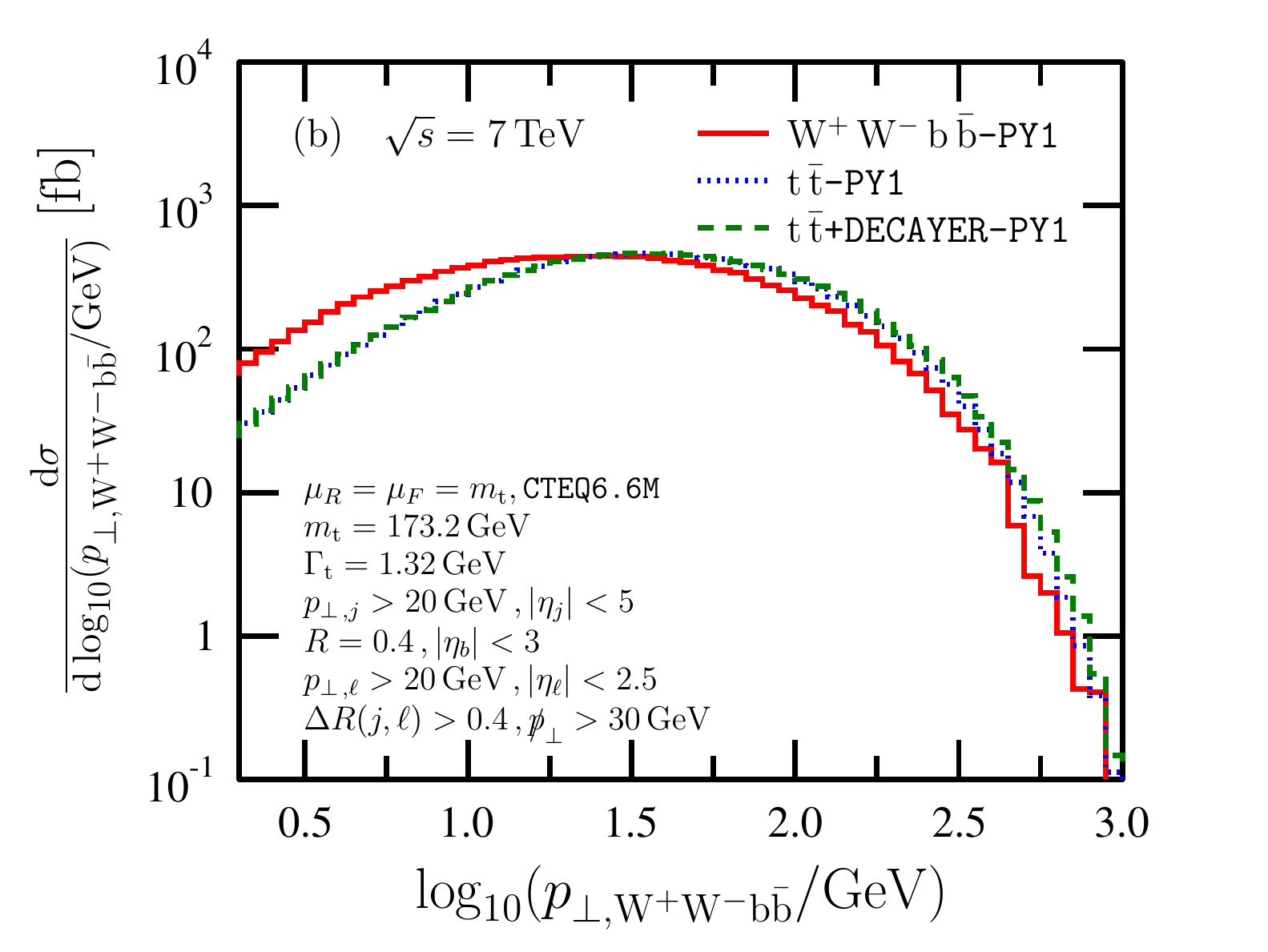}
\caption{\label{fig:ptj1-py} Distribution of the transverse momentum of
a) the hardest non b-jet and
b) the \WWbB-system.
The lower inset shows
the ratio of the predictions with decays of the t-quarks in the DCA and
\decayer\ compared to the complete \WWbB\ computation.}
\end{center}
\end{figure}

We discuss two more, closely related, distributions for these
physically most interesting predictions.  The first one is the
\pt-distribution of the hardest non-b jet in \fig{fig:ptj1-py}.a. Here we
observe a difference up to 50\,\% between the distributions from
\WWbB-production and the other two cases.  The scalar sum of the
(\WWbB) transverse momenta for the events that pass the cuts (1--6) at
the hadron level, is plotted in \fig{fig:ptj1-py}.b. Here we see the
Sudakov suppression in the low \pt-region clearly, which is different
for case 1 and cases 2, 3 for similar reason as in the \pt-distribution
of the non-b jet.  Clearly, the differences seen in \fig{fig:ptj1-py}
have nothing to do with spin correlations (cases 2 and 3 give the same)
or non-resonant Feynman graphs present in the \WWbB-calculation (case
1).  The probability of emitting the hardest non-b jet at a given
transverse momentum from the initial state is similar in the case of
\tT\ and \WWbB\ production.  However, in the case of \tT-pair
production the probability of a hard jet from the t-quarks is much
larger than from the b-quarks in the \WWbB\ final state, the
latter being dominated by soft and collinear emissions that generally
contribute to the b-jet. As a result the $p_{\bot,j_1}$-distribution is
much softer for case 1.

\begin{figure}
\begin{center}
\includegraphics[width=0.49\textwidth]{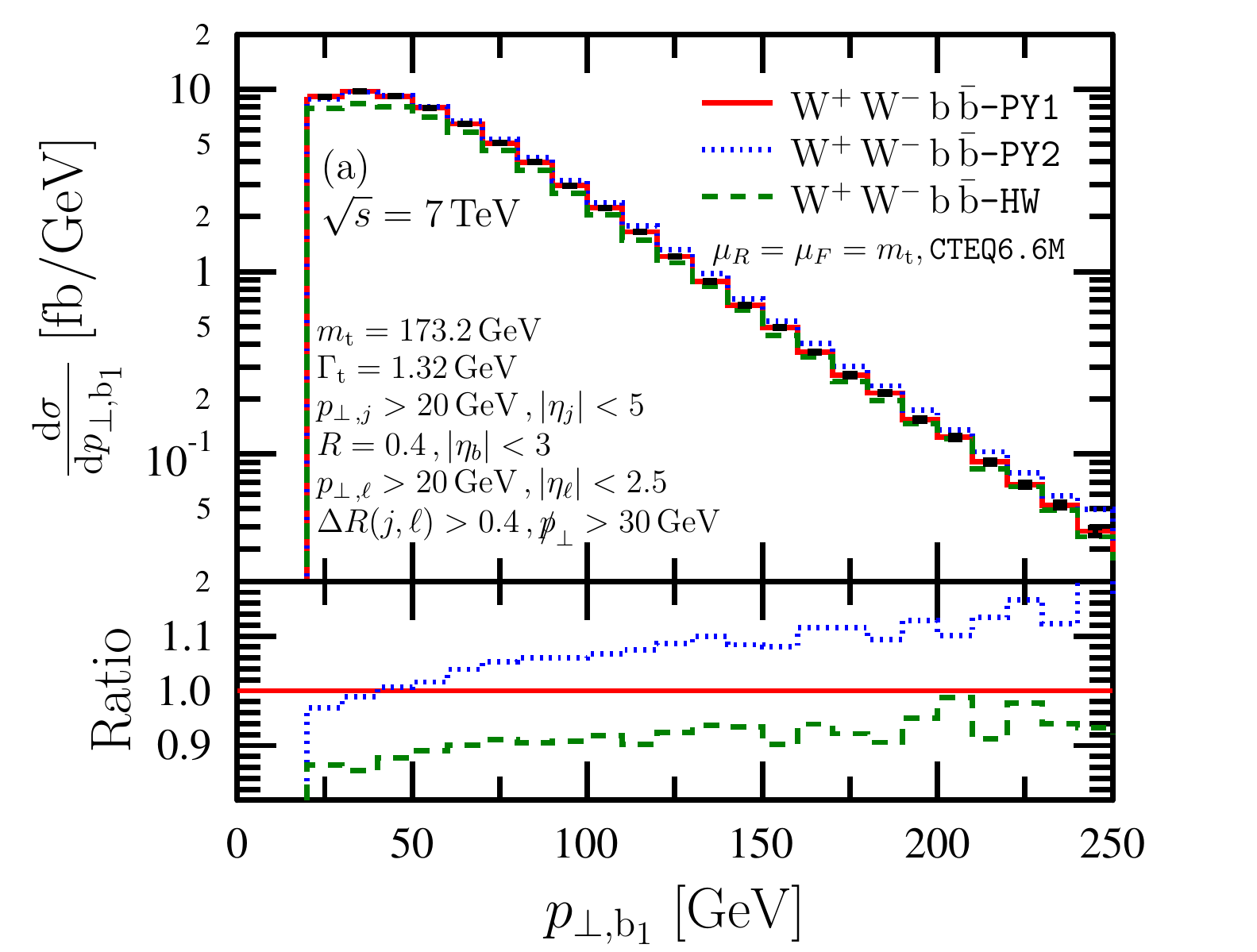}
\includegraphics[width=0.49\textwidth]{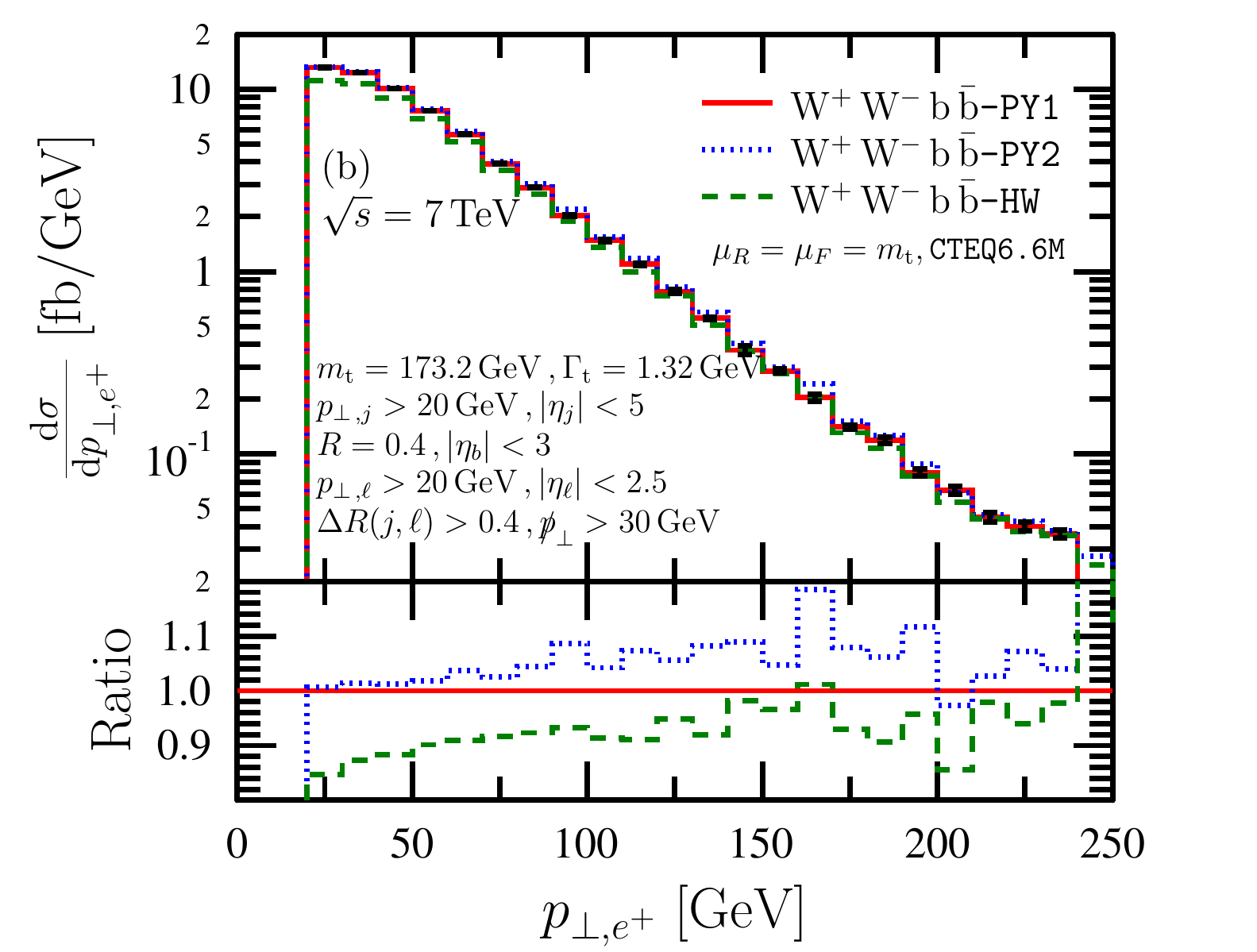}
\caption{\label{fig:pt-smc1} 
a) $p_{\bot,\bq_1}$-distribution and
b) $p_{\bot,e^+}$-distribution
after full SMC with three different SMC codes. The lower inset shows
the ratio of the predictions with \py2 and \hw\ to that with \py1.}
\end{center}
\end{figure}

Finally, we check the consistency among three different SMC codes: \py1,
\py2 and \hw\ in \figss{fig:pt-smc1}{fig:pt-smc3}. The lower insets in these plots show
the ratios of the predictions with \py2 and \hw\ to that with \py1.
Concerning the dependence on the SMC code, in general we find that
\begin{itemize}
\itemsep=-2pt
\item
in most of the phase space, the three SMC codes give predictions within 10\,\%;
\item
the shapes predicted with the \py2 and \hw\ codes are very similar, but
their normalization differs;
\item
the shapes for \pt-distributions obtained with \py2 and \hw\ are
slightly harder than those predicted by the \py1 code. The
pseudorapidity and angular distributions differ only in normalization.
\end{itemize}
\begin{figure}
\begin{center}
\includegraphics[width=0.49\textwidth]{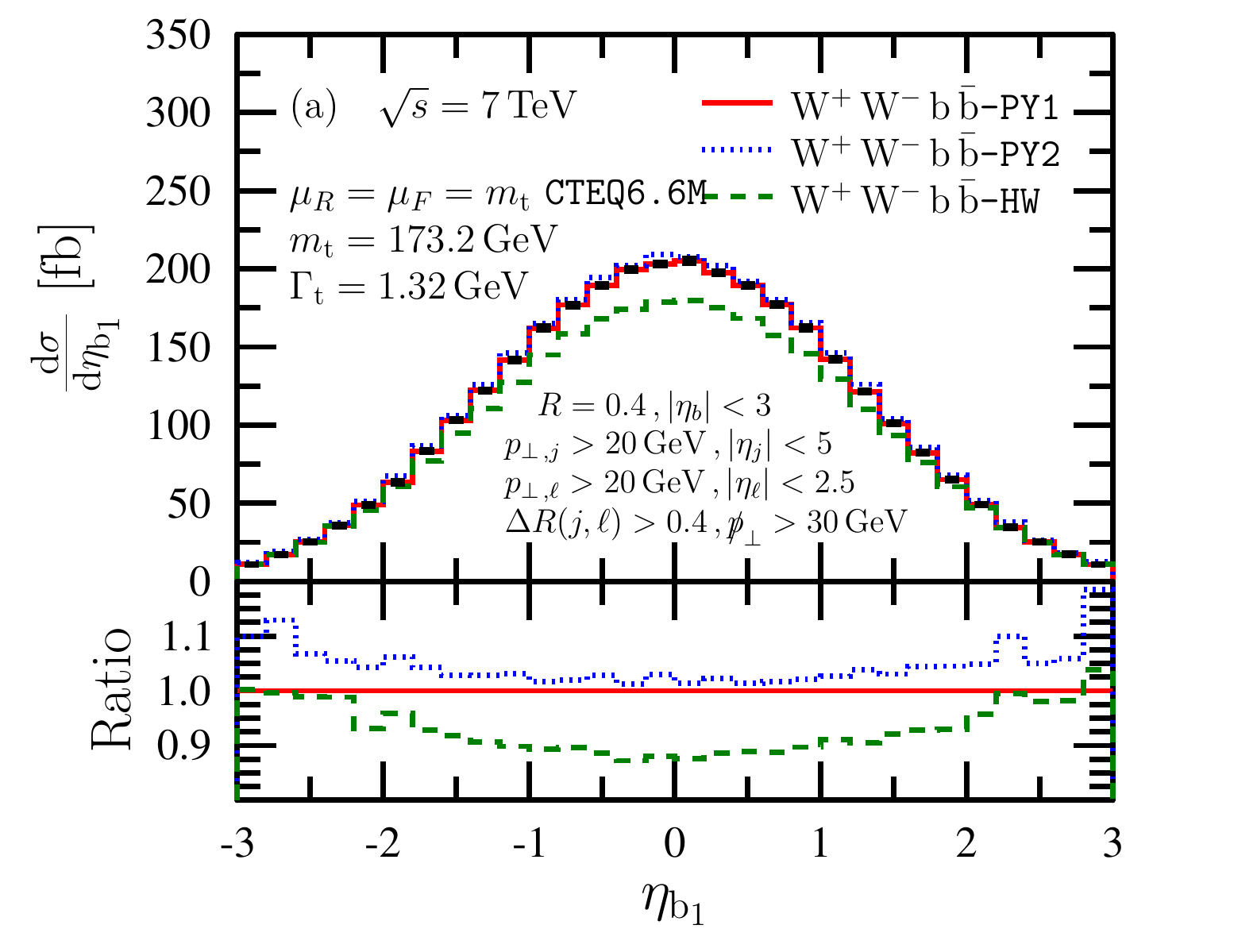}
\includegraphics[width=0.49\textwidth]{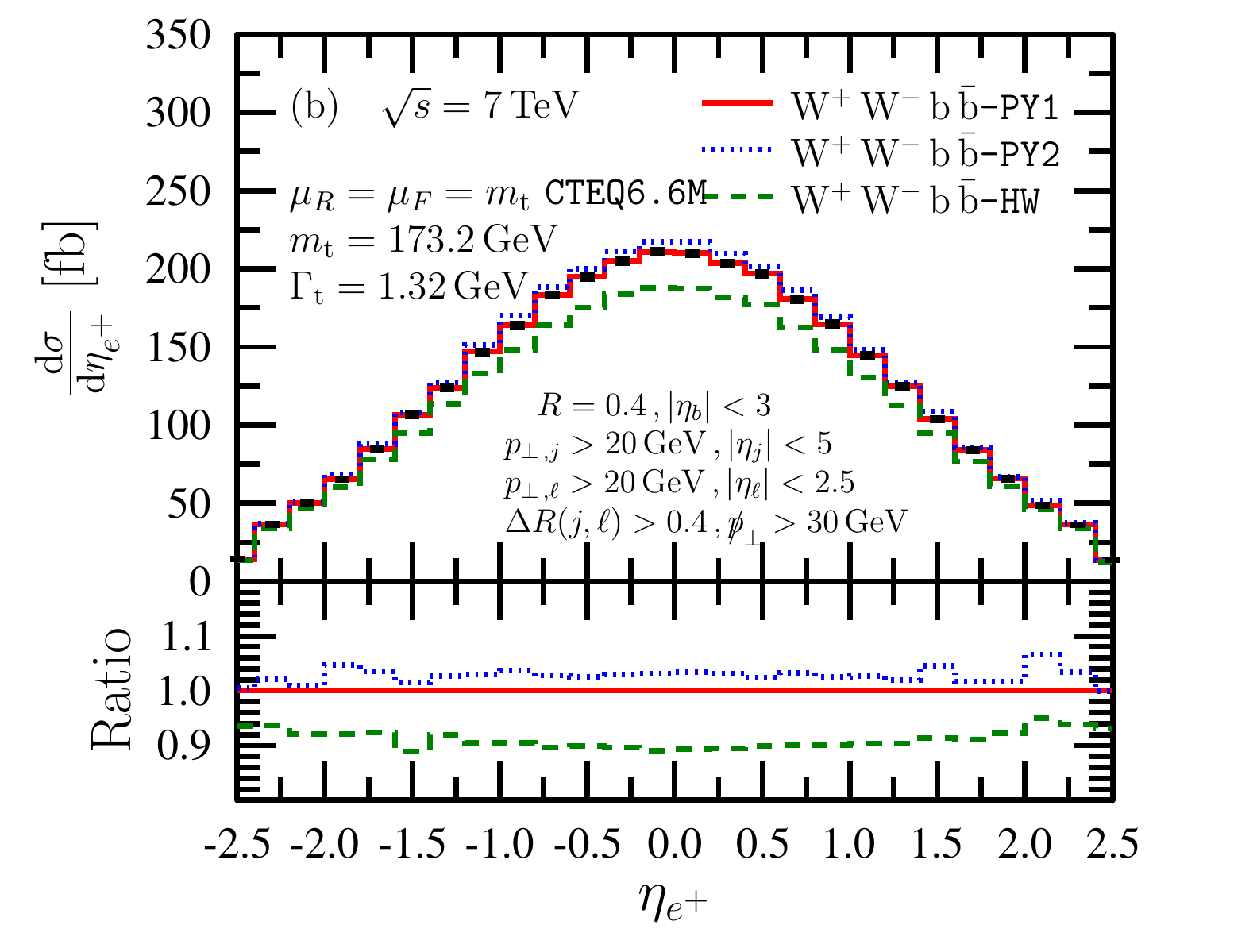}
\caption{\label{fig:pt-smc2} Same as \fig{fig:pt-smc1} as for
a) the $\eta_{\bq_1}$-distribution and
b) the $\eta_{e^+}$-distribution.
}
\end{center}
\end{figure}
\begin{figure}
\begin{center}
\includegraphics[width=0.49\textwidth]{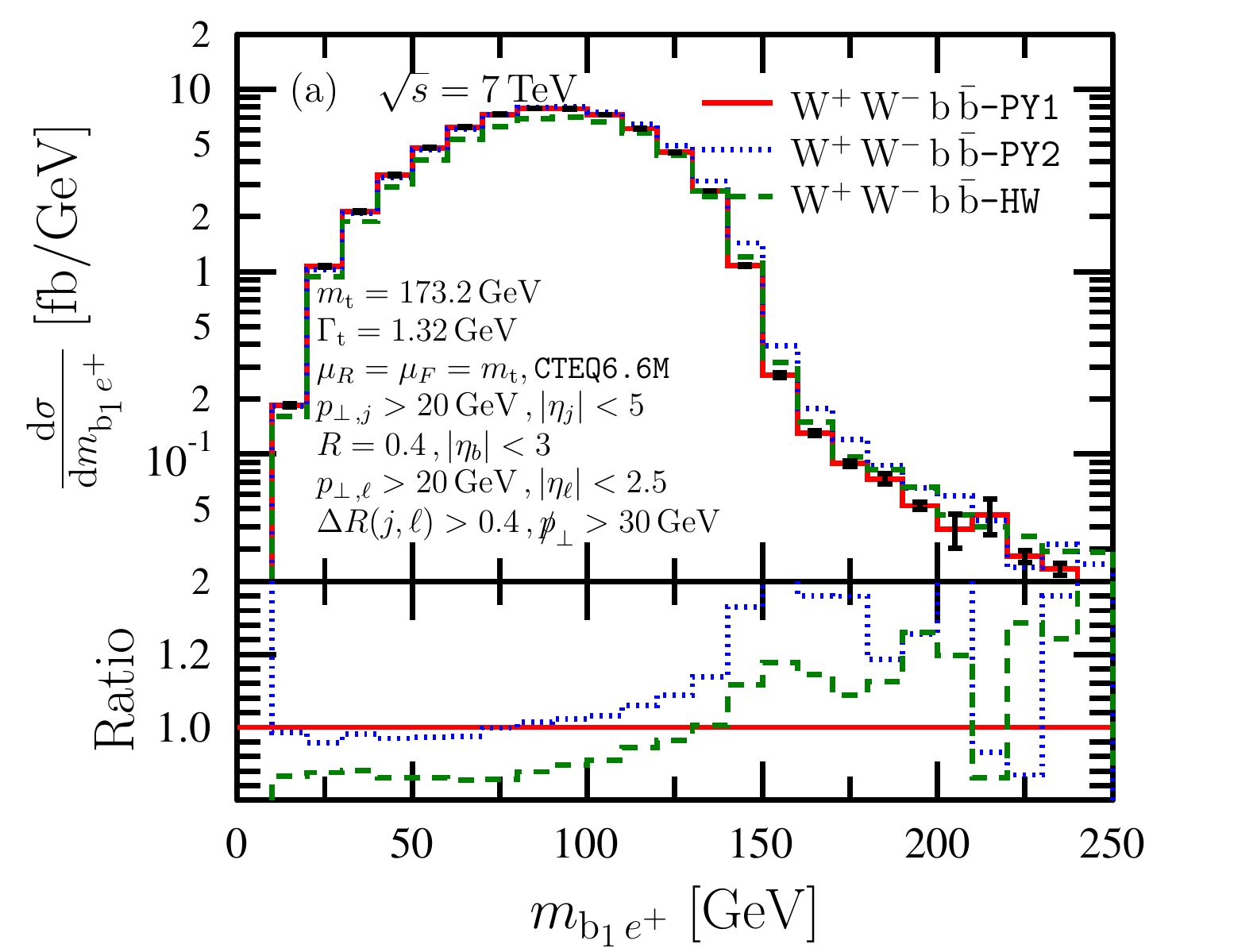}
\includegraphics[width=0.49\textwidth]{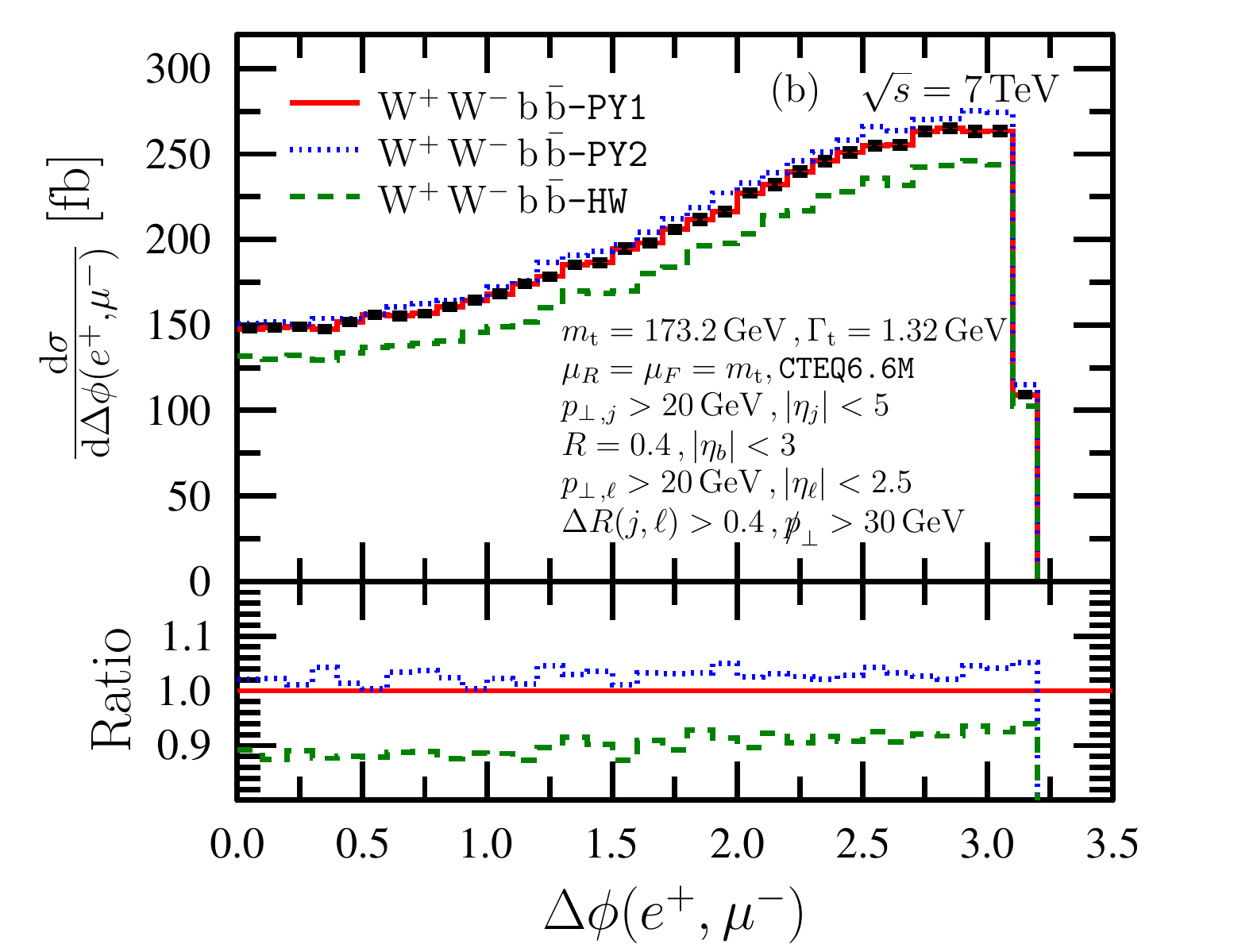}
\caption{\label{fig:pt-smc3} Same as \fig{fig:pt-smc1} as for
a) the $m_{\bq_1 e^+}$-distribution and 
b) the distribution of the azimuthal separation between the hardest isolated positron and muon.
}
\end{center}
\end{figure}

\subsection{Predictions at the TeVatron \label{sec:Tevatron}}

\begin{figure}
\begin{center}
\includegraphics[width=0.49\textwidth]{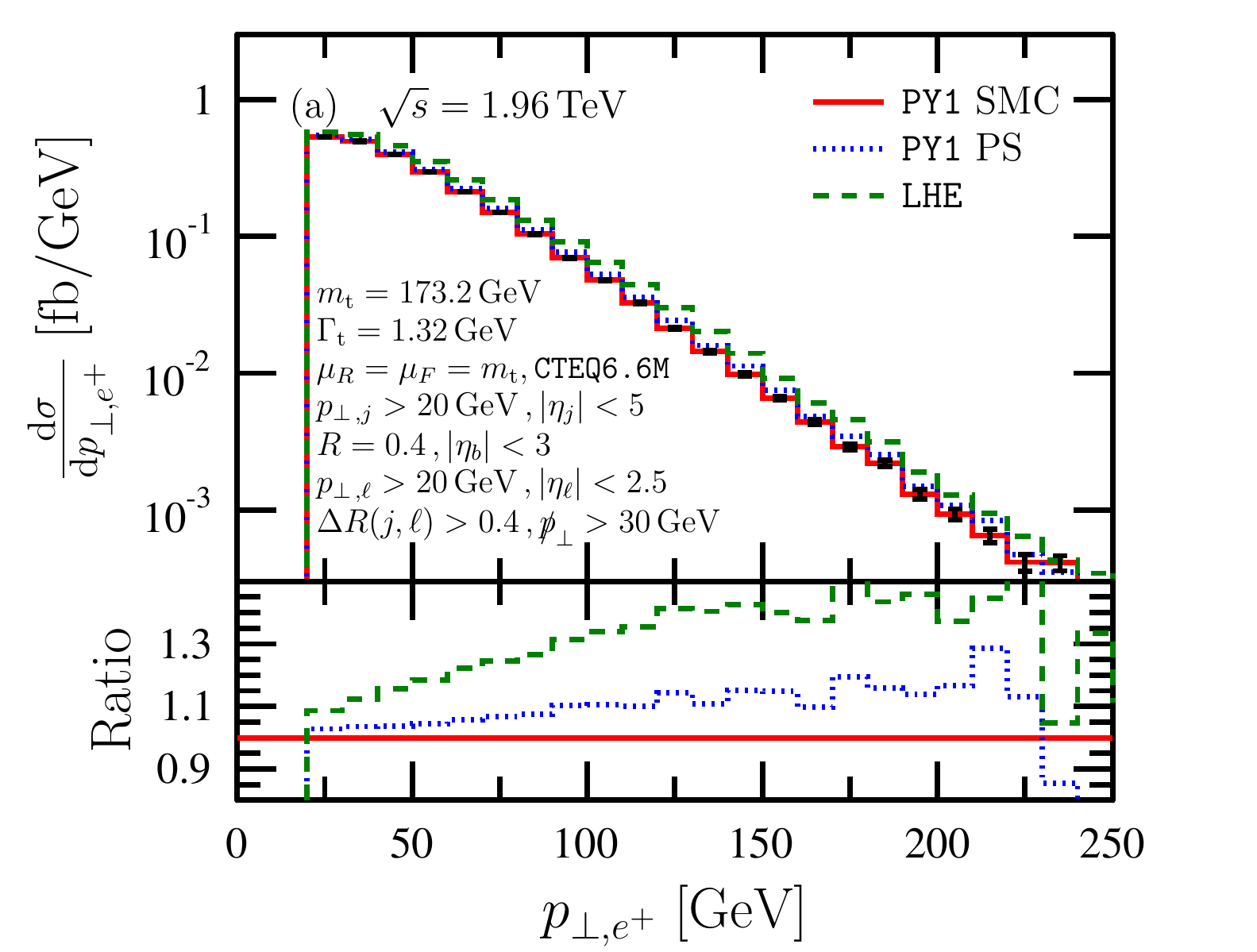}
\includegraphics[width=0.49\textwidth]{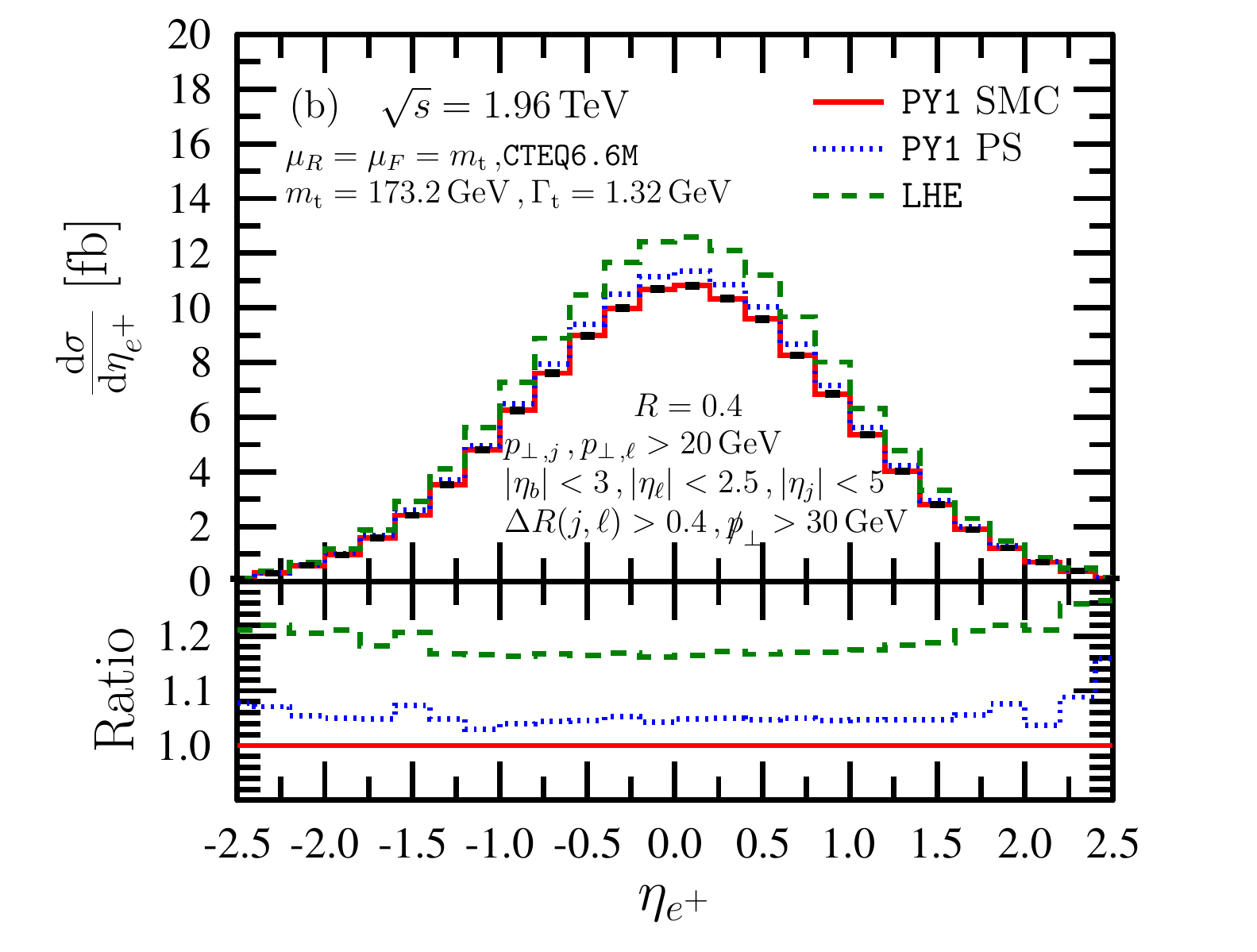}
\caption{\label{fig:tev-ep-jveto} Distributions of
a) transverse momentum and
b) pseudorapidity of the hardest isolated positron 
from the LHEs, after PS and after full SMC with \py1 at the TeVatron.
The lower inset shows the ratio of the predictions to the full SMC case.}
\end{center}
\end{figure}

\begin{figure}
\begin{center}
\includegraphics[width=0.49\textwidth]{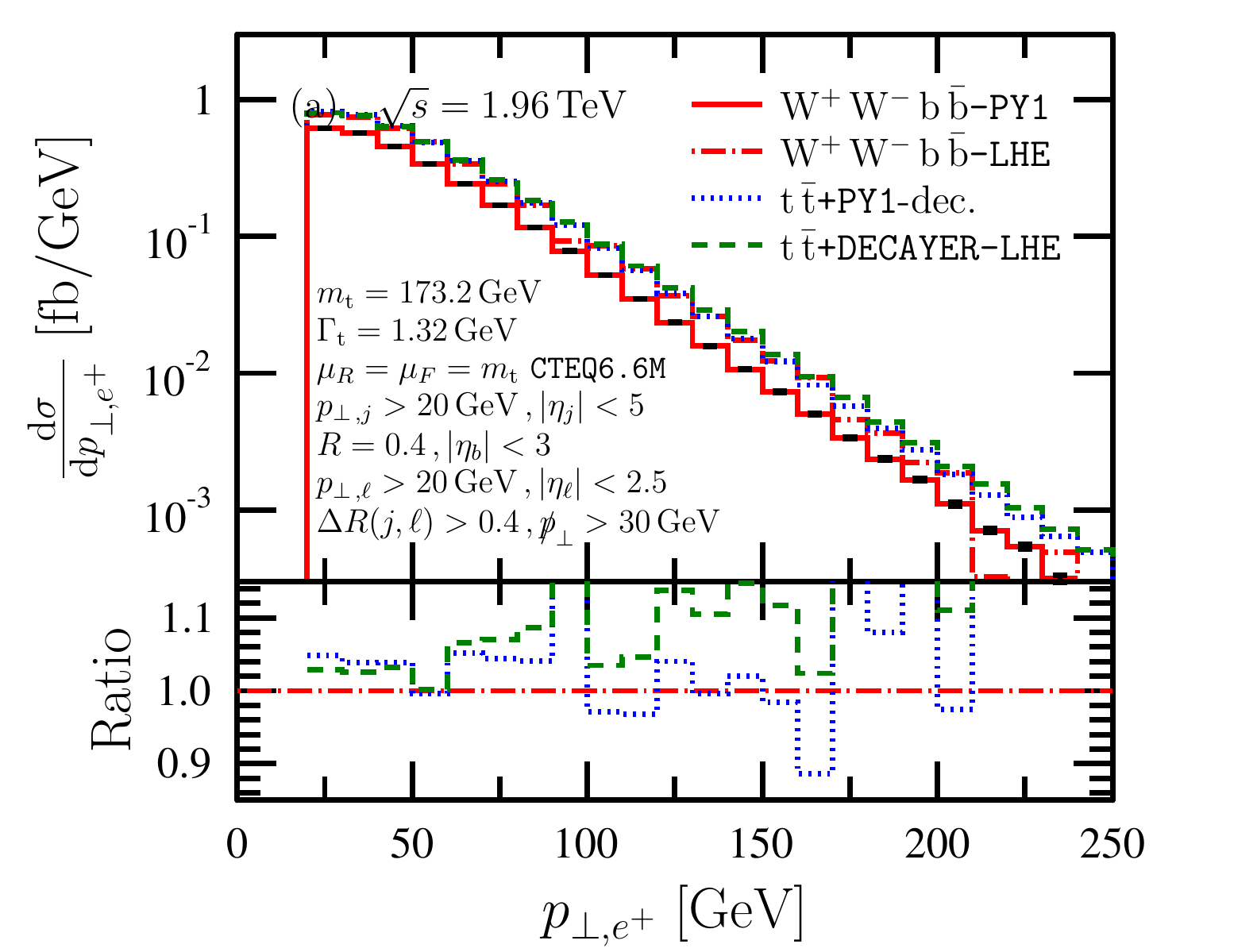}
\includegraphics[width=0.49\textwidth]{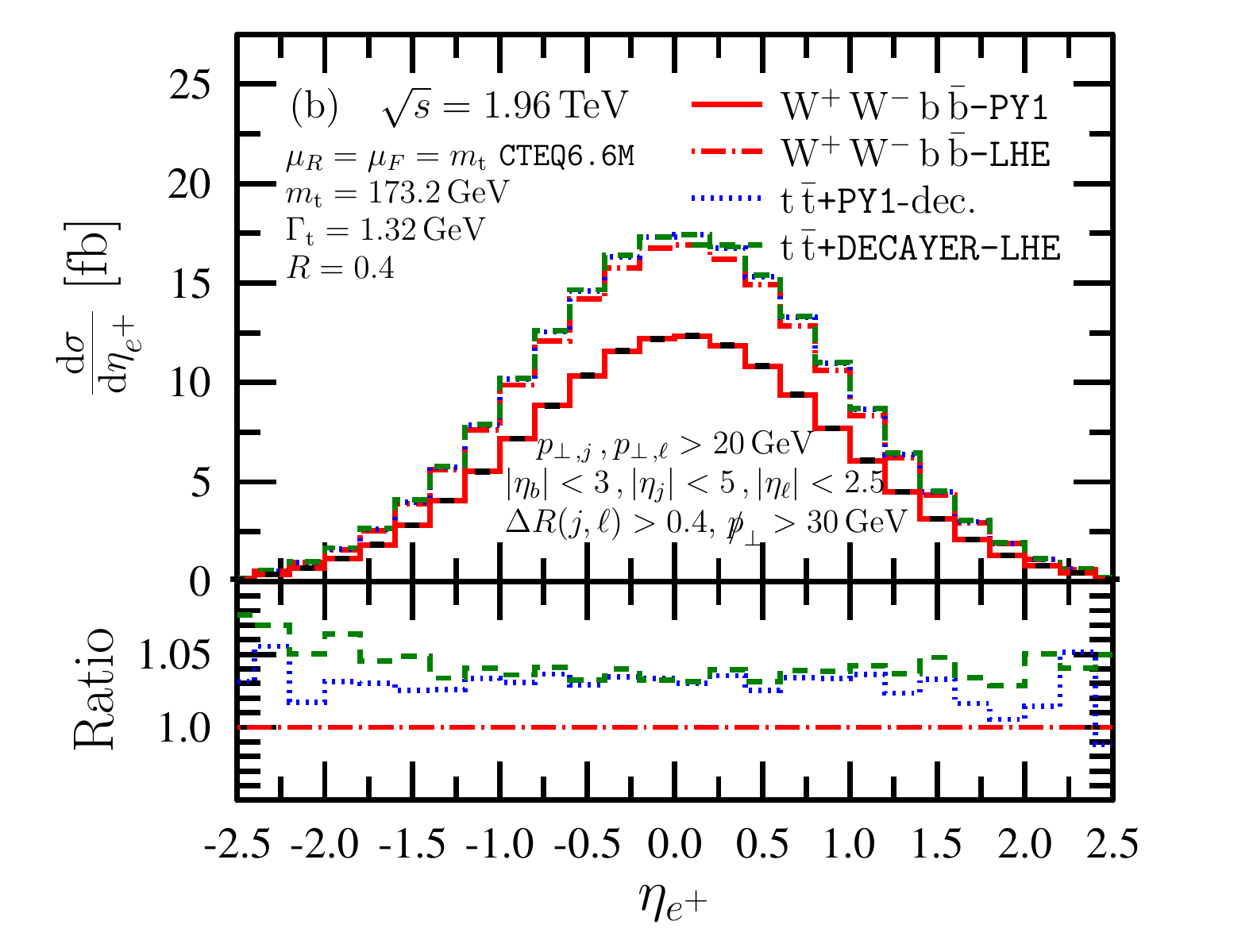}
\caption{\label{fig:tev-ep-lhe} Distributions of
a) transverse momentum and
b) pseudorapidity of the hardest isolated positron 
from the LHEs for the three cases the TeVatron.
The lower inset shows the ratio of the predictions with decay
to the \WWbB-prediction.}
\end{center}
\end{figure}

\begin{figure}
\begin{center}
\includegraphics[width=0.49\textwidth]{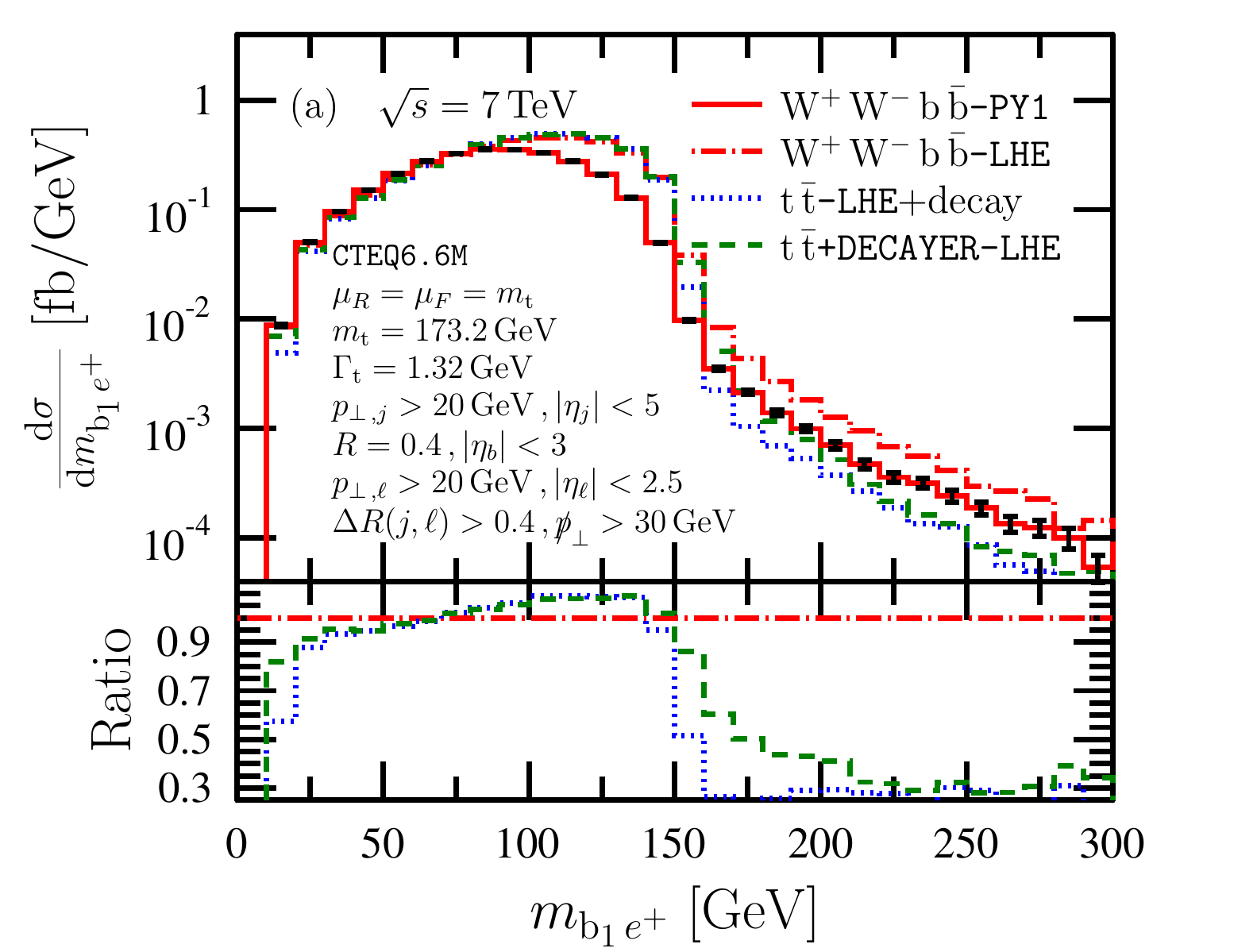}
\includegraphics[width=0.49\textwidth]{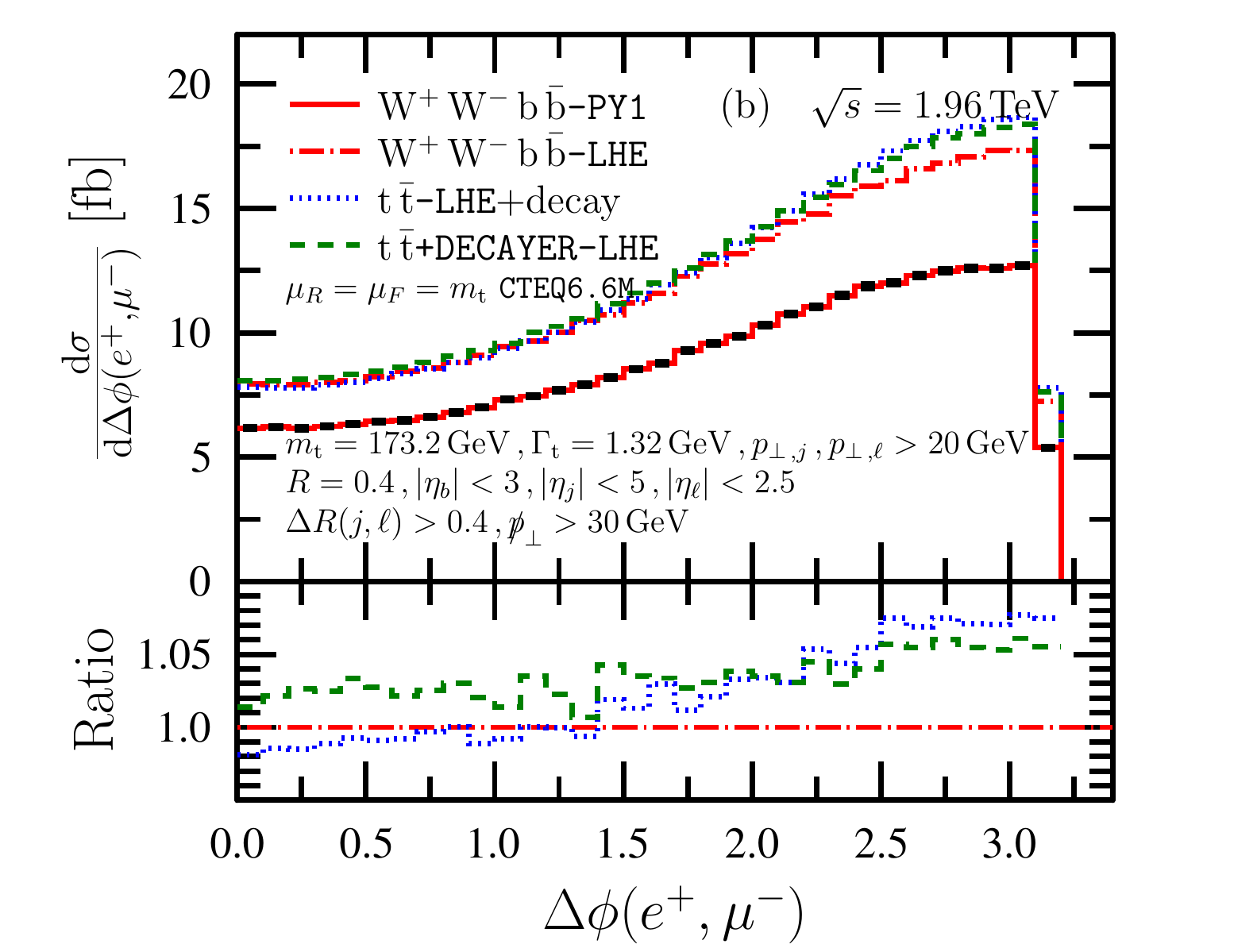}
\caption{\label{fig:tev-lhe} Distributions of
a) invariant mass of hardest b-jet and the hardest isolated positron and
b) azimuthal separation between the hardest isolated positron and muon
from the LHEs for the three cases the TeVatron.
The lower inset shows the ratio of the predictions with decay
to the \WWbB-prediction.}
\end{center}
\end{figure}

\begin{figure}
\begin{center}
\includegraphics[width=0.49\textwidth]{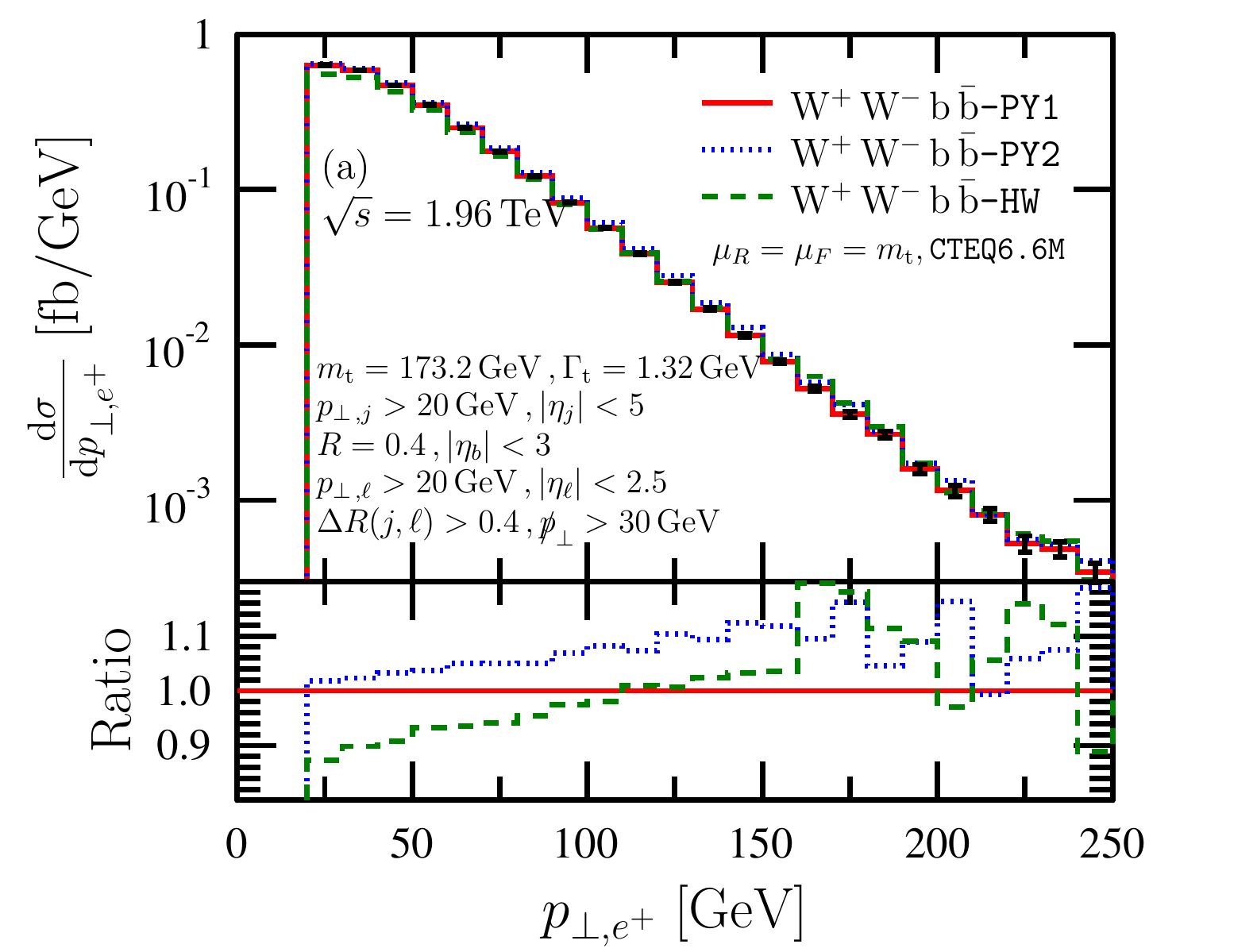}
\includegraphics[width=0.49\textwidth]{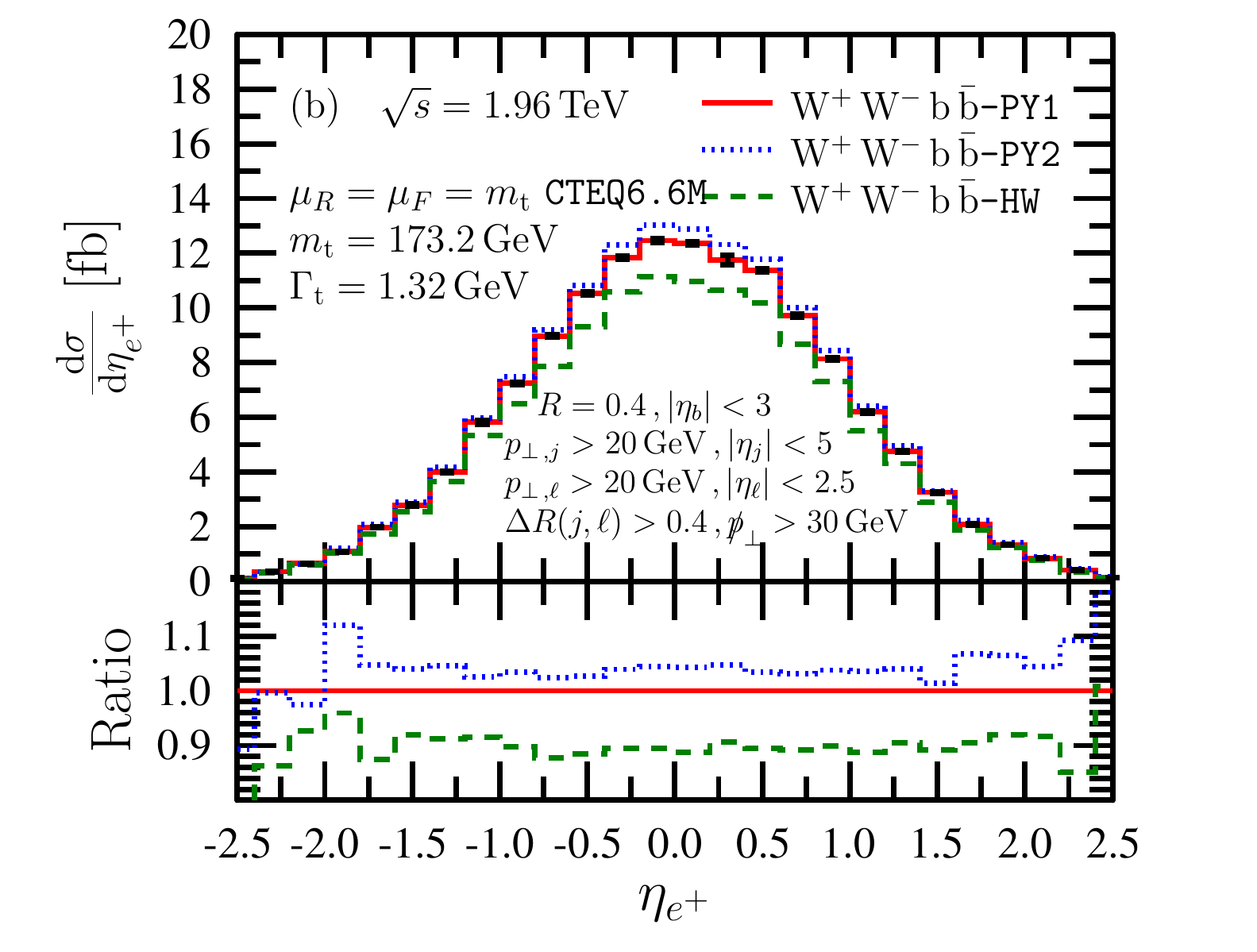}
\caption{\label{fig:tev-ep-smc} Distributions of
a) transverse momentum and
b) pseudorapidity of the hardest isolated positron 
after full SMC with three different SMC codes. The lower inset shows
the ratio of the predictions with \py2 and \hw\ to that with \py1.}
\end{center}
\end{figure}

We also studied the same process at the TeVatron. We do not repeat the
whole analysis as some of the conclusions are the same or very similar
to those drawn for the LHC, only present examples to illustrate the
following general observations:
\begin{itemize}
\itemsep=-2pt
\item
The effect of PS and hadronization on the shapes of the distributions
is very similar as found at the LHC, only their sizes are larger, which
is expected due to the smaller colliding energies
(examples are shown in \fig{fig:tev-ep-jveto}). As for the inclusive cross
section after cuts, we found the cross section values given in
\tab{tab:sigmaTeV}, obtained with \py1 SMC.
\item
Similarly to the LHC case, the singly- and non-resonant contributions
have a small effect (about 1\,\%), but this time negative (the NWA
predictions being larger) \cite{Denner:2012mx}. Having this in mind
and looking at \figs{fig:tev-ep-lhe}{fig:tev-lhe}.b, we find that
predictions with decays describe the kinematic distributions
better at the TeVatron than at the LHC, the effect of the missing NLO
corrections in the decays being a fairly uniform decrease of about 4\,\%.
There are some exceptions, one shown in \fig{fig:tev-lhe}.a. Comparing
\fig{fig:mb1ep-lhe}.a and \fig{fig:tev-lhe}.a, we see that the
resonant contributions and NLO corrections in the decay are similar
at the TeVatron as at the LHC in the hard tail of the
$m_{\bq_1 e^+}$-distribution. 
\item
The dependence of the predictions on the three SMC codes, \py1, \py2
and \hw\ is very similar as found at the LHC (for instance, compare
the two plots in \fig{fig:tev-ep-smc} to \fig{fig:pt-smc1}.b and
\fig{fig:pt-smc2}.b).
\end{itemize}
\begin{table}
\begin{center}
\begin{tabular}{|c|c|c|}
\hline
\hline
 & cuts (1--6) & cuts (1--6) + jet veto \\
\hline
$\sigma_\lhe$ (fb) & $37.4\pm 0.3$ & $26.7\pm 0.3$ \\
$\sigma_{\rm PS}$ (fb) & $30.7\pm 0.3$ & $23.8\pm 0.3$ \\
$\sigma_{\rm SMC}$ (fb) & $28.0\pm 0.3$ & $23.0\pm 0.3$ \\
\hline
\hline
\end{tabular}
\caption{\label{tab:sigmaTeV}Cross-sections from the LHEs, after PS and
at the hadron level at the TeVatron after cuts (1--6) (first column) and
after an additional jet veto (second column).
The quoted uncertainties are statistical only.
The PS and SMC predictions are obtained with the \py1 SMC.}
\end{center}
\end{table}

Thus we focus on making predictions at the hadron level.  For this kind
of predictions the selection cuts (1--6) were applied after shower,
hadronization, hadron decay and the application of jet algorithms. The
integrated cross-sections after cuts, in the three cases are collected
in \tab{tab:table2}, for different jet sizes (anti-\kt\ with $R=0.4$
versus anti-\kt\ with $R=1.2$).
\begin{table}
\begin{center}
\begin{tabular}{|c|c|c|c|}
\hline
\hline
R/case & \WWbB\ & DCA & \decayer\ \\
\hline
$\sigma(R=0.4)$ (fb) & $28.0\pm 0.3$& $27.44\pm 0.04$ & $27.83\pm 0.06$\\
$\sigma(R=1.2)$ (fb) & $13.5\pm 0.3$& $13.02\pm 0.04$ & $13.13\pm 0.06$\\
\hline
\hline
\end{tabular}
\caption{\label{tab:table2}Cross-sections at the hadron level at the
TeVatron after cuts (1--6) for the three cases as a function of the
$R$ parameter of the anti-\kt\ algorithm used to identify jets. The
quoted uncertainties are statistical only.
The predictions are obtained with the \py1 SMC.}
\end{center}
\end{table}
\begin{figure}
\begin{center}
\includegraphics[width=0.49\textwidth]{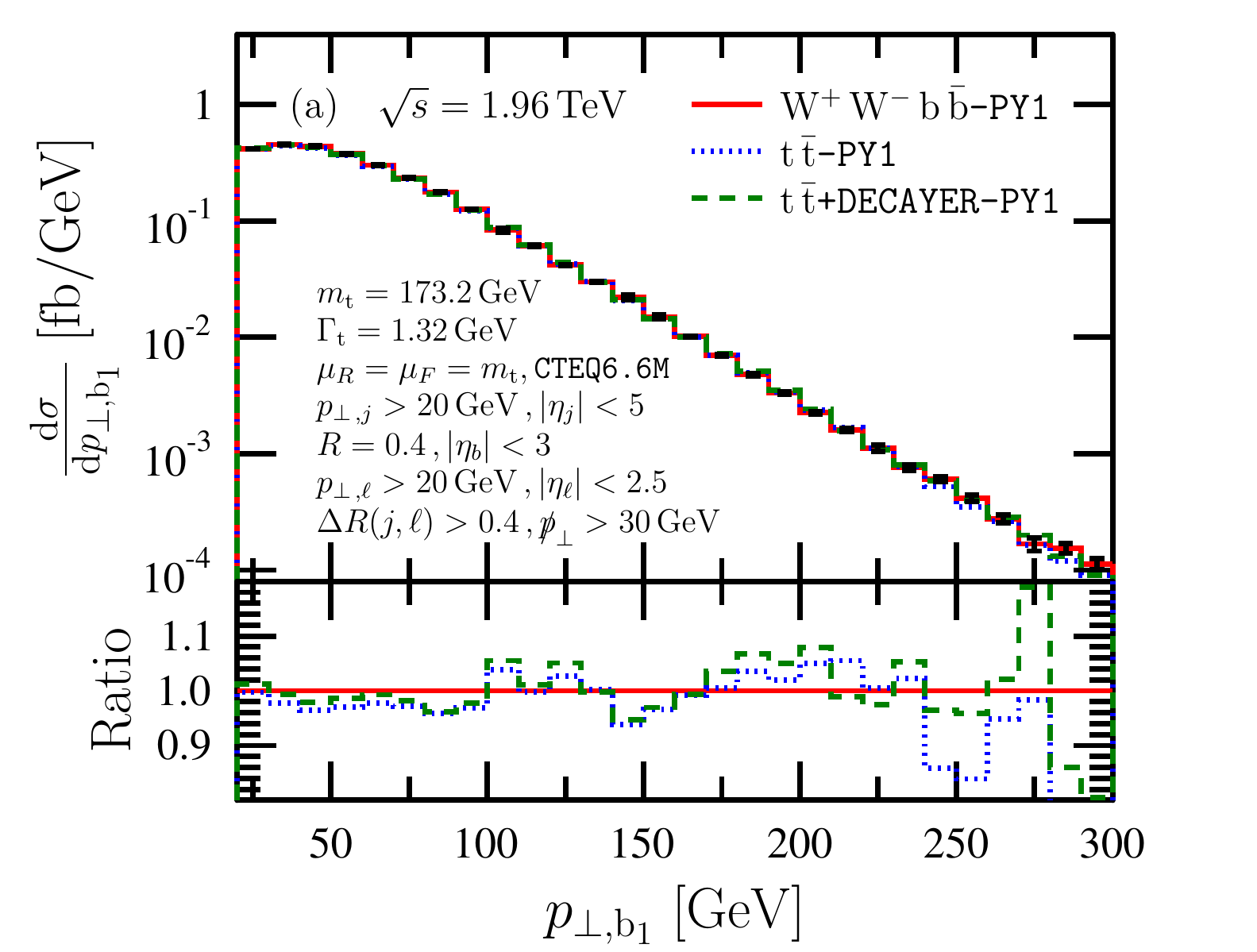}
\includegraphics[width=0.49\textwidth]{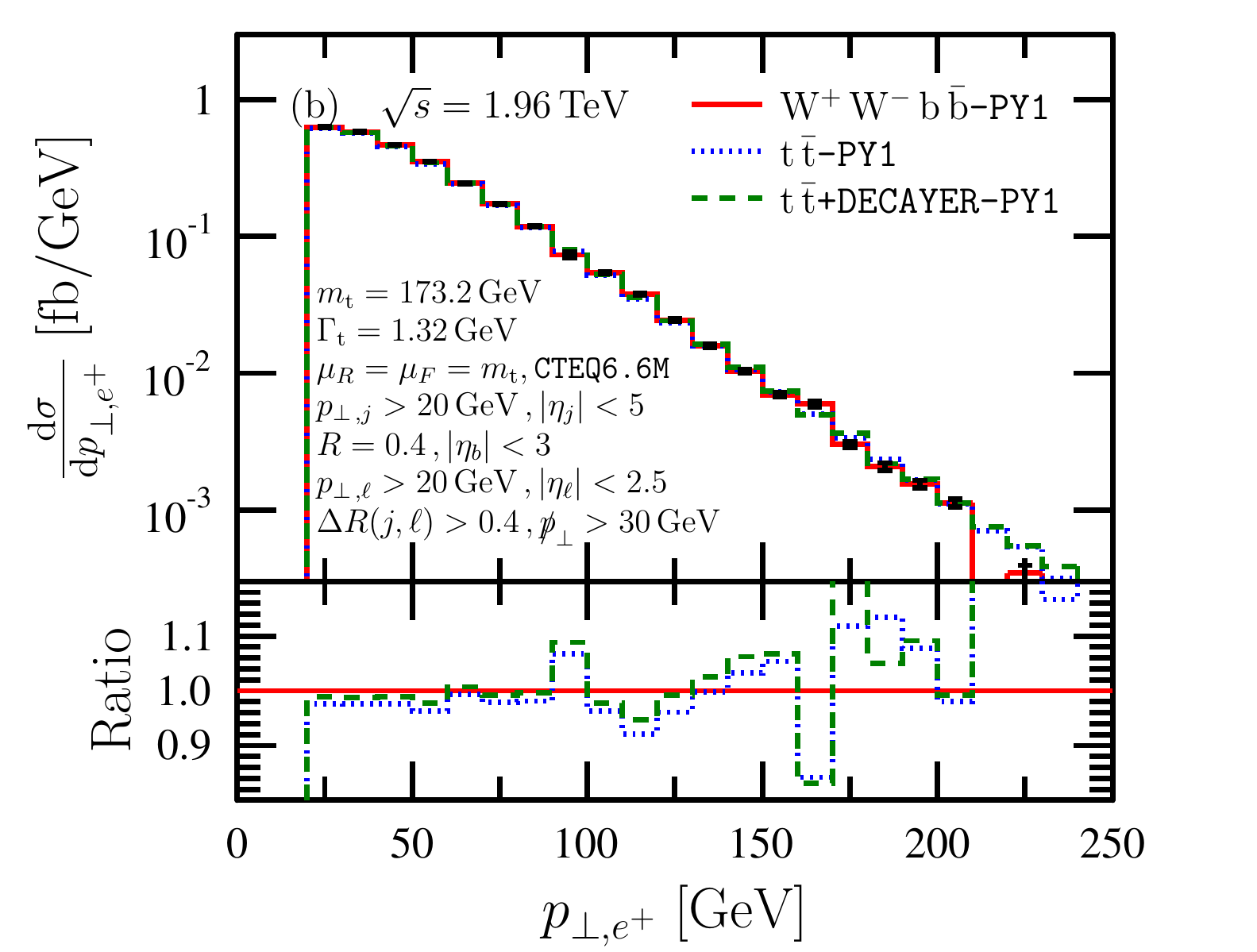}
\caption{\label{fig:tev-pt-py} Distributions of transverse momentum  of
a) the hardest b-jet and
b) the hardest isolated positron
after full SMC for the three cases.  The lower inset shows the ratio of
the predictions with decays of the t-quarks in DCA and \decayer\
compared to the complete \WWbB\ computation.}
\end{center}
\end{figure}

Turning to distributions, we show our standard selection in
\figss{fig:tev-pt-py}{fig:tev-mb1ep-py}, presented similarly to the LHC
plots.  We see in \figs{fig:tev-pt-py}{fig:tev-eta-py}
the general trend that \decayer\ and DCA give very similar predictions
both in shape and normalization for \pt- and $\eta$-distributions,
while the full \WWbB-computation followed by the SMC differs, but
the difference is smaller than in the case of LHC, usually
within 5\,\%. Thus the features seen in the distributions from the LHEs
are kept after full SMC.

\begin{figure}
\begin{center}
\includegraphics[width=0.49\textwidth]{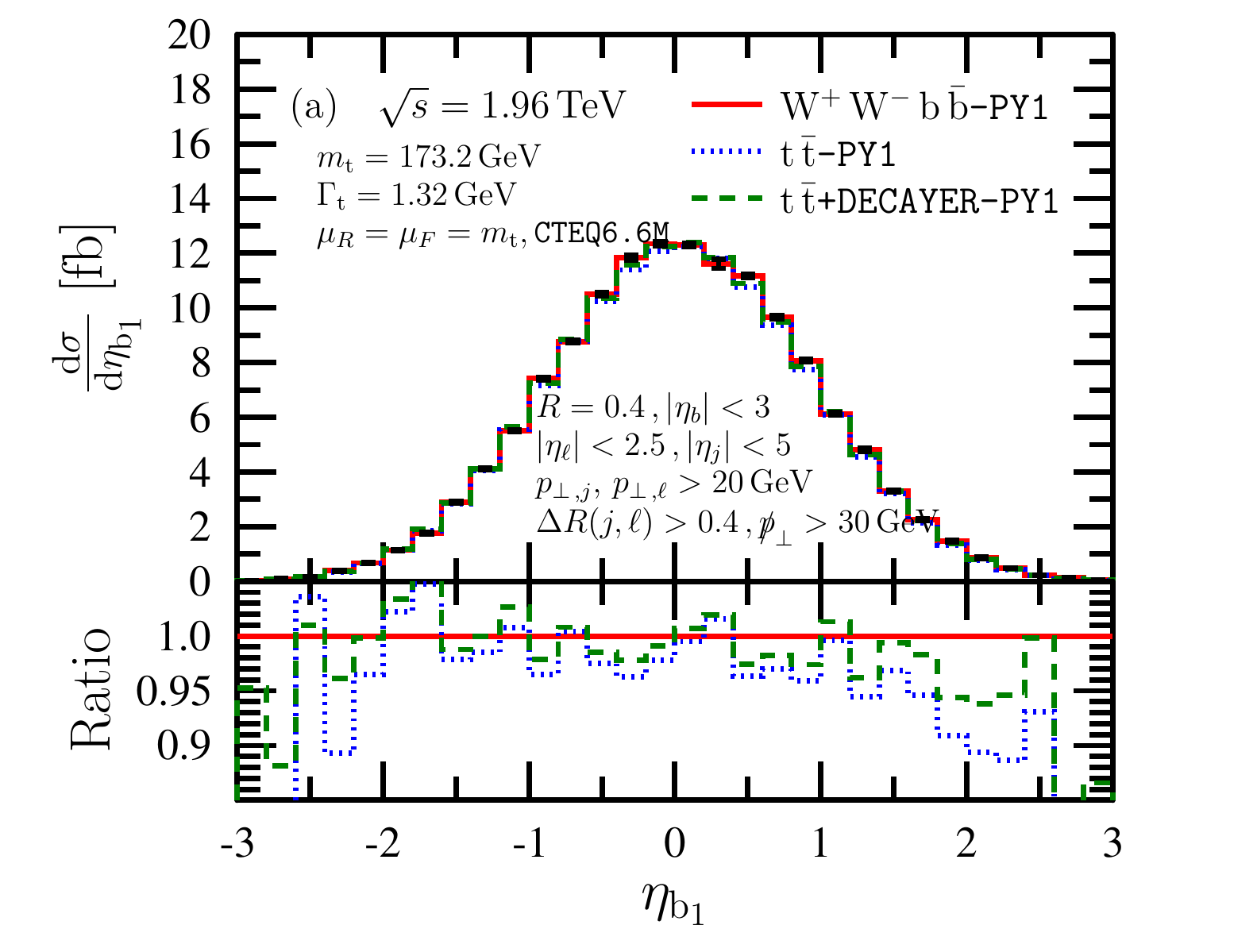}
\includegraphics[width=0.49\textwidth]{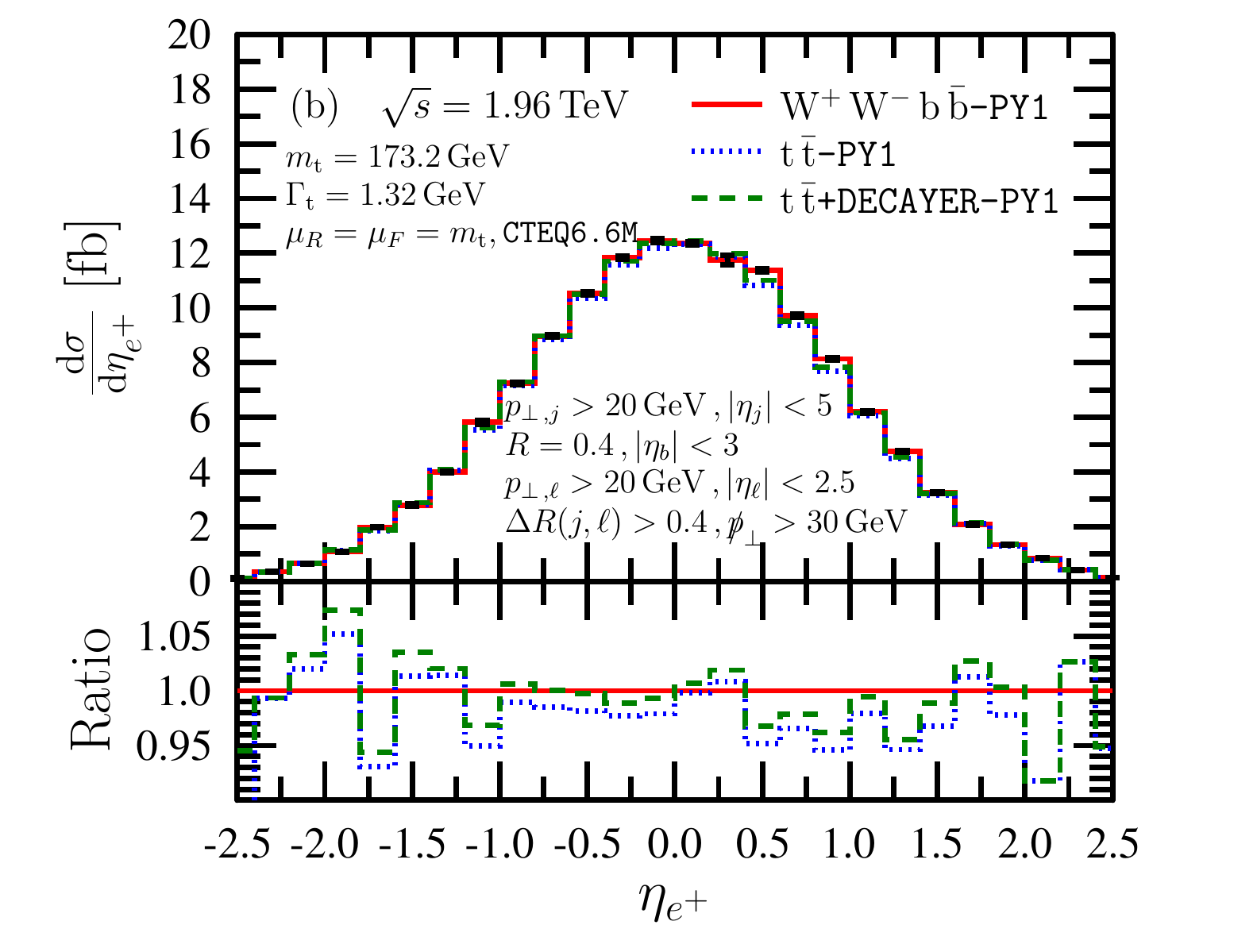}
\caption{\label{fig:tev-eta-py} Pseudorapidity distributions of
a) the hardest b-jet and
b) the hardest isolated positron
after full SMC for the three cases.  The lower inset shows the ratio of
the predictions with decays of the t-quarks in DCA and \decayer\
compared to the complete \WWbB\ computation.}
\end{center}
\end{figure}

The pseudorapidity distributions of the hardest b-jet and of the positron,
shown in \fig{fig:tev-eta-py}, exhibit a forward-backward asymmetry
computed by all three methods. However, the size of the asymmetry is
significantly bigger for \WWbB-production than that obtained from the
DCA or \decayer, although the increase is
predicted partially by \decayer.

\begin{figure}
\begin{center}
\includegraphics[width=0.49\textwidth]{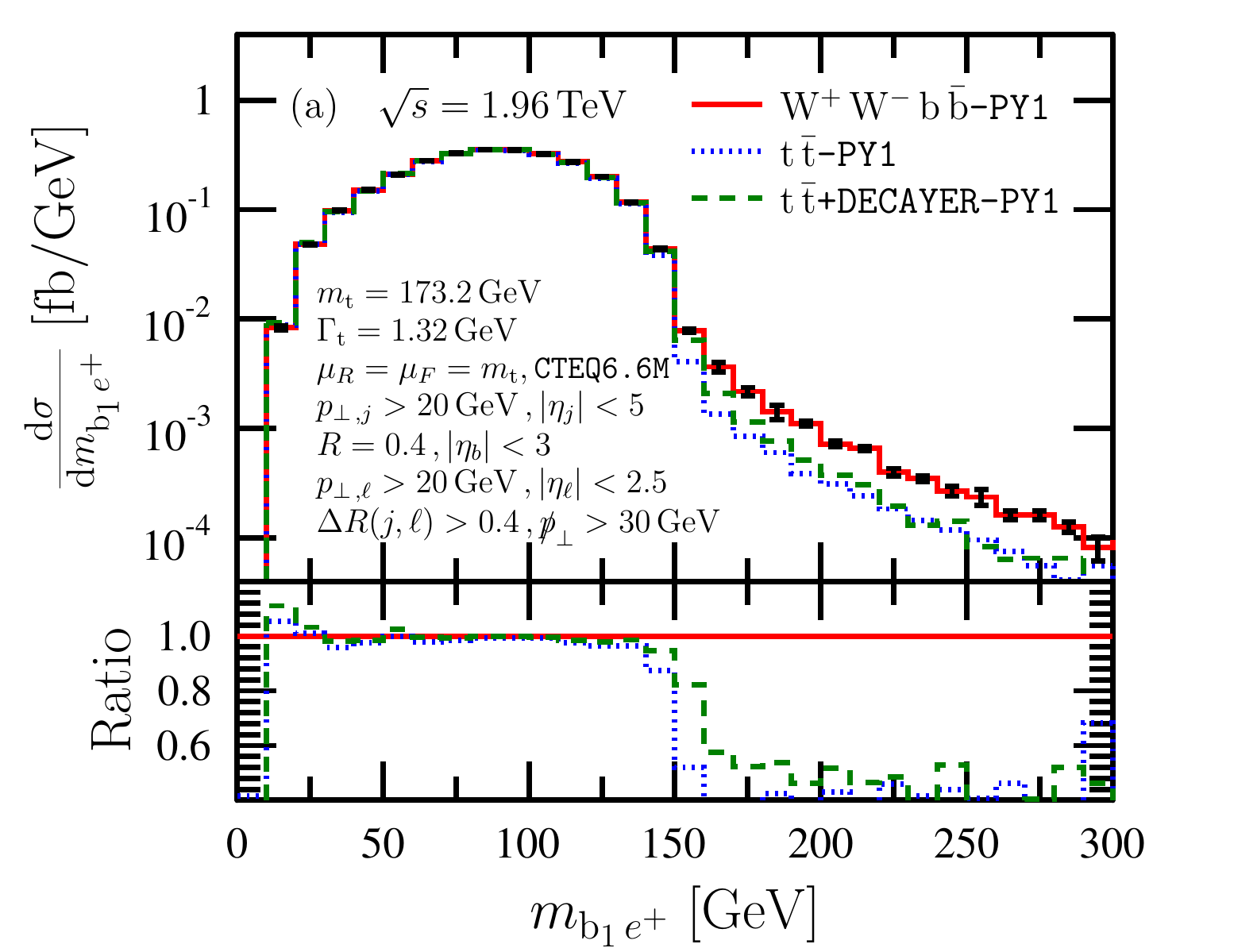}
\includegraphics[width=0.49\textwidth]{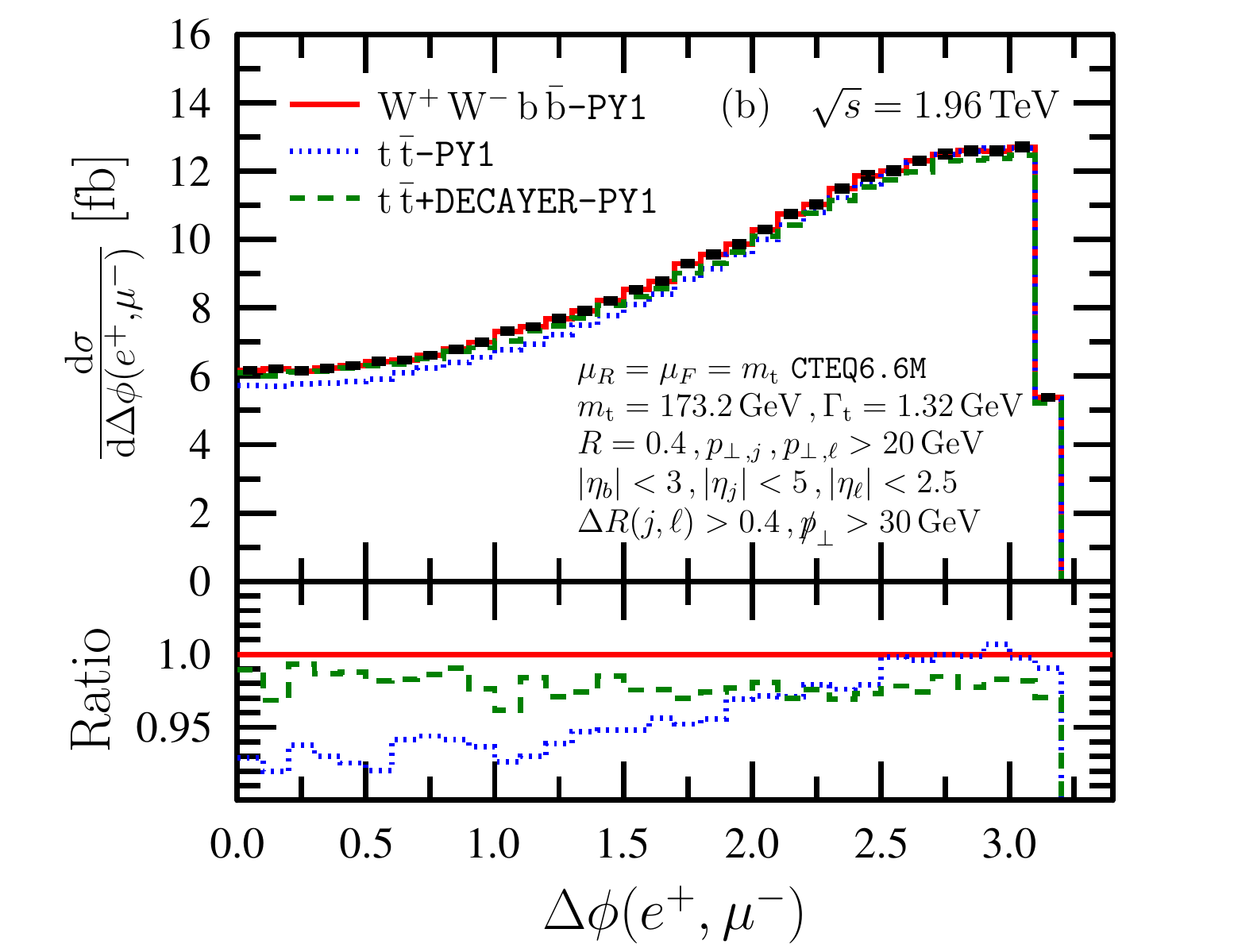}
\caption{\label{fig:tev-mb1ep-py} Distributions of
a) invariant mass of hardest b-jet and the hardest isolated positron and of
b) azimuthal separation between the hardest isolated positron and muon
after full SMC for the three cases.  The lower inset shows the ratio of
the predictions with decays of the t-quarks in DCA and \decayer\
compared to the complete \WWbB\ computation.}
\end{center}
\end{figure}

Looking at the $m_{\bq_1 e^+}$-distribution in \fig{fig:tev-mb1ep-py}.a,
we find that in the hard tail above the kinematic limit at 150\,GeV
\decayer\ and DCA approximations give very similar predictions, while that
of the full calculation is almost twice as large. Thus the trend observed
at the level of the LHE (see \fig{fig:tev-lhe}.a) survives the SMC, being
only slightly attenuated.
We present the $\Delta \phi_{e^+\mu^-}$-distribution in
\fig{fig:tev-mb1ep-py}.b.  For this plot the differences are again much
smaller than observed for the LHC (cf.~with \fig{fig:mb1ep-py}.b).

\section{Conclusions}

In this paper we presented the first study of \WWbB\ hadroproduction at
NLO accuracy matched with SMC. For the matching we used the POWHEG method
as implemented in the \powhegbox\ within the \powhel\ framework. This
framework allows the generation of events including first radiation that
can be processed further with the SMC. The events are stored in event
files according to the Les Houches accord, and are freely available for
experimental analyses. We included all Feynman-graphs in the NLO
computation, the doubly-resonant as well as singly- and non-resonant
ones.

We checked the validity of our computations by comparing cross sections
obtained from the LHEs to the NLO predictions available in the
literature and found agreement within the expected accuracy. We also
studied the effect of the parton shower and hadronization separately. The
PS decreases the cross sections by 10-20\,\% at the LHC (depending on the
observable) fairly uniformly. The hadronization results in a further
decrease, which however depends on the observable. For \pt-distribution
of hadronic objects significant amount of hadronic energy is transformed
into electromagnetic and missing energy during hadronization,
resulting in large corrections due to a softening shift of the
spectrum. This effect is not observed for other observables, such as
\pt-distributions of leptons or pseudorapidity distributions. The
effect of the hadronization on the inclusive cross section after cuts
remains below 5\,\% both for the LHC and the TeVatron, with somewhat
larger effect at the latter machine.

At the end of the SMC the events contain stable hadrons and leptons. Such
a final state can also be obtained from LHE files containing events with a
\tT-pair (including first radiation), and let the SMC decay the heavy
quarks. This much simpler computation gives predictions in the decay
chain approximation for the original heavy particles of the hard
scattering event. With a little more effort we can also perform the
decay of the heavy particles including off-shell effects of the
t-quarks and spin correlations, implemented in a new program \decayer.
An important goal in this paper was to understand the validity of these
approximations, with emphasis on the predictions after full SMC. 

We found that standard transverse-momentum and pseudorapidity
distributions are in general described by the predictions of \decayer\
and in DCA equally well, although the predictions of the former show the
off-shell effects in certain kinematic domains. Both approximations lack
the NLO corrections in the decays. The main effect of the latter is
to increase the cross section by about 10\,\% at the LHC and about
5\,\% at the TeVatron. The obvious remedy to correct for these
inaccuracies is to include the NLO corrections in the decays. However,
such a project is rather non-trivial in the NLO+PS matching framework
and is beyond our present scope.

We observe three types of deviations from the general trend: one that
does not require the inclusion of the NLO effects in the decays and two
that would not be cured even by decays at NLO accuracy.

The first kind of deviation is related to the different treatment of
spin correlations, which is fairly well described by \decayer, but the
DCA fails, as expected. The example we discussed is the azimuthal
difference between the positron and the muon where the effect can reach
up to 20\,\% at the LHC, while it is smaller at the TeVatron. Any
related observable (e.g.~$\Delta R$-separation, angular separation
between these leptons) shows the same effect. 

The second deviation is related to the effect of singly-, and
non-resonant contributions in regions of the phase space that are less
reachable by the doubly-resonant graphs. Typical example is the
$m_{\bq_1 e^+}$-distribution above $\sqrt{m_\tq^2-m_W^2}\simeq 153$\,GeV
where there is a sharp fall of the cross section in the fixed order
predictions due to kinematic reason.

The third kind of deviation is related to the different probability of
emitting a hard jet from a t-quark and from a b-quark (treated massless
in our computation). The emission of a parton from a b-quark is dominated
by soft and collinear emissions. Such emissions usually become part of the
b-jets. As a result  the \pt-distribution of the hardest non-b jets is
much softer for \WWbB-production. This effect is very significant both
at the LHC and at the TeVatron and can be observed in related
distributions that use the transverse momentum of the non-b jets.

Sets of events, both for LHC and for TeVatron, obtained with the
same parameters as used for preparing the distributions in
\sect{sec:pheno} of this paper can be downloaded from
{\texttt{http://www.grid.kfki.hu/twiki/bin/view/DbTheory/WwbbProd}}.

\section{Acknowledgements}
This research was supported by
the Hungarian Scientific Research Fund grant K-101482,
the SNF-SCOPES-JRP-2014 grant ``Preparation for and exploitation of the CMS data taking at the next LHC run'', the European Union and the European Social Fund through Supercomputer, the national virtual lab TAMOP-4.2.2.C-11/1/KONV-2012-0010
and the LHCPhenoNet network PITN-GA-2010-264564 projects.
We are grateful to G. Bevilacqua, M. Worek and A. van Hameren for useful 
discussions and help with \helacnlo{} and \oneloop. We also thank P. Skands and
G. Rodrigo for sharing their insights at the very beginning of this
project. 

\appendix

\section{The \decayer\ program}

In \sect{subsec:offshell} we presented a general method how to
reinstate off-shell effects to decays of massive particles. This
section is devoted to the special case of \tT\ production. To
reinstate off-shellness and to obtain a decay kinematics the approach
is applied to LHEs. These events are generated by \powhel\ and they
can be Born-like or after first emission, the final state being a
\tT-pair or a \tT-pair plus one extra parton, respectively.  To have a
better understanding of this decay approximation it will be illustrated
on the process $\qq\,\qaq\to\tT\ g$. For this flavor structure the
undecayed phase space can be written as
\begin{align}
\ud\Phi_\mathrm{u} = \ud\Phi_3(Q;k_\tq,k_\taq,p_g)
\,,
\end{align}
while the decayed phase space is
\begin{align}
\ud\Phi_\mathrm{d} =
\ud\Phi_7(Q;k_\bq,k_{e^+},k_{\nu_e},k_\baq,k_{\mu^-},k_{\bar{\nu}_\mu},p_g)
\,.
\end{align}
For a decaying top, two consecutive  decays have to be iterated, as
the top decays into a \bq{} and a \wp{}, followed by the decay of this
massive \wp{} into a lepton-antilepton pair (in the leptonic decay
channel). Hence the decayed phase space can take the following form
\begin{align}
\ud\Phi_\mathrm{d} &= \ud\Phi_3(Q;k_\tq,k_\taq,p_g)
\frac{\ud t_\tq}{2\pi}\ud\Phi_2(t_\tq;k_\bq,k_\wp)
\frac{\ud t_\taq}{2\pi}\ud\Phi_2(t_\taq;k_\baq,k_\wm)
\cdot
\nonumber\\
&\cdot
\frac{\ud t_\wp}{2\pi}\ud\Phi_2(t_\wp;k_{e^+},k_{\nu_e})
\frac{\ud t_\wm}{2\pi}\ud\Phi_2(t_\wm;k_{\mu^-},k_{\bar{\nu}_\mu})
\,.
\end{align}
The exact formula for the two-particle phase space can be found in
\eqn{eqn:twopartPS}.  To unweight over the decayed phase space we have
to calculate $\mathcal{F}_1$ and $\mathcal{F}_2$ of
\eqns{eq:f1factor}{eq:f2factor}, respectively. While the expression for
the latter remains
the same as \eqn{eq:f2factor}, the former is written explicitly as
\begin{align}
\mathcal{F}_1 &= \frac{E_\tq\ E_\taq\ E_\wp\ E_\wm}
{\tilde{E}_\tq\ \tilde{E}_\taq\ \tilde{E}_\wp\ \tilde{E}_\wm}\cdot
\frac{|\tilde{\boldsymbol{k}}_\bq^{(\tq)}|\ |\tilde{\boldsymbol{k}}_\bq^{(\taq)}\ \tilde{\boldsymbol{k}}_{e^+}^{(\wp)}|\ |\tilde{\boldsymbol{k}}_{\mu^-}^{(\wm)}|}
{|\boldsymbol{k}_\bq^{(\tq)}|\ |\boldsymbol{k}_\bq^{(\taq)}|\ 
 |\boldsymbol{k}_{e^+}^{(\wp)}|\ |\boldsymbol{k}_{\mu^-}^{(\wm)}|}\cdot
\frac{\mt^4\ \mw^4}{t_\tq^2\ t_\taq^2\ t_\wp^2\ t_\wm^2}
\,,
\end{align}
where $\boldsymbol{k}_i^{(j)}$ is the 3-momentum of the $i$th particle
in the rest frame of the $j$th one with on-shell kinematics, while
$\tilde{\boldsymbol{k}}_i^{(j)}$ is the same quantity but with
off-shell kinematics. Making connection with \eqn{eq:f1factor} is
possible by  using \eqn{eqn:twopartPSmom}. By going from the on-shell
kinematics to the off-shell one, we keep all 3-momenta fixed, and the
four virtualities are generated according to the Breit-Wigner
distribution. In doing so, we have to set lower and upper limits of the
virtualities.  We employed the following constraints:
\beq
\mw^2\, \leq\, t_\tq^2\,,t_\taq^2\, \leq\, 2\mt^2 \,,
\qquad
0\, \leq\, t_\wp^2\, \leq\, t_\tq^2
\,,
\qquad
0\, \leq\, t_\wm^2\, \leq\, t_\taq^2
\,.
\eeq
These values are suggested by \powhegbox.

We introduced $\mathcal{F}_1$ and $\mathcal{F}_2$ to unweight over the
new decayed phase space. To limit the change introduced by the decaying
and off-shellness, the maximal value of $\mathcal{F}_1\cdot\mathcal{F}_2$
is set to 2 (this value was also suggested by \powhegbox).

By this procedure a decayed phase space can be obtained, where
off-shellness is assigned to the decaying massive particles. One
further step can be taken by the introduction of spin-correlations to
the decay. The previous phase space production was only kinematic one.
By the unweighting each kinematic configuration becomes equally
favorable. However, spin-correlations affect the probability of these
configurations. To include spin-correlations, tree-level matrix
elements are needed for the undecayed and decayed configurations. The
uncorrelated squared matrix element is
\begin{align}
\bsp
|\mathcal{M}_{\mathrm{uncorr}}|^2 &=  
\frac{1}{2 s}|\mathcal{M}_{\mathrm{u}}|^2\cdot
|\mathcal{M}_{\tbWp}|^2\cdot|\mathcal{M}_{\tbWm}|^2\cdot
\\[2mm]
&\cdot|\mathcal{M}_{\Wpln}|^2\cdot|\mathcal{M}_{\Wmln}|^2\cdot
\mathcal{F}_{\mathrm{BW}}(s_\tq,\mt,\Gamma_\tq)\cdot
\mathcal{F}_{\mathrm{BW}}(s_\taq,\mt,\Gamma_\taq)\cdot
\\[2mm]
&\cdot
\mathcal{F}_{\mathrm{BW}}(s_\wp,\mw,\Gamma_\wp)\cdot
\mathcal{F}_{\mathrm{BW}}(s_\wm,\mw,\Gamma_\wm)
\,,
\esp
\end{align} 
where $\mathcal{M}_{\mathrm{u}}$ is the matrix element for
$\tT\, g$ production, $\mathcal{M}_{\tbWp}$, $\mathcal{M}_{\tbWm}$,
$\mathcal{M}_{\Wpln}$ and $\mathcal{M}_{\Wmln}$ are the matrix elements
for t, anti-t, \wp{} and \wm{} decays, respectively.
The original undecayed phase space is used for computing
$\mathcal{M}_{\mathrm{u}}$, while the matrix elements concerning the
decays use the appropriate momenta from the decayed phase space.
The partonic flux factor is included in this matrix element.
According to \Ref{Frixione:2007zp} there exists a
$\mathcal{U}_\mathrm{max}$ upper bound, such that
\beq
\frac{|\mathcal{M}_{\mathrm{corr}}|^2}{|\mathcal{M}_{\mathrm{uncorr}}|^2}
\, \leq\, \mathcal{U}_\mathrm{max}
\,,
\eeq
where $\mathcal{M}_{\mathrm{corr}}$ is the matrix element corresponding to 
$\bq\ \baq\ e^+\ \mu^-\ \nu_e\ \bar{\nu}_\mu\ g$ production, with 
$\bq\ e^+\ \nu_e$ and $\baq\ \mu^-\ \bar{\nu}_\mu$ coming from t- and
\taq-decays, respectively. The upper bound can be calculated
analytically, or, like in our case, the decay is started with a pre-defined
$\mathcal{U}_\mathrm{max}$
value and, if the event under consideration is proved to have a larger
value, the upper bound is replaced by this larger one.  All matrix
elements needed to perform the decays are taken from
\helacphegas.  


\bibliographystyle{JHEP}
\bibliography{wwbb}

\end{document}